\documentclass[11pt]{article}
\usepackage{color}
\usepackage{amssymb}   %basic
\usepackage{amsthm}    %theoremstyle
\usepackage{amsmath}   %\mathcal
\usepackage{stmaryrd}  %\interleave
\usepackage{titletoc}  %\ttl
\usepackage{mathrsfs}  %\mathscr
\usepackage{graphicx}
\usepackage{multicol}
\usepackage{multirow}
\usepackage{subcaption}
\usepackage{caption}
\usepackage{float}
\usepackage{hyperref}
\usepackage{cleveref}
\usepackage{xcolor}
\usepackage{hyperref}
\usepackage[english]{babel}
\addto\extrasenglish{
  \def\sectionautorefname{Section}
}
\addto\extrasenglish{
  \def\tableautorefname{Table}
}
\addto\extrasenglish{
  \def\figautorefname{Figure}
}
\usepackage{comment}

\hypersetup{
  colorlinks   = true, %Colours links instead of ugly boxes
  urlcolor     = red, %Colour for external hyperlinks
  linkcolor    = blue, %Colour of internal links
  citecolor   = blue %Colour of citations
}
\usepackage[round]{natbib}   % omit 'round' option if you prefer square brackets
\newlength{\defbaselineskip}
\newcommand{\setlinespacing}[1]%
           {\setlength{\baselineskip}{#1 \defbaselineskip}}
\newcommand{\doublespacing}{\setlength{\baselineskip}%
                           {2.0 \defbaselineskip}}
\newcommand{\singlespacing}{\setlength{\baselineskip}{\defbaselineskip}}

\renewcommand{\baselinestretch}{1.1}

\renewcommand{\theequation}{\thesection.\arabic{equation}}
\makeatletter\@addtoreset{equation}{section} \makeatother
\DeclareMathOperator*{\argmin}{arg\,min}
\DeclareMathOperator*{\argmax}{arg\,max}
 \allowdisplaybreaks
\begin{document}
\title{Variable selection in the joint frailty model of recurrent and terminal events using Broken Adaptive Ridge regression}

\author{Christian Chan$^1$  \and Fatemeh Mahmoudi$^2$ \and Chel Hee Lee$^3$ \and Quan Long$^{4,1}$ \and Xuewen Lu$^1$ \\
\small{$^1$ Department of Mathematics and Statistics, University of Calgary} \\
\small{$^2$ Department of Mathematics and Computing, Mount Royal University} \\
\small{$^3$ Department of Community Health Sciences, University of Calgary} \\ 
\small{$^4$ Department of Biochemistry and Molecular Biology, University of Calgary } \\
}

\maketitle

\begin{abstract}
We introduce a novel method to simultaneously perform variable selection and estimation in the joint frailty model of recurrent and terminal events using the Broken Adaptive Ridge Regression penalty. The BAR penalty can be summarized as an iteratively reweighted squared $L_2$-penalized regression, which approximates the $L_0$-regularization method. Our method allows for the number of covariates to diverge with the sample size. Under certain regularity conditions, we prove that the BAR estimator implemented under the model framework is consistent and asymptotically normally distributed, which are known as the oracle properties in the variable selection literature. In our simulation studies, we compare our proposed method to the Minimum Information Criterion (MIC) method. We apply our method on the Medical Information Mart for Intensive Care (MIMIC-III) database, with the aim of investigating which variables affect the risks of repeated ICU admissions and death during ICU stay.
\end{abstract}


\section{Introduction}
In biomedical studies that follow patients over a period of time, recurrent occurrences of the same event may be observed. Examples include multiple hospitalizations, repeated asthma attacks, and multiple opportunistic infections in HIV/AIDS studies. Correlation exists among the event times for the same subject, and discounting it leads to biased estimates \citep{lawless1995some}. Since then, there have been many important papers published on the analysis of recurrent event data \citep{lin1998accelerated, zeng2007semiparametric, liu2014accelerated}. Furthermore, the sequence of recurrent events could be stopped by a terminal event, like death. Therefore, recurrent event data is subject to either a dependent terminal event or an informative dropout, which has a non-negligible impact on the recurrent events \citep{ghosh2002marginal, cook2007statistical}. An example of this phenomena is patients experiencing multiple admissions into the hospital or ICU, which may be stopped by death or censoring. To model the dependence between the recurrent event history to the terminal event, we use the joint frailty model of recurrent and terminal events \citep{liu2004shared}, which models the dependence between the recurrent and terminal events through a shared random frailty term. Furthermore, many variables are measured in biomedical studies. Hence, selecting only the important variables becomes an important task in order to improve the interpretability and efficiency of the model.

One of the early variable selection techniques is the best subset selection (BSS) method, which uses the $L_0$ penalty. However, due to discrete counting nature of the $L_0$ penalty which penalizes the cardinality of the model, the BSS method becomes computationally expensive for even a moderately large set of variables. The $L_0$ penalty is non-convex, implying it is computationally difficult to find a global solution. To solve this issue, \citet{tibshirani1996regression} introduced the LASSO penalty, which is an $L_1$-norm penalty and reformulates the discrete optimization problem into a convex optimization problem. Since then, many other penalty functions have been introduced, such as the Adaptive LASSO \citep{zou2006adaptive}, Elastic Net \citep{zou2005regularization} and SCAD \citep{fan2001variable}. 

A variable selection method recently introduced to the variable selection literature is the minimum information criterion (MIC). The MIC method approximates the $L_0$-norm by using a re-parameterized hyperbolic tangent function, such that sparsity is enforced at some zero regression coefficients. First implemented under the Cox model for right-censored data \citep{su2016sparse}, it has also been implemented in GLM \citep{su2018sparse}, the random two-part model \citep{han2019variable}, and the joint frailty model of recurrent and terminal events \citep{han2020variable}. However, the MIC method has its drawbacks. Although the performance of the MIC method was good as demonstrated in \citet{han2020variable}, the authors did not provide asymptotic properties for using the MIC method from a theoretical perspective. The authors also only considered the case when the number of covariates is fixed. Therefore, these two reasons motivate us to consider a different variable selection method and investigate its theoretical properties.

Another recent variable selection method introduced to the already comprehensive literature is the Broken Adaptive Ridge (BAR) penalty method. First introduced by \citet{liu2016efficient}, the BAR penalty can be summarized as a reweighted squared $L_2$-penalized regression which approximates the $L_0$-norm, where the estimator is taken at the limit of the algorithm. Since then, many papers have investigated the BAR method for different models and data types, including the linear model \citep{dai2018broken}, the Cox PH model with large-scale right-censored survival data \citet{kawaguchi2020surrogate}, the additive hazards model with recurrent event data \citep{zhao2018variable}, the Cox PH model with interval-censored data \citep{zhao2019simultaneous}, the partly linear Cox PH model with right-censored data \citep{wu2020variable}, and the accelerated failure time model with right-censored data \citep{sun2022broken}, among others. More recently, \citet{mahmoudi2022penalized} incorporated the BAR method for semi-competing risks data under the illness-death model, and \citet{chan2023broken} proposed the BAR method for generalized partly linear models. Previous work \citep{dai2018broken,zhao2018variable,kawaguchi2020surrogate} have established that the BAR method possesses desired large-sample properties: consistency for variable selection, sparsity and asymptotic normality, which are collectively called oracle properties in the literature.  

Our motivation for this study is from the MIMIC-III (Medical Information Mart for Intensive Care) database \citep{johnson2016mimic}, which contains de-identified health-related data associated with over forty thousand patients who stayed in critical care units of the Beth Israel Deaconess Medical Center between 2001 and 2012. During the study period, patients may have been re-admitted into the critical care units of the aforementioned hospital, and some patients died during their hospital stay. The database contains information about demographics, vital-sign measurements taken at every hour, laboratory test results, procedures, medications, length of stay at each hospitalization and intensive care unit (ICU) admission, and mortality. The database also contains information regarding multiple hospitalization and ICU admissions.

The contributions of our work in this article is as follows. \textbf{First}, we implement the BAR method under the joint frailty model of recurrent and terminal events for the case when the number of covariates diverges with the sample size. \textbf{Second}, under certain conditions, we prove that the oracle properties hold for the BAR penalty implemented in the joint frailty model of recurrent and terminal events. To the best of our knowledge, \citet{han2020variable} is the only published work on variable selection in joint frailty models of recurrent and terminal events, where the MIC penalty was used but without any theoretical justification such as asymptotic properties. The MIC penalty is proven to have oracle properties in \citet{su2016sparse}, but only under the fixed number of covariates case. \textbf{Third}, in our extensive simulation studies, we show that the BAR method performs better than the MIC method. Moreover, we demonstrate that the performance of the BAR method is not sensitive to the choice of initial values of the parameters, while the MIC method is. \textbf{Fourth}, we apply our method on the MIMIC-III database, with the aim of discovering relevant variables that affect the risks of dying during hospital stay and recurrent ICU admissions. We make our code for the simulation study publicly available at \url{https://github.com/chrischan94/Broken-Adaptive-Ridge-Joint-frailty-model}.

The rest of this article is organized as follows. In Section \ref{MM1}, we introduce the necessary notation and framework of the joint frailty of recurrent and terminal events model and it's likelihood function.  We give a detailed outline of our proposed algorithm by substituting the likelihood function with a least-squares approximation. We also establish the oracle properties of the BAR variable selection method. In Section \ref{SS1}, we present the results of our simulation studies, comparing the BAR method to the MIC method, under a few realistic scenarios. In particular, we examine the sensitivity of both methods to different choices of initial values. In Section \ref{rda1}, we apply our method on the MIMIC data and interpret the results. Finally, we present our discussions and conclude our findings. 

\section{Model and Methods} \label{MM1}

\subsection{Notation, model and likelihood}
Suppose there are $n$ subjects in a given study, and each subject may experience recurrences of the same event. Consider $T_{i1} < T_{i2} < \dots < T_{in_i}$ to be the sequence of recurrent event times of any given subject $i$, for $i=1,\ldots,n$. In the sequence, $n_i$ represents the total number of observed recurrent events by subject $i$, where $n_i$ is a non-negative integer. The sequence of recurrent event times is stopped by either the end of follow-up or the terminal event. Let $d_1$ and $d_2$ represent the number of covariates for the recurrent event submodel and terminal event submodel, respectively. Let $C_i$ and $D_i$ be the censoring and terminal event times, respectively. For any given subject $i$, there exists two sets of covariates, $\textbf{Z}_{i,1} = (Z_{i1,1},\ldots,Z_{id_1,1})^\top$ and $\textbf{Z}_{i,2} = (Z_{i1,2},\ldots,Z_{id_2,2})^\top$. We assume $C_i$ and $D_i$ are independent given $\textbf{Z}_{i,1}$ and $\textbf{Z}_{i,2}$. Let $Y_i = \min(C_i, D_i)$ be the observed survival time. We also denote $\delta_i = I(D_i \leq C_i)$ as the terminal event indicator. To model the dependence between the terminal event and the recurrent event history, denote $u_i$ as the random frailty term. The random frailty term $u_i$ is shared by all the events of subject $i$, and $u_i$ is commonly assumed to either follow a Normal distribution or a log-Gamma distribution, i.e., $u_i \sim \text{N}(0, \phi^2) \; \text{or} \; u_i \sim \log \Gamma(1/\phi, 1/\phi)$. Thus, the complete set of observed data is $\{(T_{i1},\ldots,T_{in_i},\delta_i, Y_i,\textbf{Z}_{i,1}, \textbf{Z}_{i,2}), i=1,\ldots,n \}$. The last recurrent event time of subject $i$ after time $T_{in_i}$ is always censored by either the terminal event or the end of follow-up, and this information is implied in the observed data.

Let $r_i(t)$ and $h_i(t)$ be the hazard function of the recurrent events and terminal event for subject $i$, respectively. Then, the joint frailty model of recurrent and terminal events is defined as
\begin{equation}\label{JFRT}
\begin{split}
r_i(t) & = r_0(t) \exp( \boldsymbol{\beta}^\top_1 \textbf{Z}_{i,1} + u_i), \\
h_i(t) & = h_0(t) \exp( \boldsymbol{\beta}^\top_2 \textbf{Z}_{i,2} + \gamma u_i),
\end{split}
\end{equation}
where $r_0(t)$ and $h_0(t)$ are the unknown baseline hazard functions of the recurrent events and terminal event, respectively. Model \eqref{JFRT} contains two sets of regression parameters: $\boldsymbol{\beta}_1 = (\beta_{1,1},\ldots,\beta_{1,d_1} )^\top$ and $\boldsymbol{\beta}_2 = (\beta_{2,1},\ldots,\beta_{2,d_2} )^\top$. In addition to the random frailty term, $\gamma$ models potential different impact the random frailty term may have on the terminal event hazard function.

Because of the presence of the random frailty term in model \eqref{JFRT}, estimation will be done using the marginal likelihood function. For the parameter vector of a full set of parameters $\boldsymbol{\theta} = (\boldsymbol{\beta}^\top_1,\boldsymbol{\beta}^\top_2,h_0(\cdot),r_0(\cdot),\gamma,\phi)^\top$, the marginal likelihood function of model \eqref{JFRT} is formulated as
\begin{equation}\label{marg}
\mathcal{L}_n(\boldsymbol{\theta}) = \prod^n_{i=1} \int^\infty_{-\infty} g_1(Y_i | u_i) g_2(Y_i | u_i) f_\phi(u_i) \; du_i,
\end{equation}
where
$$
 g_1(Y_i | u_i) =\big[ h_0(Y_i)\exp( \boldsymbol{\beta}^\top_2 \textbf{Z}_{i,2} + \gamma u_i) \big]^{\delta_i} \exp \left\{-\int^{Y_i}_0 h_0(t)\exp( \boldsymbol{\beta}^\top_2 \textbf{Z}_{i,2} + \gamma u_i) \; dt \right\}
$$
is the likelihood of the terminal event $D_i$, and 
$$
g_2(Y_i | u_i) = \left\{\prod^{n_i}_{k=1}  r_0(T_{ik})\exp(\boldsymbol{\beta}^\top_1 \textbf{Z}_{i,1} + u_i) \right\} \exp \left\{ -\int^{Y_i}_0 r_0(t) \exp( \boldsymbol{\beta}^\top_1 \textbf{Z}_{i,1} + u_i) \; dt \right\}
$$
is the likelihood of the recurrent events. A parametric approximation of the unknown and non-parametric baseline hazard functions $h_0(\cdot)$ and $r_0(\cdot)$ is required, as it creates a problem to derive an analytical solution to \eqref{marg} without it. We choose the piecewise constant functions as the approximation. The observed follow-up times are divided into $Q$ intervals. Let $t^d_q$ be the $q^{th}$ percentile of the observed follow-up time with $q=1,\ldots,Q$. The piecewise constant approximation of $h_0(t)$ is
\begin{equation}\label{ht}
\widetilde{h}_0(t) = \sum^Q_{q=1} h_q I(t^d_{q-1} \leq t <t^d_q),
\end{equation}
where $h_q > 0$ for $q=1,\ldots,Q$. Likewise, the observed recurrent event times are divided into $Q$ intervals. The piecewise constant approximation of $r_0(t)$ is
\begin{equation}\label{rt}
\widetilde{r}_0(t) = \sum^Q_{q=1} r_q I(t^r_{q-1} \leq t <t^r_q),
\end{equation}
where $r_q > 0$ for $q=1,\ldots,Q$. Let $\textbf{h} = (h_1,\ldots,h_Q )^\top$ and $\textbf{r} = (r_1,\ldots,r_Q )^\top$, the unknown baseline hazard functions in \eqref{marg} are replaced by \eqref{ht} and \eqref{rt}. Hence, for the full set of parameters $\boldsymbol{\theta}^{\ast} = (\boldsymbol{\beta}^\top_1, \boldsymbol{\beta}^\top_2,\textbf{h}^\top, \textbf{r}^\top,\gamma,\phi )^\top$ is
\begin{equation}\label{Approx}
\widetilde{\mathcal{L}}_n(\boldsymbol{\theta}^*) = \prod^n_{i=1} \int^\infty_{-\infty} \widetilde{g}_1(Y_i | u_i) \widetilde{g}_2(Y_i | u_i) f_\phi(u_i) \; du_i,
\end{equation}
where 
\begin{equation*}
\widetilde{g}_1(Y_i |u_i) = \big[ \widetilde{h}_0(Y_i)\exp( \boldsymbol{\beta}^\top_2 \textbf{Z}_{i,2} + \gamma u_i) \big]^{\delta_i} \exp\big\{ -\widetilde{H}_0(Y_i)\exp( \boldsymbol{\beta}^\top_2 \textbf{Z}_{i,2} + \gamma u_i) \big\},
\end{equation*}
where $\widetilde{H}_0(Y_i) = \sum^Q_{q=1} h_q \max\{0,\min(t^d_q - t^d_{q-1}, Y_i - t^d_{q-1}) \}$ is the approximated cumulative baseline hazard for terminal event, and
$$
\widetilde{g}_2(Y_i | u_i) = \bigg\{ \prod^{n_i}_{k=1} \widetilde{r}_0(T_{ik})\exp(\boldsymbol{\beta}^\top_1 \textbf{Z}_{i,1} + u_i) \bigg\} \exp \big\{ -\widetilde{R}_0(Y_i) \exp( \boldsymbol{\beta}^\top_1 \textbf{Z}_{i,1}+ u_i)  \big\},
$$
where $\widetilde{R}_0(Y_i) = \sum^Q_{q=1} r_q \max\{0,\min(t^r_q - t^r_{q-1}, Y_i - t^r_{q-1}) \}$ is the approximated cumulative baseline hazard for recurrent events. Following from \eqref{Approx}, the log-likelihood is
\begin{equation}\label{Approx2}
    \log \widetilde{\mathcal{L}}_n(\boldsymbol{\theta}^{\ast}) = \ell_n(\boldsymbol{\theta}^{\ast}) = \sum^n_{i=1} \log \int^\infty_{-\infty} \widetilde{g}_1(Y_i | u_i) \widetilde{g}_2(Y_i | u_i) f_\phi(u_i) \; du_i.
\end{equation}
Numerical integration techniques are still needed to obtain a solution to $\boldsymbol{\theta}^{\ast}$, as there is no closed-form solution to the integral in \eqref{Approx2}. We choose to use the Gauss-Hermite Quadrature \citep{liu1994note} to approximate \eqref{Approx2}, where more details of it are given in the Appendix.

\subsection{Simultaneous variable selection and estimation procedure} \label{BARsection}
To implement simultaneous variable selection and estimation under the joint frailty model framework, we consider the approach that minimizes the penalized likelihood function
\begin{equation*}
\ell_{pp}(\boldsymbol{\beta} | \check{\boldsymbol{\beta}}) = -2 \ell_p(\boldsymbol{\beta}) + \sum^2_{j=1} \sum^{d_j}_{k=1} P(|\beta_{j,k}|; \lambda_n) = -2 \ell_p(\boldsymbol{\beta}) + \lambda_n \sum^2_{j=1}\sum^{d_j}_{k=1} \frac{\beta^2_{j,k}}{(\check{\beta}_{j,k})^2},
\end{equation*}
where $\ell_p(\boldsymbol{\beta}) = \max_{(\textbf{h}, \textbf{r}, \gamma, \phi)} \ell_n(\boldsymbol{\theta}^*)$ is the profile log-likelihood function, $\lambda_n$ is the non-negative tuning parameter, and $\check{\boldsymbol{\beta}}$ is a consistent estimator of $\boldsymbol{\beta}$ with all non-zero components. For a given initial estimate, where $\boldsymbol{\beta} = (\boldsymbol{\beta}^\top_1, \boldsymbol{\beta}^\top_2)^\top$, the update is obtained by the following reweighted squared $L_2$-penalized regression
\begin{equation}\label{BARupdate1}
g(\check{\boldsymbol{\beta}}) = \argmin_{\boldsymbol{\beta}} \left\{ -2 \ell_p(\boldsymbol{\beta}) + \lambda_n \sum^2_{j=1}\sum^{d_j}_{k=1} \frac{\beta^2_{j,k}}{(\check{\beta}_{j,k})^2} \right\}. 
\end{equation}
To implement the BAR penalty under the framework of model \eqref{JFRT}, the log-likelihood function is approximated by the least-squares function, along with using the Newton-Raphson method to obtain updates of the regression parameters $\boldsymbol{\beta}$. Let $\boldsymbol{\phi} = ( \textbf{h}^\top, \textbf{r}^\top,\gamma,\phi)^\top$ be the vector of nuisance parameters, then the vector containing the full set of parameters can be decomposed into the vector of regression parameters and the vector of nuisance parameters, i.e., $\boldsymbol{\theta}^* = (\boldsymbol{\beta}^\top, \boldsymbol{\phi}^\top )^\top$. Let
$$
\Dot{\ell}_n(\boldsymbol{\beta} \vert \boldsymbol{\phi}) = \frac{\partial \ell_n(\boldsymbol{\beta}, \boldsymbol{\phi})}{\partial \boldsymbol{\beta}} \quad \text{and} \quad \Ddot{\ell}_n(\boldsymbol{\beta} \vert \boldsymbol{\phi}) = \frac{\partial^2 \ell_n(\boldsymbol{\beta}, \boldsymbol{\phi})}{\partial \boldsymbol{\beta} \partial \boldsymbol{\beta}^\top}
$$
be the gradient vector and the Hessian matrix, respectively. Suppose there exists $(\widetilde{\boldsymbol{\beta}}, \widetilde{\boldsymbol{\phi}})$ that satisfies $\Dot{\ell}_n(\widetilde{\boldsymbol{\beta}} \vert \widetilde{\boldsymbol{\phi}}) = \boldsymbol{0}$. Then, the second-order Taylor expansion of the log-likelihood $\ell_n(\widetilde{\boldsymbol{\beta}}, \widetilde{\boldsymbol{\phi}})$ around $\boldsymbol{\beta}$ given $\widetilde{\boldsymbol{\phi}}$ is
$$
\ell_p(\boldsymbol{\beta}) \approx \frac{1}{2} \left[\Dot{\ell}_n(\boldsymbol{\beta} \vert \widetilde{\boldsymbol{\phi}}) \right]^\top \left[\Ddot{\ell}_n(\boldsymbol{\beta} \vert \widetilde{\boldsymbol{\phi}}) \right]^{-1} \left[ \Dot{\ell}_n(\boldsymbol{\beta} \vert \widetilde{\boldsymbol{\phi}}) \right] + c_2,
$$
where $c_2$ is a constant independent of $\boldsymbol{\beta}$. Let the pseudo-design matrix $\textbf{X}(\boldsymbol{\beta})$ be an upper triangular matrix that is obtained through the Cholesky decomposition of $-\Ddot{\ell}_n(\boldsymbol{\beta} | \widetilde{\boldsymbol{\phi}}) = \textbf{X}^\top(\boldsymbol{\beta}) \textbf{X}(\boldsymbol{\beta})$. And let $\textbf{Y}(\boldsymbol{\beta}) = \big[\textbf{X}^\top(\boldsymbol{\beta}) \big]^{-1}[\Dot{\ell}_n(\boldsymbol{\beta} \vert \widetilde{\boldsymbol{\phi}}) - \Ddot{\ell}_n(\boldsymbol{\beta} \vert \widetilde{\boldsymbol{\phi}}) \boldsymbol{\beta}]$ be the pseudo-response vector. Then, we have
$$
\vert \vert \textbf{Y}(\boldsymbol{\beta}) - \textbf{X}(\boldsymbol{\beta}) \boldsymbol{\beta} \vert \vert^2 = -  \left[\Dot{\ell}_n(\boldsymbol{\beta} \vert \widetilde{\boldsymbol{\phi}}) \right]^\top \left[\Ddot{\ell}_n(\boldsymbol{\beta} \vert \widetilde{\boldsymbol{\phi}}) \right]^{-1} \left[ \Dot{\ell}_n(\boldsymbol{\beta} \vert \widetilde{\boldsymbol{\phi}}) \right],
$$
where $||\cdot||$ represents the Euclidean norm. Define $\check{\boldsymbol{\beta}} = (\check{\beta}_{1,1},\ldots,\check{\beta}_{1,d_1},\check{\beta}_{2,1},\ldots,\check{\beta}_{2,d_2})^\top$ as a vector of fixed values, then \eqref{BARupdate1} is asymptotically equivalent to
\begin{equation*}
\begin{split}
g(\check{\boldsymbol{\beta}}) &= \argmin_{\boldsymbol{\beta}} \left\{ \vert \vert \textbf{Y}(\check{\boldsymbol{\beta}}) - \textbf{X}(\check{\boldsymbol{\beta}}) \boldsymbol{\beta} \vert \vert^2 + \lambda_n \sum^2_{j=1} \sum^{d_j}_{k=1} \frac{\beta^2_{j,k}}{\check{\beta}^2_{j,k}} \right\} \\
& = \left\{\textbf{X}^\top(\check{\boldsymbol{\beta}})\textbf{X}(\check{\boldsymbol{\beta}}) + \lambda_n \textbf{D}(\check{\boldsymbol{\beta}})\right\}^{-1}\textbf{X}^\top(\check{\boldsymbol{\beta}})\textbf{Y}(\check{\boldsymbol{\beta}}) \\
& = \left\{\boldsymbol{\Omega}_n(\check{\boldsymbol{\beta}}) + \lambda_n \textbf{D}(\check{\boldsymbol{\beta}}) \right\}^{-1} \textbf{v}_n(\check{\boldsymbol{\beta}}),
\end{split}
\end{equation*}
where $\textbf{D}(\check{\boldsymbol{\beta}}) = \text{diag}(\check{\beta}^{-2}_{1,1},\ldots,\check{\beta}^{-2}_{1,d_1}, \check{\beta}^{-2}_{2,1},  \ldots, \check{\beta}^{-2}_{2,d_2}), \; \textbf{X}(\check{\boldsymbol{\beta}}) = \textbf{X}(\boldsymbol{\beta}) \vert_{\boldsymbol{\beta} = \check{\boldsymbol{\beta}}}$, and $\textbf{Y}(\check{\boldsymbol{\beta}}) = \textbf{Y}(\boldsymbol{\beta}) \vert_{\boldsymbol{\beta} = \check{\boldsymbol{\beta}}}$. For a fixed value of $\lambda_n$, the proposed iterative BAR regression follows as below. \\
\textbf{Step 1:} At $m=0$, obtain initial estimates $ (\widehat{\boldsymbol{\beta}}^{(0)}, \widehat{\boldsymbol{\phi}}^{(0)})$. Good initial estimates are obtained by simply maximizing the un-penalized log-likelihood function, when $d_1 + d_2 < n$. \\
\textbf{Step 2:} For subsequent iterations $m \geq 1$, compute $\Dot{\ell}_n(\boldsymbol{\beta} \vert \widehat{\boldsymbol{\phi}}^{(m)}) \vert_{\boldsymbol{\beta}=\widehat{\boldsymbol{\beta}}^{(m)}}$ and $\Ddot{\ell}_n(\boldsymbol{\beta} \vert \widehat{\boldsymbol{\phi}}^{(m)})\vert_{\boldsymbol{\beta}=\widehat{\boldsymbol{\beta}}^{(m)}}$. The gradient vector is
$$
\Dot{\ell}_n(\boldsymbol{\beta} \vert \boldsymbol{\phi}) = \begin{bmatrix} \Dot{\ell}^{(1)}_n(\boldsymbol{\beta} | \boldsymbol{\phi}) \\ \Dot{\ell}^{(2)}_n(\boldsymbol{\beta} | \boldsymbol{\phi}) \end{bmatrix},
$$
where
\begin{equation*}
\Dot{\ell}^{(1)}_n(\boldsymbol{\beta} | \boldsymbol{\phi}) = \frac{\partial \ell_n(\boldsymbol{\beta}, \boldsymbol{\phi})}{\partial \boldsymbol{\beta}_1} = \sum^n_{i=1} \textbf{Z}_{i,1} \frac{\int^\infty_{-\infty} \widetilde{g}_1(Y_i | u_i) \widetilde{g}_2(Y_i | u_i) [n_i - \widetilde{R}_i(Y_i | u_i)]f_\phi(u_i) \; du_i}{\widetilde{\mathcal{L}}_{ni}(\boldsymbol{\beta}, \boldsymbol{\phi})}
\end{equation*}
and
\begin{equation*}
\Dot{\ell}^{(2)}_n(\boldsymbol{\beta} | \boldsymbol{\phi}) = \frac{\partial \ell_n(\boldsymbol{\beta}, \boldsymbol{\phi})}{\partial \boldsymbol{\beta}_2} = \sum^n_{i=1} \textbf{Z}_{i,2} \frac{\int^\infty_{-\infty} \widetilde{g}_1(Y_i | u_i) \widetilde{g}_2(Y_i | u_i) [\delta_i - \widetilde{H}_i(Y_i | u_i)]f_\phi(u_i) \; du_i}{\widetilde{\mathcal{L}}_{ni}(\boldsymbol{\beta}, \boldsymbol{\phi})},
\end{equation*}
where $\widetilde{R}_i(Y_i | u_i) = \widetilde{R}_0(Y_i)\exp(\boldsymbol{\beta}^\top_1 \textbf{Z}_{i,1}  + u_i)$ and $\widetilde{H}_i(Y_i | u_i) = \widetilde{H}_0(Y_i)\exp(\boldsymbol{\beta}^\top_2 \textbf{Z}_{i,2}  + \gamma u_i)$ are the conditional cumulative hazard functions of the recurrent and terminal events, respectively. Let $\widetilde{\mathcal{L}}_{ni}(\boldsymbol{\beta}, \boldsymbol{\phi})$ be the individual likelihood contribution of subject $i$, where $\widetilde{\mathcal{L}}_n(\boldsymbol{\beta}, \boldsymbol{\phi}) = \prod^n_{i=1} \widetilde{\mathcal{L}}_{ni}(\boldsymbol{\beta}, \boldsymbol{\phi})$. The Hessian matrix composes of four submatrices
$$
\Ddot{\ell}_n(\boldsymbol{\beta} \vert \boldsymbol{\phi}) = \begin{bmatrix} \Ddot{\ell}^{(1)}_n(\boldsymbol{\beta} | \boldsymbol{\phi}) & \Ddot{\ell}^{(12)}_n(\boldsymbol{\beta} | \boldsymbol{\phi}) \\ \Ddot{\ell}^{(21)}_n(\boldsymbol{\beta} | \boldsymbol{\phi}) & \Ddot{\ell}^{(2)}_n(\boldsymbol{\beta} | \boldsymbol{\phi}) \end{bmatrix}.
$$
The entries of the Hessian matrix are
\begin{equation*}
\begin{split}
 \Ddot{\ell}^{(1)}_n(\boldsymbol{\beta} | \boldsymbol{\phi}) = \frac{\partial^2 \ell_n(\boldsymbol{\beta}, \boldsymbol{\phi})}{\partial \boldsymbol{\beta}_1 \partial \boldsymbol{\beta}^{\top}_1} &= \sum^n_{i=1} \textbf{Z}_{i,1} \textbf{Z}^\top_{i,1} \left\{ \frac{\int^{\infty}_{-\infty} m_4(u_i; Y_i) \; du_i  \cdot \widetilde{\mathcal{L}}_{ni}(\boldsymbol{\beta}, \boldsymbol{\phi}) - [\widetilde{g}_3(Y_i)]^2}{[\widetilde{\mathcal{L}}_{ni}(\boldsymbol{\beta}, \boldsymbol{\phi})]^2} \right\}, \\
 \Ddot{\ell}^{(12)}_n(\boldsymbol{\beta} | \boldsymbol{\phi}) = \frac{\partial^2 \ell_n(\boldsymbol{\beta}, \boldsymbol{\phi})}{\partial \boldsymbol{\beta}_1 \partial \boldsymbol{\beta}^{\top}_2} &= \sum^n_{i=1}\textbf{Z}_{i,1} \textbf{Z}^\top_{i,2} \left\{ \frac{\int^{\infty}_{-\infty} m_6(u_i; Y_i) \; du_i \cdot \widetilde{\mathcal{L}}_{ni}(\boldsymbol{\beta}, \boldsymbol{\phi}) - \widetilde{g}_3(Y_i)\widetilde{g}_4(Y_i) }{[\widetilde{\mathcal{L}}_{ni}(\boldsymbol{\beta}, \boldsymbol{\phi})]^2} \right\}, \\
 \Ddot{\ell}^{(21)}_n(\boldsymbol{\beta} | \boldsymbol{\phi}) = \frac{\partial^2 \ell_n(\boldsymbol{\beta}, \boldsymbol{\phi})}{\partial \boldsymbol{\beta}_2 \partial \boldsymbol{\beta}^{\top}_1} &= \sum^n_{i=1}\textbf{Z}_{i,2} \textbf{Z}^\top_{i,1} \left\{ \frac{\int^{\infty}_{-\infty} m_6(u_i; Y_i) \; du_i \cdot \widetilde{\mathcal{L}}_{ni}(\boldsymbol{\beta}, \boldsymbol{\phi}) - \widetilde{g}_3(Y_i)\widetilde{g}_4(Y_i) }{[\widetilde{\mathcal{L}}_{ni}(\boldsymbol{\beta}, \boldsymbol{\phi})]^2} \right\}, \\
 \Ddot{\ell}^{(2)}_n(\boldsymbol{\beta} | \boldsymbol{\phi}) = \frac{\partial^2 \ell_n(\boldsymbol{\beta}, \boldsymbol{\phi})}{\partial \boldsymbol{\beta}_2 \partial \boldsymbol{\beta}^\top_2} &= \sum^n_{i=1} \textbf{Z}_{i,2} \textbf{Z}^\top_{i,2} \left\{ \frac{ \int^{\infty}_{-\infty} m_5(u_i; Y_i) \; du_i \cdot \widetilde{\mathcal{L}}_{ni}(\boldsymbol{\beta}, \boldsymbol{\phi}) - [\widetilde{g}_4(Y_i)]^2}{[\widetilde{\mathcal{L}}_{ni}(\boldsymbol{\beta}, \boldsymbol{\phi})]^2} \right\},
\end{split}
\end{equation*}
where  
\begin{equation*}
\begin{split}
\widetilde{g}_3(Y_i) &= \int^{\infty}_{-\infty} \widetilde{g}_1(Y_i | u_i) \widetilde{g}_2(Y_i | u_i) [\delta_i - \widetilde{H}_i(Y_i | u_i)] f_\phi(u_i) \; du_i, \\
\widetilde{g}_4(Y_i) &= \int^{\infty}_{-\infty} \widetilde{g}_1(Y_i | u_i) \widetilde{g}_2(Y_i | u_i) [n_i - \widetilde{R}_i(Y_i | u_i)] f_\phi(u_i) \; du_i,
\end{split}
\end{equation*}
and
\begin{equation*}
\begin{split}
m_4(u_i; Y_i) &= \widetilde{g}_1(Y_i | u_i) \widetilde{g}_2(Y_i | u_i) \left\{[n_i - \widetilde{R}_i(Y_i | u_i)]^2 - \widetilde{R}_i(Y_i | u_i) \right\} f_\phi(u_i), \\
m_5(u_i; Y_i) &= \widetilde{g}_1(Y_i | u_i) \widetilde{g}_2(Y_i | u_i) \left\{[\delta_i - \widetilde{H}_i(Y_i | u_i)]^2 - \widetilde{H}_i(Y_i | u_i) \right\} f_\phi(u_i), \\
m_6(u_i; Y_i) &= \widetilde{g}_1(Y_i | u_i) \widetilde{g}_2(Y_i | u_i) \Big[n_i - \widetilde{R}_i(Y_i | u_i) \Big] \Big[\delta_i - \widetilde{H}_i(Y_i | u_i) \Big] f_\phi(u_i).
\end{split}
\end{equation*}

\noindent
\textbf{Step 3:} For $m \geq 1$, update the estimates of $\boldsymbol{\beta}$ by
$$
\widehat{\boldsymbol{\beta}}^{(m+1)} = \big\{ \boldsymbol{\Omega}_n(\widehat{\boldsymbol{\beta}}^{(m)}) + \lambda_n \textbf{D}(\widehat{\boldsymbol{\beta}}^{(m)})\big\}^{-1} \textbf{v}_n(\widehat{\boldsymbol{\beta}}^{(m)}),
$$
where $\textbf{D}(\widehat{\boldsymbol{\beta}}^{(m)}) = \text{diag}((\widehat{\beta}^{(m)}_{1,1})^{-2},\ldots,(\widehat{\beta}^{(m)}_{1,d_1})^{-2}, (\widehat{\beta}^{(m)}_{2,1})^{-2},\ldots,(\widehat{\beta}^{(m)}_{2,d_2})^{-2})$, $\boldsymbol{\Omega}_n(\widehat{\boldsymbol{\beta}}^{(m)}) = -\Ddot{\ell}_n(\boldsymbol{\beta} \vert \widehat{\boldsymbol{\phi}}^{(m)}) \vert_{\boldsymbol{\beta} = \widehat{\boldsymbol{\beta}}^{(m)}}$ and $\textbf{v}_n(\widehat{\boldsymbol{\beta}}^{(m)}) = \Dot{\ell}_n(\boldsymbol{\beta} \vert \widehat{\boldsymbol{\phi}}^{(m)})  \vert_{\boldsymbol{\beta} = \widehat{\boldsymbol{\beta}}^{(m)}} -\Ddot{\ell}_n(\boldsymbol{\beta} \vert \widehat{\boldsymbol{\phi}}^{(m)}) \vert_{\boldsymbol{\beta} = \widehat{\boldsymbol{\beta}}^{(m)}} \widehat{\boldsymbol{\beta}}^{(m)}$. To ensure numerical stability of the calculation of $\textbf{D}(\widehat{\boldsymbol{\beta}}^{(m)})$ in each iteration, a small positive constant is added to the diagonal entries of $\textbf{D}(\widehat{\boldsymbol{\beta}}^{(m)})$, i.e.,
$$
\textbf{D}(\widehat{\boldsymbol{\beta}}^{(m)}) = \text{diag}\left( \frac{1}{(\widehat{\beta}^{(m)}_{1,1})^2 + \eta^2}, \ldots, \frac{1}{(\widehat{\beta}^{(m)}_{1,d_1})^2 + \eta^2}, \frac{1}{(\widehat{\beta}^{(m)}_{2,1})^2 + \eta^2}, \ldots, \frac{1}{(\widehat{\beta}^{(m)}_{2,d_2})^2 + \eta^2}\right)
$$
for some $\eta > 0$.
\\
\textbf{Step 4:} Given $\widehat{\boldsymbol{\beta}}^{(m+1)}$, the updated estimates of the nuisance parameters $\widehat{\boldsymbol{\phi}}^{(m+1)}$ are updated by equating $\partial \ell_n(\boldsymbol{\phi} | \widehat{\boldsymbol{\beta}}^{(m+1)})/\partial \boldsymbol{\phi} = 0$.\\
\textbf{Step 5:} Return to Step 2. Repeat the algorithm until convergence, i.e.,
$$
\widehat{\boldsymbol{\beta}} = \lim_{m \longrightarrow \infty} \widehat{\boldsymbol{\beta}}^{(m)}.
$$
In the proposed iterative method described above, the Cholesky decomposition of $-\Ddot{\ell}_n(\boldsymbol{\beta} | \widetilde{\boldsymbol{\phi}})$ is actually not needed. Only the calculation of $\boldsymbol{\Omega}_n(\boldsymbol{\beta})$ and $\textbf{v}_n(\boldsymbol{\beta})$ is required. In Step 1, common non-linear numerical optimization such as the Nelder-Mead algorithm \citep{nelder1965simplex} can be used to obtain good estimates of $\boldsymbol{\beta}$ and $\boldsymbol{\phi}$. In Step 4, given $\widehat{\boldsymbol{\beta}}^{(m+1)}$, there are no closed form updates to $\boldsymbol{\phi}$. Therefore, we utilize non-linear numerical optimization methods to find the updates of $\boldsymbol{\phi}$. It is important to note that the initial estimator $\widehat{\boldsymbol{\beta}}^{(0)}$ and subsequent updates $\widehat{\boldsymbol{\beta}}^{(m)}, \; \text{for} \; m \geq 1$, do not yield any zero coefficient, so that they can be used in the denominator of the BAR penalty. The full derivation of the gradient vector and Hessian matrix is deferred to the Appendix.

\subsection{Generalized cross-validation}
Variable selection methods usually are subjected to choosing a tuning parameter, which greatly affects the number of variables retained in the model. For likelihood-based methods, one popular method is to select $\lambda_n$ by using the AIC \citep{akaike1974new}, BIC criterion \citep{schwarz1978estimating}, or generalized cross-validation (GCV) \citep{wahba1990spline}. One of the most popular methods is by having a pre-determining a sequence of the tuning parameter $\lambda_n$, and by doing a ``grid search" to obtain the value that optimizes a criterion of choice.  To increase computational efficiency, we use the generalized cross-validation (GCV) method to select the optimal tuning parameter $\lambda_n$. We use the following procedure similar to \citet{cai2020group} in order to utilize the GCV method. We define
$$
\Sigma_{\lambda_n}(\boldsymbol{\beta}) = \lambda_n \text{diag}\left(\frac{1}{s(\beta_{1,1})},\ldots,\frac{1}{s(\beta_{1,d_1})}, \frac{1}{s(\beta_{2,1})},\ldots,\frac{1}{s(\beta_{1,d_2})}\right),
$$
with 
$$
s(\beta_{j,k}) = 
\begin{cases}
|\beta_{j,k}|, \; &\text{if} \; |\beta_{j,k}| \neq 0, \\
\epsilon, \; &\text{otherwise},
\end{cases}
$$
where $\epsilon$ is an arbitrarily small positive number. Let $\widehat{\boldsymbol{\beta}}_{\lambda_n}$ be the unbiased estimator of $\boldsymbol{\beta}$, define $d(\lambda_n) = \text{tr}[(\textbf{D}(\widehat{\boldsymbol{\beta}}_{\lambda_n}) + \Sigma_{\lambda_n}(\widehat{\boldsymbol{\beta}}_{\lambda_n}))^{-1}\textbf{D}(\widehat{\boldsymbol{\beta}}_{\lambda_n})]$. Then, the GCV criterion is
\begin{equation}\label{GCV}
\text{GCV}(\lambda_n, \widehat{\boldsymbol{\beta}}_{\lambda_n}) = -\frac{\ell_n(\widehat{\boldsymbol{\beta}}_{\lambda_n})}{n\{ 1- d(\lambda_n)/n\}^2}.
\end{equation}
Since \eqref{GCV} works only with unbiased estimators of $\boldsymbol{\beta}$, we use the following process to obtain unbiased estimators. First, we obtain the penalized BAR estimators from our algorithm described in Section \ref{BARsection}. Then, we omit the unimportant covariates, and re-estimate the remaining regression parameters in the joint frailty model, by minimizing the un-penalized likelihood function, to obtain the unbiased estimates. Finally, the optimal $\lambda_n$ is the value that minimizes \eqref{GCV} with respect to $\lambda_n$.
 
\subsection{Oracle properties of BAR estimator}
Denote $\boldsymbol{\beta}_{s0} = (\beta_{s0,1},\ldots,\beta_{s0,p_n})^\top$ as the true values of $\boldsymbol{\beta}$ with dimension $p_n = d_1 + d_2$, where $p_n$ diverges to infinity but $p_n < n$. We decompose $\boldsymbol{\beta}_{s0} = (\boldsymbol{\beta}^\top_{s01}, \boldsymbol{\beta}^\top_{s02})^\top$. Without loss of generality, we assume $\boldsymbol{\beta}_{s01}$ contains the non-zero components of $\boldsymbol{\beta}_{s0}$ with dimension $q_n$, and $\boldsymbol{\beta}_{s02}$ contains the zero components of $\boldsymbol{\beta}_{s0}$ with dimension  $p_n - q_n$. \textit{Note: To distinguish the original true parameter vector $\boldsymbol{\beta}_0$ in model \eqref{JFRT}, the subscript s in $\boldsymbol{\beta}_{s0}$ denotes the true values of $\boldsymbol{\beta}$ after grouping the non-zero and zero coefficients, respectively, where $\boldsymbol{\beta}_{s0}$ is partitioned into the vector of non-zero and zero components.} The following conditions are required to prove the oracle properties:

\noindent \textbf{C1.} (i) The set $\mathcal{B}$ is a compact subset of $\mathbb{R}^{p_n}$ and $\boldsymbol{\beta}_{s0}$ is an interior point of $\mathcal{B}$. (ii) Let $\textbf{Z}$ be a corresponding $q_n$-dimensional covariate vector. There exists a value $z_0$, $z_0 > 0$, such that $P(||\textbf{Z}|| \leq z_0) =1$, i.e., $\textbf{Z}$ is bounded, and the matrix $E(\textbf{Z}\textbf{Z}^\top)$ is non-singular. \\
\noindent \textbf{C2.} $\int^\tau_0 h_0(t) \; dt < \infty$ and $\int^\tau_0 r_0(t) \; dt < \infty$ for some constant $\tau$. \\
\noindent \textbf{C3.} The baseline cumulative hazard function $H_{0}(\cdot)$ is continuously differentiable up to order $r$ in $[u,v]$ and satisfy $h_0^{-1} < H_{0}(u) < H_{0}(v) < h_0$. Similarly, the baseline cumulative hazarrd function $R_{0}(\cdot)$ is also continuously differentiable up to order $r$ in $[u,v]$ and satisfy $b_0^{-1} < R_{0}(u) < R_{0}(v) < b_0$.  \\
\noindent \textbf{C4.} For $\boldsymbol{\Omega}_n (\boldsymbol{\beta}) = -\Ddot{\ell}_n(\boldsymbol{\beta} | \widetilde{\boldsymbol{\phi}})$, there exists a compact neighbourhood $\mathcal{B}_0$ of the true value $\boldsymbol{\beta}_{s0}$ such that 
$$
\sup_{\boldsymbol{\beta} \in \mathcal{B}_0} || n^{-1} \boldsymbol{\Omega}_n (\boldsymbol{\beta}) - \textbf{I}(\boldsymbol{\beta}) || \longrightarrow 0,
$$
where $\textbf{I}(\boldsymbol{\beta})$ is a $p_n \times p_n$ positive-definite matrix. \\
\noindent \textbf{C5.} Define $\lambda_{\min}(\boldsymbol{\beta}) = \lambda_{\min}(n^{-1} \boldsymbol{\Omega}_n (\boldsymbol{\beta}))$ and $\lambda_{\max}(\boldsymbol{\beta}) = \lambda_{\max}(n^{-1} \boldsymbol{\Omega}_n (\boldsymbol{\beta}))$, where $\lambda_{\min}(\cdot)$ and $\lambda_{\max}(\cdot)$ denote the smallest and largest eigenvalues of the matrix. There exists a constant $c > 0$, for $\mathcal{B}_0$ given in \textbf{C4.}, such that
$$
c^{-1} < \inf_{\boldsymbol{\beta} \in \mathcal{B}_0} \{ \lambda_{\min}(
\boldsymbol{\beta})\} \leq \sup_{\boldsymbol{\beta} \in \mathcal{B}_0} \{ \lambda_{\max}(\boldsymbol{\beta}) \} < c
$$
for a sufficiently large $n$. \\
\noindent \textbf{C6.} As $n \longrightarrow \infty, \; p_n q_n / \sqrt{n} \longrightarrow 0, \; \lambda_n/\sqrt{n} \longrightarrow 0, \text{and} \; \lambda^2_n/(p_n \sqrt{n}) \longrightarrow \infty$. \\
\noindent \textbf{C7.} There exists positive constants $a_0$ and $a_1$ such that $a_0 \leq |\beta_{s0,j}| \leq a_1$, for $1 \leq j \leq q_n$. \\
\noindent \textbf{C8.} The initial estimator $\widehat{\boldsymbol{\beta}}^{(0)}$  satisfies $||\widehat{\boldsymbol{\beta}}^{(0)} -  \boldsymbol{\beta}_{s0}|| = O_p(\sqrt{p_n/n})$.\\
\noindent \textbf{C9.} For every $n$, the observations $\{v_{ni}, \; i=1,\ldots,n \}$ are independent and identically distributed with the probability density $f_n(v_{ni}, \boldsymbol{\beta}, \boldsymbol{\phi})$ which has a common support and the model is identifiable. The parameter space is $\boldsymbol{\theta}^{\ast} = \{ \vartheta: \vartheta =(\boldsymbol{\beta}, \boldsymbol{\phi}) \in \mathcal{B} \times \boldsymbol{\Phi} \}$, $\boldsymbol{\beta}_{s0}$ is an interior point of $\mathcal{B}$, then for almost all $v_{ni}$, the density $f_n$ admits all third derivatives $\partial^3 \log f_n(v_{ni}, \boldsymbol{\beta}, \boldsymbol{\phi}) /\partial \beta_j \partial \beta_k \partial \beta_h$ for all $\boldsymbol{\beta} \in \mathcal{B}$. Furthermore, there are functions $M_{njkh}(\cdot)$ such that 
$$
\left\vert \frac{\partial^3 \log f_n(v_{ni}, \boldsymbol{\beta}, \boldsymbol{\phi})}{\partial \beta_j \partial \beta_k \partial \beta_h} \right\vert \leq M_{njkh}(v_{ni})
$$
for all $\boldsymbol{\beta} \in \mathcal{B}$ and $\boldsymbol{\phi}$, and
$$
\mathbb{E}_{\boldsymbol{\beta}, \boldsymbol{\phi}} \{ M^2_{njkh}(v_{ni})\} < M_d < \infty.
$$

Let $\boldsymbol{\Omega}^{(1)}_n(\boldsymbol{\beta})$ is the leading submatrix of $\boldsymbol{\Omega}_n(\boldsymbol{\beta})$, and $\textbf{v}^{(1)}_n(\boldsymbol{\beta})$ is the vector consisting of the first $q_n$ components of $\textbf{v}_n(\boldsymbol{\beta})$. That is, $\boldsymbol{\Omega}_n(\boldsymbol{\beta})$ and $\textbf{v}_n(\boldsymbol{\beta})$ can be written as
$$
\boldsymbol{\Omega}_n(\boldsymbol{\beta}) = \begin{pmatrix} \boldsymbol{\Omega}^{(1)}_n(\boldsymbol{\beta}) & \boldsymbol{\Omega}^{(12)}_n(\boldsymbol{\beta}) \\ \{\boldsymbol{\Omega}^{(12)}_n(\boldsymbol{\beta})\}^\top & \boldsymbol{\Omega}^{(2)}_n(\boldsymbol{\beta})\end{pmatrix} \;\; \text{and} \;\; \textbf{v}_n(\boldsymbol{\beta}) = \begin{pmatrix} \textbf{v}^{(1)}_n(\boldsymbol{\beta}) \\ \textbf{v}^{(2)}_n(\boldsymbol{\beta}) \end{pmatrix}, \;\; \text{respectively},
$$
where $\boldsymbol{\Omega}^{(1)}_n(\boldsymbol{\beta})$ is a $q_n \times q_n$ matrix, $\boldsymbol{\Omega}^{(12)}_n(\boldsymbol{\beta})$ is a $q_n \times (p_n - q_n)$ matrix, $\boldsymbol{\Omega}^{(2)}_n(\boldsymbol{\beta})$ is a $(p_n - q_n) \times (p_n - q_n)$ matrix, $\textbf{v}^{(1)}_n(\boldsymbol{\beta})$ is a $q_n$-vector, and $\textbf{v}^{(2)}_n(\boldsymbol{\beta})$ is a $p_n - q_n$-vector. \\
\noindent \textbf{Theorem 1. \textit{(Oracle properties)}} Under conditions \textbf{C1} - \textbf{C9}, with probability tending to 1, the BAR estimator $\widehat{\boldsymbol{\beta}} = ((\widehat{\boldsymbol{\beta}}_{s1})^\top, (\widehat{\boldsymbol{\beta}}_{s2})^\top)^\top$ has the following properties: 

\noindent (1) $\widehat{\boldsymbol{\beta}}_{s2}=0$. \\
\noindent (2) $\widehat{\boldsymbol{\beta}}_{s1}$ exists and is the unique fixed point of the equation $\boldsymbol{\beta}_{s1} = \{\boldsymbol{\Omega}^{(1)}_n(\boldsymbol{\beta}_{s1}) + \lambda_n \textbf{D}_1(\boldsymbol{\beta}_{s1}) \}^{-1} \textbf{v}^{(1)}_n(\boldsymbol{\beta}_{s1})$, where $\textbf{D}_1(\boldsymbol{\beta}_{s1}) = \text{diag}(\beta^{-2}_{s1,1}, \ldots, \beta^{-2}_{s1, q_n} )$, and $\beta_{s1,j}, \; j=1,\ldots,q_n$, represents the non-zero elements with dimension $q_n$. \\
\noindent (3) For any $\textbf{b}_n$ being a $q_n$-vector, assume $||\textbf{b}_n||=1$, then $\sqrt{n} \textbf{b}^\top_n \boldsymbol{\Sigma}^{-1/2} (\widehat{\boldsymbol{\beta}}_{s1} - \boldsymbol{\beta}_{s01}) \xrightarrow{d} N(0,1)$, where 
\begin{equation*}
\boldsymbol{\Sigma} = (\textbf{I}^{(1)}(\boldsymbol{\beta}_{s0}))^{-1},
\end{equation*}
where $\textbf{I}^{(1)}(\boldsymbol{\beta}_{s0})$ is the leading $q_n \times q_n$ submatrix of $\textbf{I}(\boldsymbol{\beta}_0)$. That is, informally, we can say $\widehat{\boldsymbol{\beta}}_{s1}$ is asymptotically normal with asymptotic variance $\boldsymbol{\Sigma}/n$. Note: we have showed a different result to \citet{zhao2019simultaneous}. Proof of Theorem 1 is explained in detail in the Appendix.

\section{Simulation Studies} \label{SS1}
In this section, we examine the performance of the BAR method under three scenarios, and compare our results with the MIC penalty and the Oracle method, which assumes the true model is known. For all scenarios, without loss of generality, we set $d_1 = d_2$, and $\textbf{Z}_{i} = \textbf{Z}_{i,1} = \textbf{Z}_{i,2}$. We provide the details of the simulation set-up and report the results of Scenarios 1 and 2 in this section. The simulation set-up and results of Scenario 3, and additional results of Scenarios 1 and 2, are provided in the Appendix.

\subsection{Scenario 1: Fixed dimension of covariates $p$} \label{S1inSS1}
In Scenario 1, we investigate the performance of the competing methods when the number of covariates is fixed. Let the total number of covariates $p = d_1 + d_2$, the number of non-zero regression parameters $q_n = 4$, and we set $d_1 = d_2 = 10$. We consider two types of covariates, continuous covariates and binary covariates. The continuous covariates are generated from the multivariate normal distribution with mean zero and covariance $\boldsymbol{\Sigma}$. The $(i,j)^{th}$ element of the covariance matrix $\boldsymbol{\Sigma}$ is defined as $\rho^{|i-j|}$, where the correlation coefficient $\rho = 0.25$. The binary covariates are generated from marginal Bernoulli distributions with the same probability of success of 0.5, where the pairwise correlation is $\text{cor}(\textbf{Z}_{i}, \textbf{Z}_{j}) = 0.25^{|i-j|}$. The total number of covariates is split equally into continuous and binary covariates. 

We set the true values of $\boldsymbol{\beta}_1$ and $\boldsymbol{\beta}_2$ to be 
\begin{align}
  \boldsymbol{\beta}_{01} &= (1,0,\ldots,0, -1)^\top, \label{b01} \\
  \boldsymbol{\beta}_{02} &= (1,-0.5,0,\ldots,0)^\top, \label{b02}
\end{align}
respectively. For the baseline hazards, we consider the linear case for both hazard functions, where $h_0(t) = 5+0.2t$ and $r_0(t) = 8+0.2t$. For the frailty term, we use positive and negative values of $\gamma$, i.e., $\gamma = 1$ and $\gamma = -0.6$. We generate the random frailty term $u_i$ from the standard normal distribution, i.e., $u_i \sim N(0, \phi^2)$, and the dispersion parameter $\phi=1$. 

To simulate terminal event times, we couple the random uniform sampling technique with the inverse cumulative hazard function. That is, we generate $u$ from the standard uniform distribution, i.e., $u \sim U(0, 1)$. Then, given the frailty $u_i$, $u \; \text{and} \; \textbf{Z}_i$ the terminal event time $D_i$ is sampled from 
$$
H^{-1}_0(-\log(u)\exp(-\boldsymbol{\beta}^\top_2 \textbf{Z}_i + \gamma u_i)),
$$ 
where $H_0(\cdot)$ is the cumulative baseline hazard function for terminal events. To generate recurrent event times $T_{im}, \; m > 0$, assume $T_{i0} = 0$, we can use the following recursive method. Assume $T_{im}$ is generated, $m > 0$, then we can generate $T_{i(m+1)}$ by the probability inversion method based on the following identity. In fact, it can be shown that 
$$
F(T_{i(m+1)} | T_{i(m+1)} > T_{im}) = 1 - \exp(\{ R_0(T_{im}) - R_0(T_{i(m+1)}) \} \exp(\boldsymbol{\beta}^\top_2 \textbf{Z}_i)]),
$$
where $F(\cdot)$ is the cumulative distribution function of $T_{i(m+1)} | T_{i(m+1)} > T_{im}$, and $R_0(\cdot)$ is the cumulative baseline hazard function for recurrent events. We simulate the censoring time $C_i$ from the uniform distribution $U(0,2)$. Following the above set-up, there are on average 0.8 to 1.2 recurrent events per subject when $\gamma = 1$, and 2 to 3 recurrent events when $\gamma = -0.6$. The right-censoring rate has a range between 15\% and 25\%. For Scenario 1, we consider two values of the sample size $n$, $n=300$ and $n=500$. We use $B=200$ Monte-Carlo replications to summarize our results in Tables \ref{TS1} - \ref{gammaphiS1}.

In our simulation study, to improve the approximation of the numerical integration method and numerical stability, we decide to increase the number of quadrature points of the Gauss-Hermite Quadrature, contrary to the recommended 10 quadrature points by \citet{han2020variable}. Generally, we increase the number of quadrature points when the number estimated parameters also increases. We utilize the \verb|fastGHQuad| R package \citep{blocker2011fastghquad} to perform Gauss-Hermite Quadrature numerical integration. We also utilize the \verb|mvtnorm| R package \citep{genz2021package} and the \verb|mipfp| R package \citep{barthelemy2018mipfp} to draw random samples from the multivariate normal distribution and correlated Bernoulli distributions, respectively. We calculate two measures to evaluate the variable selection accuracy for the competing methods. First is true positives (TP), which is the average number of correctly estimated non-zero regression coefficients. Second is false positives (FP), which is the average number of falsely estimated non-zero regression parameters. We also calculate the similarity measure (SM) and frequency of true model selected (TM). SM has the formula 
$$
SM = \frac{|\widehat{S} \cap S|_0}{\sqrt{|\widehat{S}|_0 |S|_0}},
$$
where $\widehat{S}$ is the estimated set of regression parameters, $S$ is the true set of regression parameters, and $|\cdot|_0$ denotes the model size. To evaluate estimation accuracy, we use the mean squared error (MSE), which has the formula 
$$
\text{MSE} = \frac{1}{B} \sum^{B}_{k=1} ||\widehat{\boldsymbol{\beta}}_k - 
\boldsymbol{\beta}_0||^2_2,
$$
where $B$ is the number of replications, $\boldsymbol{\beta}_0 = (\boldsymbol{\beta}^\top_{01}, \boldsymbol{\beta}^\top_{02})^\top$, and $\widehat{\boldsymbol{\beta}}_k$ is the estimator of $\boldsymbol{\beta}$ of the $k^{th}$ simulated dataset. For the BAR method, we use a simple grid search between $\lambda_n = 2$ and $\lambda_n = 4$. We use the GCV criterion described in \eqref{GCV} to select the optimal tuning parameter $\lambda_n$.

Non-linear optimization methods are sensitive to the initial input values. To investigate the sensitivity to initial values of the competing methods, we choose three different initial values. The three different initial values are defined as
\begin{align}
  \boldsymbol{\theta}_{initial1} &= \boldsymbol{\theta}^{\ast}_0 + \boldsymbol{\epsilon}_1, \quad \text{elements of $\boldsymbol{\epsilon}_1 \sim N(0, 0.1^2)$}, \label{init1} \\
  \boldsymbol{\theta}_{initial2} &= \boldsymbol{\theta}^{\ast}_0 + \boldsymbol{\epsilon}_2, \quad \text{elements of $\boldsymbol{\epsilon}_2 \sim N(0, 0.25^2)$}, \label{init2} \\
  \boldsymbol{\theta}_{initial3} &= \boldsymbol{\theta}^{\ast}_0 + \boldsymbol{\epsilon}_3, \quad \text{elements of $\boldsymbol{\epsilon}_3 \sim N(0, 0.4^2)$}, \label{init3}
\end{align}
where $\boldsymbol{\theta}^\ast_0$ represents the true value of $\boldsymbol{\theta}^\ast = (\boldsymbol{\beta}^\top_1, \boldsymbol{\beta}^\top_2, \textbf{h}^\top, \textbf{r}^\top, \gamma, \phi)^\top$. There are no true values to $(\textbf{h}^\top, \textbf{r}^\top)$ as they are parameters from the piecewise function approximation. Therefore, we choose the starting values to be $\textbf{h}_0 = (5,\ldots,5)^\top$ and $\textbf{r}_0 = (8,\ldots,8)^\top$. The initial values \eqref{init1}, \eqref{init2} and \eqref{init3} are denoted by (*), (**) and (***), respectively, in Tables \ref{TS1} and \ref{gammaphiS1}. We also consider a fourth set of initial values denoted by (****), where the regression estimates obtained from fitting marginal semi-parametric regression models for the recurrent event data and terminal event data respectively are used as the initial values of the regression parameters, and a normally distributed random error term centered at the origin with a standard deviation of 0.25 is added to $\boldsymbol{\phi}_0 = (\textbf{h}^\top_0, \textbf{r}^\top_0, \gamma, \phi)^\top$. We fit the marginal semi-parametric regression models using the R package \verb|reReg| \citep{chiou2023regression}.

From Table \ref{TS1}, we observe that, for all different choices of initial values, the average number of FP is larger for the MIC, and the MIC method selects the true model at a lower frequency as compared to the BAR method. Additionally, one can also observe from Table \ref{TS1} that the initial values affect the performance of the MIC more than the BAR method. When the initial values of the parameters are farther away from the true values, the variable selection and estimation results of the MIC become worse. However, the performance of the BAR procedure is largely insensitive to the input of initial values. We also report the estimation results of $\gamma$ and $\phi$ in Table \ref{gammaphiS1}, where we observe that the estimation of $\phi$ is more accurate for the case when a negative value of $\gamma$ is considered.

\begin{table}[H]
\centering
\caption{Summary of variable selection and estimation results in Scenario 1. TP: the average number of true positives FP: the average number of false positives; SM: similarity measure; TM: frequency of true model selected; MSE: mean squared error.} \label{TS1}
\scalebox{0.97}{\begin{tabular}{p{1.25cm}|p{3cm}p{2.85cm}p{1.8cm}p{1.8cm}p{1.8cm}p{1.8cm}}
\hline
$\gamma$ & Method & MSE(SD) & TP & FP & SM & TM \\
\hline
\multicolumn{7}{c}{$n=300, \; p=20, \; q_n = 4$} \\
\hline
\multirow{9}{*}{1} & BAR (*) & 0.173(0.199) & 3.97 & 0.14 & 0.98 & 85\% \\
&  BAR (**) & 0.180(0.206) & 3.96 & 0.13 & 0.98 & 86\% \\
&  BAR (***) & 0.190(0.227) & 3.97 & 0.13 & 0.98 & 87\% \\
&  BAR (****) & 0.178(0.177)& 3.97 & 0.14 & 0.98 & 86\%\\
&  MIC (*) & 0.116(0.132) & 4 & 0.17 & 0.98 & 86\%\\
&  MIC (**) & 0.138(0.145) & 3.98 & 0.28 & 0.97 & 79\%\\
&  MIC (***) & 0.148(0.168) & 3.96 & 0.38 & 0.96 & 68\%\\
&  MIC (****) & 0.207(0.231) & 3.83 & 0.58 & 0.92 & 58\%\\
&  Oracle & 0.102(0.093) & 4 & 0 & 1 & 100\% \\
\hline
\multirow{9}{*}{-0.6} & BAR (*) & 0.204(0.131) & 3.99 & 0.07 & 0.99 & 94\% \\
&  BAR (**) & 0.203(0.151) & 3.99 & 0.05 & 0.99 & 95\% \\
&  BAR (***) & 0.206(0.156) & 3.99 & 0.06 & 0.99 & 94\% \\
&  BAR (****) & 0.197(0.165) & 4 & 0.09 & 0.99 & 92\%\\
&  MIC (*) & 0.132(0.141) & 3.99 & 0.11 & 0.99 & 90\%\\
&  MIC (**) & 0.151(0.221) & 3.98 & 0.22 & 0.97 & 81\%\\
&  MIC (***) & 0.179(0.227) & 3.97 & 0.44 & 0.95 & 68\%\\
&  MIC (****) & 0.159(0.147)& 3.99 & 0.37 & 0.96 & 72\%\\
&  Oracle & 0.119(0.077)& 4 & 0 & 1 & 100\% \\
\hline
\multicolumn{7}{c}{$n=500, \; p=20, \; q_n = 4$} \\
\hline
\multirow{9}{*}{1} & BAR (*) & 0.091(0.078) & 4 & 0.09 & 0.99 & 92\%\\
&  BAR (**) & 0.090(0.079) & 4 & 0.09 & 0.99 & 92\% \\
&  BAR (***) & 0.124(0.125) & 3.99 & 0.09 & 0.99 & 92\% \\
&  BAR (****) & 0.088(0.091) & 4 & 0.10 & 0.99 & 91\%\\
&  MIC (*) & 0.078(0.062) & 4 & 0.14 & 0.99 & 88\% \\
&  MIC (**) & 0.085(0.073) & 3.99 & 0.32 & 0.97 & 77\% \\
&  MIC (***) & 0.097(0.135) & 4 & 0.47 & 0.95 & 69\%\\
&  MIC (****) & 0.094(0.090) & 3.98 & 0.37 & 0.96 & 73\%\\
&  Oracle & 0.067(0.053)& 4 & 0 & 1 & 100\% \\
\hline
\multirow{9}{*}{-0.6} & BAR (*) & 0.127(0.131) & 4 & 0.05 & 0.99 & 95\%\\
&  BAR (**) & 0.123(0.094) & 4 & 0.02 & 1 & 98\%\\
&  BAR (***) & 0.126(0.088) & 4 & 0.04 & 1 & 97\%\\
&  BAR (****) & 0.119(0.131)& 4 & 0.04 & 1 & 97\% \\
&  MIC (*) & 0.089(0.063) & 4 & 0.10 & 0.99 & 92\% \\
&  MIC (**) & 0.098(0.074) & 3.99& 0.19 & 0.98 & 86\% \\
&  MIC (***) & 0.127(0.110) & 3.98 & 0.43 & 0.96 & 68\% \\
&  MIC (****) & 0.103(0.088) & 4 & 0.33 & 0.97 & 75\% \\
&  Oracle &  0.082(0.055) & 4 & 0 & 1 & 100\% \\
\hline
\end{tabular}}
\end{table}

\begin{table}[H]
    \centering
    \caption{Summary of the estimation results of $\gamma$ and $\phi$ in Scenario 1. $Ave()$: the sample mean of the 200 parameter estimates; $SD()$: the sample standard deviation of the 200 parameter estimates.} \label{gammaphiS1}
    \scalebox{0.97}{\begin{tabular}{p{2cm}p{2cm}p{3cm}p{1.8cm}p{1.8cm}p{1.8cm}p{1.8cm}}
    \hline
    True $\gamma$ & True $\phi$ & Method  & $Ave(\widehat{\gamma})$ & $Ave(\widehat{\phi})$ & $SD(\widehat{\gamma})$ & $SD(\widehat{\phi})$ \\
    \hline
    \multicolumn{7}{c}{$n=300, \; p = 20, \; q_n = 4$} \\
    \hline
    \multirow{7}{*}{1} & \multirow{7}{*}{1} & BAR (*) & 0.987 & 1.232 & 0.185 & 0.204\\
    & & BAR (**) & 0.963 & 1.226 & 0.176 & 0.204\\
    & & BAR (***) & 0.932 & 1.218 & 0.171 & 0.202\\
    & & BAR (****) & 0.959 & 1.229 & 0.168 & 0.206\\
    & & MIC (*) & 0.989 & 1.208 & 0.181 & 0.188\\
    & & MIC (**) & 0.950 & 1.200& 0.167 & 0.187\\
    & & MIC (***) & 0.889 & 1.191 & 0.173 & 0.187\\
    & & MIC (****) & 0.941 & 1.195 & 0.172 & 0.187\\
    & & Oracle & 0.982 & 1.209 & 0.179 & 0.182 \\
    \hline
    \multirow{7}{*}{-0.6} & \multirow{7}{*}{1} & BAR (*) & -0.766 & 1.099 & 0.156 & 0.125 \\
    & & BAR (**) & -0.759 & 1.105 & 0.146 & 0.119 \\
    & & BAR (***) & -0.740 & 1.109 & 0.137 & 0.112 \\
    & & BAR (****) & -0.763 & 1.099 & 0.155 & 0.119 \\
    & & MIC (*) & -0.752 & 1.102 & 0.132 & 0.111\\
    & & MIC (**) & -0.722 & 1.122 & 0.131 & 0.107\\
    & & MIC (***) & -0.648 & 1.168 & 0.148  &0.121 \\
    & & MIC (****) & -0.717 & 1.113 & 0.150 & 0.112\\
    & & Oracle & -0.755 & 1.115 & 0.133 & 0.115 \\
    \hline
    \multicolumn{7}{c}{$n=500, \; p=20, \; q_n =4$} \\
    \hline
    \multirow{7}{*}{1} & \multirow{7}{*}{1} & BAR (*) & 1.025 & 1.181 & 0.147 & 0.177\\
    & & BAR (**) & 0.994 & 1.174 & 0.135 & 0.176 \\
    & & BAR (***) & 0.951 & 1.170 & 0.141 & 0.189 \\
    & & BAR (****) & 0.998 & 1.176 & 0.132 & 0.178 \\
    & & MIC (*) & 1.017 & 1.165 & 0.138 & 0.161 \\
    & & MIC (**) & 0.970 & 1.154 & 0.128 & 0.157 \\
    & & MIC (***) & 0.895 & 1.147 & 0.139 & 0.159 \\
    & & MIC (****) & 0.976 & 1.154 & 0.125 & 0.161\\
    & & Oracle & 1.009 & 1.149 & 0.142 & 0.147 \\
    \hline
    \multirow{7}{*}{-0.6} & \multirow{7}{*}{1} & BAR (*) & -0.762 & 1.100 & 0.125 & 0.097\\
    & & BAR (**) & -0.750 & 1.105 & 0.115 & 0.094\\
    & & BAR (***) & -0.732 & 1.111 & 0.109 & 0.093 \\
    & & BAR (****) & -0.754 & 1.096 & 0.121& 0.094\\
    & & MIC (*) & -0.750 & 1.101 & 0.115 & 0.090\\
    & & MIC (**) & -0.718 & 1.119 & 0.106 & 0.086\\
    & & MIC (***) & -0.637 & 1.161 & 0.131 & 0.100\\
    & & MIC (****) & -0.721 & 1.101 & 0.125 & 0.085\\
    & & Oracle & -0.648 & 1.101 & 0.115 & 0.087\\
    \hline
    \end{tabular}}
\end{table}
\subsection{Scenario 2: Diverging dimension of covariates $p_n$} \label{S2inSS1}
In Scenario 2, we investigate the performance of the competing methods when the total number of covariates $p_n$ diverges with the sample size $n$. We use the same true values for $\boldsymbol{\beta}_1$ and $\boldsymbol{\beta}_2$ as in Scenario 1, defined in \eqref{b01} and \eqref{b02}. We consider a mixture of continuous and binary covariates, where the continuous and binary covariates are sampled in the same way as in Scenario 1. Additionally, we use the same baseline hazard functions as described in the previous scenario. We fix $\gamma = 1$ and $\phi = 1$. From this simulation set-up, the right-censoring rate has a range between 15\% and 25\%, with an average of 20\%.

To obtain the total number of covariates $p_n$ when the sample size $n$ varies, we set $d_k = \lfloor 5n^{1/5} \rfloor$, for $k = 1,2$, where the output of the floor function $f(x) = \lfloor x \rfloor$ is the largest integer less than or equal to $x$. We use three different sample sizes, $n = 100, \; 300, \; \text{and} \; 500$. The sample sizes and the total number of covariates are
\begin{align*}
n = 100, & \quad p_n = 12 \times 2 = 24, \\
n = 300, & \quad p_n = 15 \times 2 = 30, \\
n = 500, & \quad p_n = 17 \times 2 = 34.
\end{align*}
Additionally, we use the same initial values described in \eqref{init1}, \eqref{init2} and \eqref{init3}, which are denoted by (*), (**), and (***), respectively. We also use the fourth set of initial values denoted as (****), where the description is given in \ref{S1inSS1}. We use 200 Monte-Carlo replications to summarize the results below in Tables \ref{TS2} and \ref{gammaphiS2}, and we use the same measures used in the first scenario to assess the selection and estimation errors.

\begin{table}[H]
\centering
\caption{Summary of the variable selection and estimation results in Scenario 2 when the average censoring rate is 20\%. TP: the average number of true positives; FP: the average number of false positives; SM: similarity measure; TM: frequency of true model selected; MSE: mean squared error.} \label{TS2}
\begin{tabular}{p{3.25cm}|p{2.75cm}p{1.8cm}p{1.8cm}p{1.8cm}p{1.8cm}}
\hline
Method & MSE(SD) & TP & FP & SM & TM \\
\hline
\multicolumn{6}{c}{$n=100, \; p_n = 12 \times 2, \; q_n = 4$} \\
\hline
BAR (*) & 0.843(0.863) & 3.45 & 0.57 & 0.86 & 32\% \\
BAR (**) & 0.837(0.947)& 3.44 & 0.54 & 0.87 & 33\% \\
BAR (***) & 0.835(0.755) & 3.46 & 0.58 & 0.87 & 34\% \\
BAR (****) & 0.878(0.768) & 3.41 & 0.52 & 0.87 & 33\% \\
MIC (*) & 0.378(0.432) & 3.89 & 0.10 & 0.97 & 82\% \\
MIC (**) & 0.478(0.646) & 3.88 & 0.37 & 0.95 & 65\% \\
MIC (***) & 0.584(0.650) & 3.74 & 0.54 & 0.91 & 48\%\\
MIC (****) & 0.818(0.895) & 3.51 & 0.95 & 0.85 & 29\%\\
Oracle & 0.281(0.267) & 4 & 0 & 1 & 100\% \\
\hline
\multicolumn{6}{c}{$n=300, \; p_n = 15 \times 2, \; q_n = 4$} \\
\hline
BAR (*) & 0.160(0.164) & 3.99 & 0.13 & 0.99 & 88\%\\
BAR (**) & 0.178(0.211) & 3.98 & 0.13 & 0.98 & 87\%\\
BAR (***) & 0.176(0.163) & 4 & 0.12 & 0.99 & 89\%\\
BAR (****) & 0.165(0.177) & 3.99 & 0.14 & 0.98 & 87\%\\
MIC (*) & 0.114(0.097) & 4 & 0.18 & 0.98 & 83\% \\
MIC (**) & 0.128(0.124) & 3.98 & 0.34 & 0.96 & 76\%\\
MIC (***) & 0.186(0.216) & 3.96 & 0.71 & 0.93 & 54\%\\
MIC (****) & 0.216(0.232) & 3.92 & 0.84 & 0.91 & 51\% \\
Oracle & 0.097(0.084) & 4 & 0 & 1 & 100\%\\
\hline
\multicolumn{6}{c}{$n=500, \; p_n = 17 \times 2, \; q_n = 4$} \\
\hline
BAR (*) & 0.112(0.094) & 4 & 0.25 & 0.97 & 78\% \\
BAR (**) & 0.106(0.084) & 4 & 0.24 & 0.98 & 80\% \\
BAR (***) & 0.109(0.087) & 4 & 0.20 & 0.98 & 84\% \\
BAR (****) & 0.115(0.127)& 3.99 & 0.28 & 0.97 & 75\%\\
MIC (*) & 0.088(0.076) & 4 & 0.24 & 0.98& 79\% \\
MIC (**) &0.089(0.095) & 4 & 0.43 & 0.96& 71\% \\
MIC (***) & 0.152(0.184) & 3.97 & 1.07 & 0.90 & 47\% \\
MIC (****) & 0.130(0.147) & 3.94 & 0.66 & 0.93 & 57\%\\
Oracle & 0.069(0.067) & 4 & 0 & 1 &100\% \\
\hline
\end{tabular}
\end{table}

\begin{table}[H]
\centering
\caption{Summary of the estimation results of $\gamma$ and $\phi$ in Scenario 2 when the censoring rate is 20\%. $Ave()$: the sample mean of the 200 parameter estimates; $SD()$: the sample standard deviation of the 200 parameter estimates. The true values are $\gamma = 1$ and $\phi = 1$.} \label{gammaphiS2}
\scalebox{1}{\begin{tabular}{p{3cm}|p{2cm}p{2cm}p{2cm}p{2cm}}
\hline
Method  & $Ave(\widehat{\gamma})$ & $Ave(\widehat{\phi})$ & $SD(\widehat{\gamma})$ & $SD(\widehat{\phi})$ \\
\hline
\multicolumn{5}{c}{$n=100, \; p_n = 12\times2, \; q_n = 4$} \\
\hline
BAR (*) & 1.150 & 1.227 & 0.402 & 0.450\\
BAR (**) & 1.112 & 1.228 & 0.389 & 0.451\\
BAR (***) & 1.079 & 1.228 & 0.397 & 0.464\\
BAR (****) & 1.088 & 1.236 & 0.417 & 0.471\\
MIC (*) & 1.085 & 1.221 & 0.325 & 0.448 \\
MIC (**) & 1.031 & 1.215 & 0.307 & 0.449 \\
MIC (***) & 0.956 & 1.226 & 0.341 & 0.464\\
MIC (****) & 1.011 & 1.194 & 0.371 & 0.445 \\
Oracle & 1.099 & 1.189 & 0.372 & 0.368\\
\hline
\multicolumn{5}{c}{$n=300, \; p_n = 15\times2, \; q_n = 4$} \\
\hline
BAR (*) & 1.021 & 1.204 & 0.181 & 0.207\\
BAR (**) &  0.983 & 1.201 & 0.172 & 0.209 \\
BAR (***) & 0.940 & 1.190 & 0.160 & 0.204 \\
BAR (****) & 0.983 & 1.200 & 0.162 & 0.209\\
MIC (*) & 1.005 & 1.185 & 0.163 & 0.191 \\
MIC (**) & 0.938 & 1.174 & 0.154 & 0.187 \\
MIC (***) & 0.861 & 1.173 & 0.163 & 0.190 \\
MIC (****) & 0.945 & 1.166 & 0.156 & 0.190 \\
Oracle & 1.006 & 1.182 & 0.164 & 0.186 \\
\hline
\multicolumn{5}{c}{$n=500, \; p_n = 17\times2, \; q_n = 4$} \\
\hline
BAR (*) & 0.973 & 1.209 & 0.137 & 0.186\\
BAR (**) & 0.941 & 1.201 & 0.127 & 0.184 \\
BAR (***) & 0.899 & 1.195 & 0.119 & 0.183 \\
BAR (****) & 0.949 & 1.205 & 0.125 & 0.186\\
MIC (*) & 0.972 & 1.196 & 0.121 & 0.171\\
MIC (**) & 0.906 & 1.182 & 0.1118 & 0.168 \\
MIC (***) & 0.818 & 1.182 & 0.138 & 0.173 \\
MIC (****) & 0.930 & 1.180 & 0.115 & 0.170\\
Oracle & 0.967& 1.188 & 0.117 & 0.120 \\
\hline
\end{tabular}}
\end{table}
The variable selection results of Scenario 2 are summarized in Table \ref{TS2}, we observe that as $p_n$ diverges from $n$, the average number of FP of the BAR method reduces, which resulted in the true model being selected at a higher frequency. Conversely, the average number of FP of the MIC method increases as $p_n$ diverges from $n$, which resulted in the true model being selected at a lower frequency. The estimation error decreases as the sample size increases for both methods. From Table \ref{gammaphiS2}, we observe a similar trend to the results in Table \ref{gammaphiS1}, where the estimation of $\gamma$ is better than the estimation of $\phi$. We repeat the set up described in this scenario with a higher censoring rate, which has a range between 35\% to 45\%. Both methods have larger variable selection and estimation errors when the censoring rates increases, and the results are summarized in Tables \ref{TS2high} and \ref{gammaphiS2high} in the Appendix. 

\subsection*{Scenario 3: Grouped variables}
In Scenario 3, we investigate the performance of the competing methods when there exists multiple groups of highly correlated covariates. We explain the simulation set-up and the summary of the results in the Appendix. 

\section{Real Data Analysis: MIMIC-III Database} \label{rda1}
As a real data application, we apply our proposed method on the data obtained from the MIMIC-III clinical database. The MIMIC-III database integrates de-identified, highly granular and comprehensive clinical information of over forty thousand patients that were admitted to the Beth Israel Deaconess Medical Center between 2001 and 2012 in Boston, Massachusetts, USA. The MIMIC-III database is accessible to researchers worldwide, subject to a data use agreement and ethics training. To acquire the data, the authors completed the training and signed an agreement. Specifically, we use the data from MIMIC-III \citep{johnson2016mimic} as our real data application.

Information contained in the MIMIC-III database includes demographics (e.g. age and gender), vital sign information (e.g. heart rate and blood pressure level) measured every hour, date and time of each hospital admission and discharge, date and time of each ICU admission and discharge, laboratory test results, type of medications given, procedure, imaging reports, caregiver notes, and mortality. Information regarding the type of medical insurance is also given. For our real data application, we define the terminal event to be the event of death occurring during ICU stay. We define the recurrent event as subsequent ICU admission after the initial ICU admission, as all patients have at least one ICU admission. The follow-up time is measured starting from the time a patient is admitted into the ICU until the patient has either discharged or died during ICU stay. An observation is censored if the event of death is not observed when the patient is under observation in the ICU.  

For the real data analysis, we decide to include 12 variables. Of these 12 variables, ten are considered as quantitative (continuous) variables, and two are considered as qualitative (categorical) variables. The qualitative variable Race has six levels - white American, black American, Asian American, Hispanic American, Native American and unspecified. From the variable race, we create five dummy variables, where white American is used as the reference category. For the quantitative variables, we only include the baseline measurement, i.e., the first measurement taken when the patient is first admitted into the hospital. The data dictionary of the 12 variables used in our analysis is summarized in Table \ref{MIMICvariables}, where we provide the mean, standard deviation and the range of the quantitative variables, and the coding and proportion of the qualitative variables.

In the analysis, we standardize all quantitative variables, such that the quantitative variables are on the same scale. We do not standardize the binary variables. We use the subset of patients that used Medicaid insurance, where the sample size is $n=2822$, and the total number of recurrent events is $N = \sum^{2822}_{i=1} n_i = 1153$. We compare the BAR method to the MIC method on the Medicaid sub-dataset. The grid search of the tuning parameter $\lambda_n$ is between 3 and 4 for the BAR method.

For the selection of the initial values, one does not know the true values of neither the regression parameters nor the nuisance parameters, for real data analysis. To overcome this problem, we fit univariate models for each covariate for the recurrent and terminal events submodels separately, and we use the regression estimates of the univariate models as the initial values. 
\begin{table}[H]
\centering
\caption{Data dictionary about the qualitative and quantitative variables used for the real data analysis of the MIMIC-III data.} \label{MIMICvariables}
\scalebox{0.95}{\begin{tabular}{p{6.45cm}|p{4.75cm}|p{5cm}}
\hline
Quantitative variable (Unit) & Mean (SD) & Range \\
\hline
Age (years) & 49.1 (14.3) & [15, 90] \\
Weight (kg) & 82.1 (26.8) & [1, 575] \\
Heart rate (Beats per min) & 89.1 (16.5) & [35.74, 153.21] \\
Systolic blood pressure (BP) & 120 (16.5) & [66.7, 191.8] \\
Diastolic blood pressure (BP) & 65.7 (11.4) & [34.3, 125.2] \\
Respiratory rate (Breaths per min) & 18.9 (4.3) & [10, 41.8] \\
Blood oxygen saturation (spO2) & 97.3 (2.3) & [62, 100] \\
Temperature (Celsius) & 36.9 (0.7) & [32.1, 39.8] \\
Glucose (mmol/L) & 136.4 (45.5) & [43.3, 786.2]\\
Urine (mL) & 2278 (1534.4) & [0, 34235] \\
\hline
Qualitative variable & Coding & Ratio \\ 
\hline
\multirow{2}{*}{Gender} & 1 - Male & 1617 Males\\
   & 0 - Female & 1205 Females \\
\hline
\multirow{2}{*}{Black American} & 1 - Black American &  453 Black Americans \\
 & 0 - other & 2369 others \\
\hline
\multirow{2}{*}{Asian American} &  1 - Asian American & 176 Asian Americans\\
 & 0 - other & 2646 others \\ 
\hline
\multirow{2}{*}{Hispanic American} & 1 - Hispanic American & 291 Hispanic Americans \\
 & 0 - other & 2531 others \\
\hline
\multirow{2}{*}{Unknown Ethnicity} & 1 - Ethnicity not known  & 287 not specified  \\
& 0 - Ethnicity known & 2535 specified  \\
\hline
\multirow{2}{*}{Native American} & 1 - Native American & 7 Native Americans \\
 & 0 - others & 2815 others \\
\hline
\end{tabular}}
\end{table}

\begin{table}[H]
\centering
\caption{Analysis of patients using Medicaid data from the MIMIC-III database. The estimates of the regression coefficients of the recurrent event sub-model are on the left, and the estimates of the regression coefficients of the terminal event sub-model are on the right.}\label{MIMICIII}
\begin{tabular}{c|cc}
\hline
\multicolumn{3}{c}{Recurrent event submodel}\\
\hline
Variable & BAR Est & MIC Est \\
\hline
Gender & 0 & 0 \\
Age & 0 & 0\\
Weight & 0 & 0 \\
Heart rate & 0 & 0\\
Systolic BP & 0 & 0\\
Diastolic BP & 0 & 0\\
Respiratory Rate & 0 & 0 \\
spO2 & 0 & 0\\
Temperature &  0 & -0.150\\
Glucose & 0 & 0 \\
Black-American & 0.280 & 0 \\
Asian-American & 0 & -0.288\\
Hispanic-American & 0 & 0 \\
Unknown & -0.220 & -0.263\\ 
Native-American & 0 & 0\\
Urine output & 0 & 0\\
$\gamma$ & -0.492 & -0.564 \\
\hline
\end{tabular}
\quad
\begin{tabular}{c|cc}
\hline
\multicolumn{3}{c}{Terminal event submodel}\\
\hline
Variable & BAR Est & MIC Est \\
\hline
Gender & 0 & 0\\
Age & 0 & 0 \\
Weight & 0 & 0\\
Heart rate & 0.419 & 0.535\\
Systolic BP & -0.333 & 0\\
Diastolic BP & 0 & -0.270\\
Respiratory Rate & 0.283 & 0\\
spO2 & 0 & -0.076\\
Temperature &  -0.506 & -0.528\\
Glucose & 0 & 0\\
Black-American &0 & -0.237\\
Asian-American & 0& 0\\
Hispanic-American & 0& 0\\
Unknown & 0.293 & 0.271\\ 
Native-American & 0& 0\\
Urine output & -0.331 & -0.338 \\
$\phi$ & 0.945 & 1.028 \\
\hline
\end{tabular}
\end{table}
The results of the real data analysis using the BAR and MIC methods are summarized in Table \ref{MIMICIII}. We observe that the BAR method identifies fewer relevant variables that contribute to the risks of death during ICU admission and repeated ICU admissions respectively, than the MIC method.  Both methods indicate a lower temperature results in higher risks of dying during ICU stay and repeated ICU admissions. The BAR method indicates higher heart rate and respiratory rates increases the risk of death. Heart and respiratory rates were shown to be important variables in previous investigations of the MIMIC-III database \citep{li2021prediction}. Gender, age and weight are shown to not have any effect on the risks of death, or repeated ICU admissions. When the ethnicity of a patient is not known, it results in a higher risk of death during ICU stay. 

\section{Discussion and Conclusion}
In this article, we have proposed a novel simultaneous variable selection and estimation approach under the framework of joint frailty models of recurrent and terminal event. Our proposed approach uses the BAR penalty, which approximates the $L_0$-norm by an iterative reweighted squared $L_2$-penalized regression. To implement the BAR penalty under our model framework, we approximate the log-likelihood function using the least-squares function in order to get closed-form estimate updates of our regression parameters. Additionally, we use the Gauss-Hermite quadrature to tackle the computational complexity, as the integrals contained within the log-likelihood function, gradient vector, and Hessian matrix have no closed form solutions. In our simulation studies, we have observed that our proposed method outperformed the MIC. Additionally, the MIC penalty is very sensitive to the input of initial values, which was not discussed in \citet{han2020variable}. However, the performance of the BAR penalty is not affected by the choice of the initial values. We have proved that the oracle properties hold for the BAR estimator under certain regularity conditions. We also applied our proposed approach to the MIMIC-III database, where the number of relevant variables by our proposed method is fewer than the MIC method.

There are a few research directions left for future work. For example, in our simulation studies and the real data application, we only considered the case of diverging number of covariates, i.e., $p_n \rightarrow \infty$, but $p_n < n$. For the high-dimensional or ultra high-dimensional cases, i.e., $p_n > n$ or $p_n \gg n$, a screening method such as the Sure Independence Screening \citep{fan2008sure} could be used first to reduce the dimension to $p_n < n$. Then, our proposed method can be applied on the screened dataset.

\section*{Acknowledgement}
The authors would like to acknowledge New Frontiers in Research Fund to Quan Long (NFRFE-2018-00748) administered by the Canada Research Coordinating Committee, the Canada Foundation for Innovation JELF grant (36605) awarded to Quan Long, and the Discovery Grant to Xuewen Lu (RG/PIN06466-2018) administered by the National Science and Engineering Research Council of Canada. 

\section*{Declaration of Conflicting Interests} 

The author(s) declared no potential conflicts of interest with respect to the research, authorship, and/or publication of this article.

\bibliographystyle{plainnat}
\bibliography{ForArXiv}
\newpage

\section*{Appendix I: Derivation of the Gradient Vector and Hessian Matrix}
The fully-parametric marginal likelihood function is
\begin{equation}\label{margApprox}
\widetilde{\mathcal{L}}_n(\boldsymbol{\theta}^{\ast}) = \prod^n_{i=1} \int^\infty_{-\infty} \widetilde{g}_1(Y_i | u_i) \widetilde{g}_2(Y_i | u_i) f_\phi(u_i) \; du_i = \prod^n_{i=1} \int^\infty_{-\infty} m_1(u_i; Y_i) \; du_i,
\end{equation}
where $\boldsymbol{\theta}^{\ast} = (\boldsymbol{\beta}^\top_1, \boldsymbol{\beta}^\top_2, \textbf{h}^\top, \textbf{r}^\top, \gamma, \phi)^\top$. Here, the terms in the integrand in \eqref{margApprox} are
\begin{equation*}
\begin{split}
\widetilde{g}_1(Y_i |u_i) &=  \big[ \widetilde{h}_0(Y_i)\exp( \boldsymbol{\beta}^\top_2 \textbf{Z}_{i,2} + \gamma u_i) \big]^{\delta_i} \exp\big\{ -\widetilde{H}_0(Y_i)\exp( \boldsymbol{\beta}^\top_2 \textbf{Z}_{i,2} + \gamma u_i) \big\} \\
& = \big[\widetilde{h}_i(Y_i \vert u_i) \big]^{\delta_i} \cdot \exp\big\{-\widetilde{H}_i(Y_i \vert u_i) \big\}, 
\end{split}
\end{equation*}
and
\begin{equation*}
\begin{split}
\widetilde{g}_2(Y_i | u_i) &= \bigg\{ \prod^{n_i}_{k=1} \widetilde{r}_0(T_{ik})\exp(\boldsymbol{\beta}^\top_1 \textbf{Z}_{i,1} + u_i) \bigg\} \exp \big\{ -\widetilde{R}_0(Y_i) \exp( \boldsymbol{\beta}^\top_1 \textbf{Z}_{i,1} + u_i)  \big\} \\
& = \bigg\{  \prod^{n_i}_{k=1} \widetilde{r}_i(T_{ik} \vert u_i) \bigg\} \exp \big\{ -\widetilde{R}_i(Y_i \vert u_i) \big\}.
\end{split}
\end{equation*}
If $n_i = 0$, then $\prod^{n_i}_{k=1} \widetilde{r}_i(T_{ik} \vert u_i) = 1$. For frailty models in general, we assume a non-negative probability density for $\exp(u_i)$. The log-likelihood of \eqref{margApprox} is 
\begin{equation*}
\begin{split}
\ell_n(\boldsymbol{\beta}, \boldsymbol{\phi}) 
& = \log \widetilde{\mathcal{L}}_n (\boldsymbol{\beta}, \boldsymbol{\phi})  \\
& = \sum^n_{i=1} \log \left[ \int^{\infty}_{-\infty} \widetilde{g}_1(Y_i |u_i) \widetilde{g}_2(Y_i | u_i) f_\phi(u_i) \; \text{d}u_i \right]  \\
& = \sum^n_{i=1} \log \widetilde{\mathcal{L}}_{ni}(\boldsymbol{\beta}, \boldsymbol{\phi}) .
\end{split}
\end{equation*}
The gradient vector can be decomposed into two parts
\begin{equation*}
\Dot{\ell}_n(\boldsymbol{\beta} | \boldsymbol{\phi}) = \frac{\partial \ell_n(\boldsymbol{\beta}, \boldsymbol{\phi})}{\partial \boldsymbol{\beta}} = \begin{pmatrix} \Dot{\ell}^{(1)}_n(\boldsymbol{\beta} | \boldsymbol{\phi}) \\ \Dot{\ell}^{(2)}_n(\boldsymbol{\beta} | \boldsymbol{\phi}) \end{pmatrix}, 
\end{equation*}
where 
\begin{equation}\label{m1}
\begin{split}
\Dot{\ell}^{(1)}_n(\boldsymbol{\beta} | \boldsymbol{\phi}) = \frac{\partial \ell_n(\boldsymbol{\beta}, \boldsymbol{\phi})}{\partial \boldsymbol{\beta}_1} & = \sum^n_{i=1} \frac{\partial \log \widetilde{\mathcal{L}}_{ni}(\boldsymbol{\beta}, \boldsymbol{\phi})}{\partial \boldsymbol{\beta}_1 } \\
& = \sum^n_{i=1}  \frac{\partial \log \widetilde{\mathcal{L}}_{ni}(\boldsymbol{\beta}, \boldsymbol{\phi})}{\partial \widetilde{\mathcal{L}}_{ni}(\boldsymbol{\beta}, \boldsymbol{\phi})} \cdot \frac{\partial \widetilde{\mathcal{L}}_{ni}(\boldsymbol{\beta}, \boldsymbol{\phi})}{\partial \boldsymbol{\beta}_1},
\end{split}
\end{equation}
and
\begin{equation}\label{m2}
\begin{split}
\Dot{\ell}^{(2)}_n(\boldsymbol{\beta} | \boldsymbol{\phi}) = \frac{\partial \ell_n(\boldsymbol{\beta}, \boldsymbol{\phi})}{\partial \boldsymbol{\beta}_2} & = \sum^n_{i=1} \frac{\partial \log \widetilde{\mathcal{L}}_{ni}(\boldsymbol{\beta}, \boldsymbol{\phi})}{\partial \boldsymbol{\beta}_2 } \\
& = \sum^n_{i=1} \frac{\partial \log \widetilde{\mathcal{L}}_{ni}(\boldsymbol{\beta}, \boldsymbol{\phi})}{\partial \widetilde{\mathcal{L}}_{ni}(\boldsymbol{\beta} \vert \boldsymbol{\phi})} \cdot \frac{\partial \widetilde{\mathcal{L}}_{ni}(\boldsymbol{\beta}, \boldsymbol{\phi})}{\partial \boldsymbol{\beta}_2}.
\end{split}
\end{equation}
To evaluate \eqref{m1} and \eqref{m2}, we would need to use the Leibniz Integral rule. \\

\noindent
\textbf{Definition: Leibniz integral rule.} For any given bivariate function $f(x,t)$ and domain $\Omega$, where $\int_{\Omega} f(x,t) \; dx < \infty$, we can take the derivative of $\int_{\Omega} f(x,t) \; dx$ w.r.t $t$ inside the integral sign, such that
$$
\frac{\partial}{\partial t} \int_{\Omega} f(x,t) \; \text{d}x = \int_{\Omega} \frac{\partial}{\partial t} f(x,t) \; \text{d}x.
$$ 
The derivatives $\partial \widetilde{g}_1(Y_i | u_i) / \partial \boldsymbol{\beta}_2$ and $\partial \widetilde{g}_2(Y_i | u_i) / \partial \boldsymbol{\beta}_1$ are
\begin{equation}\label{dg1}
  \frac{\partial \widetilde{g}_2(Y_i | u_i)}{\partial \boldsymbol{\beta}_1} = \textbf{Z}_{i,1} \Bigg( n_i \bigg\{\prod^{n_i}_{k=1} \widetilde{r}_i(T_{ik} \vert u_i) \bigg\} - \widetilde{R}_i(Y_i \vert u_i) \exp\big[- \widetilde{R}_i(Y_i \vert u_i) \big] \Bigg)
\end{equation}
and
\begin{equation}\label{dg2}
  \frac{\partial \widetilde{g}_1(Y_i | u_i)}{\partial \boldsymbol{\beta}_2} = \textbf{Z}_{i,2} \Big( \delta_i\big[\widetilde{h}_i(Y_i \vert u_i)\big]^{\delta_i} - \widetilde{H}_i(Y_i \vert u_i) \exp\big[- \widetilde{H}_i(Y_i \vert u_i) \big] \Big),
\end{equation}
respectively. Using Leibniz integral rule and the results of \eqref{dg1} and \eqref{dg2}, the expressions of $\partial \widetilde{\mathcal{L}}_{ni}(\boldsymbol{\beta}, \boldsymbol{\phi})/ \partial \boldsymbol{\beta}_1$ and $\partial \widetilde{\mathcal{L}}_{ni}(\boldsymbol{\beta}, \boldsymbol{\phi})/ \partial \boldsymbol{\beta}_2$ are
\begin{equation*}
\frac{\partial  \widetilde{\mathcal{L}}_{ni}(\boldsymbol{\beta}, \boldsymbol{\phi})}{\partial \boldsymbol{\beta}_1} = \textbf{Z}_{i,1} \int^{\infty}_{-\infty} \widetilde{g}_1(Y_i | u_i) \widetilde{g}_2(Y_i | u_i) \big[n_i - \widetilde{R}_i(Y_i \vert u_i) \big]  f_\phi(u_i) \; du_i
\end{equation*}
and
\begin{equation*}
\frac{\partial \widetilde{\mathcal{L}}_{ni}(\boldsymbol{\beta}, \boldsymbol{\phi})}{\partial \boldsymbol{\beta}_2} = \textbf{Z}_{i,2} \int^{\infty}_{-\infty} \widetilde{g}_1(Y_i | u_i) \widetilde{g}_2(Y_i | u_i) \big[\delta_i - \widetilde{H}_i(Y_i \vert u_i) \big] f_\phi(u_i) \; du_i,
\end{equation*}
respectively. Therefore, the expressions of $\Dot{\ell}^{(1)}_n(\boldsymbol{\beta} | \boldsymbol{\phi})$ and $\Dot{\ell}^{(2)}_n(\boldsymbol{\beta} | \boldsymbol{\phi})$ are
\begin{equation*}
\Dot{\ell}^{(1)}_n(\boldsymbol{\beta} | \boldsymbol{\phi}) = \sum^n_{i=1} \textbf{Z}_{i,1} \frac{\int^{\infty}_{-\infty} \widetilde{g}_1(Y_i | u_i) \widetilde{g}_2(Y_i | u_i)  [n_i- \widetilde{R}_i(Y_i \vert u_i)] f_\phi(u_i) \; du_i}{\widetilde{\mathcal{L}}_{ni}(\boldsymbol{\beta}, \boldsymbol{\phi})}
\end{equation*}
and
\begin{equation*}
\Dot{\ell}^{(2)}_n(\boldsymbol{\beta} | \boldsymbol{\phi}) = \sum^n_{i=1} \textbf{Z}_{i,2} \frac{\int^{\infty}_{-\infty} \widetilde{g}_1(Y_i | u_i) \widetilde{g}_2(Y_i | u_i)  [\delta_i - \widetilde{H}_i(Y_i \vert u_i)] f_\phi(u_i) \; du_i}{\widetilde{\mathcal{L}}_{ni}(\boldsymbol{\beta}, \boldsymbol{\phi})},
\end{equation*}
respectively. After deriving an expression for the gradient vector, we subsequently need to find an expression of the Hessian matrix. For simplicity, we define
\begin{equation*}
\begin{split}
\widetilde{g}_3(Y_i) &= \int^\infty_{-\infty} \widetilde{g}_1(Y_i | u_i) \widetilde{g}_2(Y_i | u_i) \big[n_i - \widetilde{R}_i(Y_i \vert u_i)\big] f_\phi(u_i) \; du_i \\
& = \int^{\infty}_{-\infty} m_2(u_i; Y_i) \; du_i
\end{split}
\end{equation*}
and
\begin{equation*}
\begin{split}
\widetilde{g}_4(Y_i) &= \int^\infty_{-\infty} \widetilde{g}_1(Y_i | u_i) \widetilde{g}_2(Y_i | u_i)  \big[\delta_i - \widetilde{H}_i(Y_i \vert u_i) \big] f_\phi(u_i) \; du_i \\
& = \int^{\infty}_{-\infty} m_3(u_i; Y_i) \; du_i.
\end{split}
\end{equation*}
We can define the Hessian matrix as
\begin{equation}\label{Hessian}
\Ddot{\ell}_n(\boldsymbol{\beta} | \boldsymbol{\phi}) = \begin{pmatrix} \Ddot{\ell}^{(1)}_n(\boldsymbol{\beta}| \boldsymbol{\phi}) &  \Ddot{\ell}^{(12)}_n(\boldsymbol{\beta}| \boldsymbol{\phi}) \\ \Ddot{\ell}^{(21)}_n(\boldsymbol{\beta}| \boldsymbol{\phi}) & \Ddot{\ell}^{(2)}_n(\boldsymbol{\beta}| \boldsymbol{\phi})
 \end{pmatrix}.
\end{equation}
We can express the diagonal sub-matrices of \eqref{Hessian} as:
\begin{equation*}
\begin{split}
\Ddot{\ell}^{(1)}_n(\boldsymbol{\beta}| \boldsymbol{\phi}) = \frac{\partial^2 \ell_n(\boldsymbol{\beta}, \boldsymbol{\phi})}{\partial \boldsymbol{\beta}_1 \partial \boldsymbol{\beta}^\top_1}
& = \frac{\partial}{\partial \boldsymbol{\beta}^\top_1} \left( \frac{\partial \ell_n(\boldsymbol{\beta}, \boldsymbol{\phi})}{\partial \boldsymbol{\beta}_1} \right) \\
& = \sum^n_{i=1} \textbf{Z}_{i,1} \frac{\frac{\partial \widetilde{g}_3(Y_i)}{\partial \boldsymbol{\beta}^\top_1} \cdot \widetilde{\mathcal{L}}_{ni}(\boldsymbol{\beta}, \boldsymbol{\phi}) - \frac{\partial \widetilde{\mathcal{L}}_{ni}(\boldsymbol{\beta}, \boldsymbol{\phi})}{\partial \boldsymbol{\beta}^\top_1} \cdot \widetilde{g}_3(Y_i)}{[\widetilde{\mathcal{L}}_{ni}(\boldsymbol{\beta}, \boldsymbol{\phi})]^2}, \\
\Ddot{\ell}^{(2)}_n(\boldsymbol{\beta}| \boldsymbol{\phi}) = \frac{\partial^2 \ell_n(\boldsymbol{\beta}, \boldsymbol{\phi})}{\partial \boldsymbol{\beta}_2 \partial \boldsymbol{\beta}^\top_2}
& = \frac{\partial}{\partial \boldsymbol{\beta}^\top_2} \left( \frac{\partial \ell_n(\boldsymbol{\beta}, \boldsymbol{\phi}) }{\partial \boldsymbol{\beta}_2} \right) \\
& = \sum^n_{i=1} \textbf{Z}_{i,2} \frac{\frac{\partial \widetilde{g}_4(Y_i)}{\partial \boldsymbol{\beta}^\top_2} \cdot \widetilde{\mathcal{L}}_{ni}(\boldsymbol{\beta}, \boldsymbol{\phi}) - \frac{\partial \widetilde{\mathcal{L}}_{ni}(\boldsymbol{\beta}, \boldsymbol{\phi})}{\partial \boldsymbol{\beta}^\top_2} \cdot \widetilde{g}_4(Y_i)}{[\widetilde{\mathcal{L}}_{ni}(\boldsymbol{\beta}, \boldsymbol{\phi})]^2}. \\
\end{split}
\end{equation*}
It is clear to see that the expressions of  $\partial \widetilde{\mathcal{L}}_{ni}(\boldsymbol{\beta}, \boldsymbol{\phi}) / \partial \boldsymbol{\beta}^\top_1$ and $\partial \widetilde{\mathcal{L}}_{ni}(\boldsymbol{\beta}, \boldsymbol{\phi}) / \partial \boldsymbol{\beta}^\top_2$ are
\begin{equation*}
  \frac{\partial \widetilde{\mathcal{L}}_{ni}(\boldsymbol{\beta}, \boldsymbol{\phi})}{\partial \boldsymbol{\beta}^{\top}_1} = \textbf{Z}^\top_{i,1} \; \widetilde{g}_3(Y_i )
\end{equation*}
and 
\begin{equation*}
\frac{\partial \widetilde{\mathcal{L}}_{ni}(\boldsymbol{\beta}, \boldsymbol{\phi})}{\partial \boldsymbol{\beta}^{\top}_2} = \textbf{Z}^\top_{i,2} \; \widetilde{g}_4(Y_i),
\end{equation*}
respectively. Using the Leibniz integral rule, the expressions of $\partial \widetilde{g}_3(Y_i) / \partial \boldsymbol{\beta}^\top_1$ and $\partial \widetilde{g}_4(Y_i) / \partial \boldsymbol{\beta}^\top_2$ are
\begin{equation*}
\begin{split}
\frac{\partial \widetilde{g}_3(Y_i)}{\partial \boldsymbol{\beta}^\top_1} &= \textbf{Z}^\top_{i,1} \int^\infty_{-\infty} \widetilde{g}_1(Y_i |u_i) \widetilde{g}_2(Y_i |u_i) \Big\{ \big[n_i - \widetilde{R}_i(Y_i \vert u_i) \big]^2 - \widetilde{R}_i(Y_i \vert u_i) \Big\} f_\phi(u_i) \; du_i \\
&= \textbf{Z}^\top_{i,1} \int^\infty_{-\infty} m_4(u_i; Y_i) \; du_i
\end{split}
\end{equation*}
and
\begin{equation*}
\begin{split}
\frac{\partial \widetilde{g}_4(Y_i)}{\partial \boldsymbol{\beta}^\top_2} &= \textbf{Z}^\top_{i,2} \int^\infty_{-\infty} \widetilde{g}_1(Y_i |u_i) \widetilde{g}_2(Y_i |u_i) \Big\{ \big[\delta_i - \widetilde{H}_i(Y_i \vert u_i)\big]^2 - \widetilde{H}_i(Y_i \vert u_i) \Big\} f_\phi(u_i) \; du_i \\
&= \textbf{Z}^\top_{1,2} \int^\infty_{-\infty} m_5(u_i; Y_i) \; du_i,
\end{split}
\end{equation*}
respectively. The final expressions of the diagonal sub-matrices of \eqref{Hessian} are
\begin{equation*}
\Ddot{\ell}^{(1)}_n(\boldsymbol{\beta}| \boldsymbol{\phi}) = \sum^n_{i=1} \textbf{Z}_{i,1} \textbf{Z}^\top_{i,1} \frac{\int^{\infty}_{-\infty} m_4(u_i; Y_i) \; du_i \cdot \widetilde{\mathcal{L}}_{ni}(\boldsymbol{\beta}, \boldsymbol{\phi}) - [\widetilde{g}_3(Y_i)]^2 }{[\widetilde{\mathcal{L}}_{ni}(\boldsymbol{\beta},\boldsymbol{\phi})]^2}
\end{equation*}
and
\begin{equation*}
\Ddot{\ell}^{(2)}_n(\boldsymbol{\beta}| \boldsymbol{\phi}) = \sum^n_{i=1} \textbf{Z}_{i,2} \textbf{Z}^\top_{i,2} \frac{\int^{\infty}_{-\infty} m_5(u_i; Y_i) \; du_i \cdot \widetilde{\mathcal{L}}_{ni}(\boldsymbol{\beta}, \boldsymbol{\phi}) - [\widetilde{g}_4(Y_i)]^2}{[\widetilde{\mathcal{L}}_{ni}(\boldsymbol{\beta}, \boldsymbol{\phi})]^2}.
\end{equation*}
After finding the diagonal sub-matrices of the Hessian matrix, we need to find the off-diagonal sub-matrices. They are
\begin{equation*}
\begin{split}
\Ddot{\ell}^{(12)}_n(\boldsymbol{\beta}| \boldsymbol{\phi}) = \frac{\partial^2 \ell_n(\boldsymbol{\beta} , \boldsymbol{\phi})}{\partial \boldsymbol{\beta}_1 \partial \boldsymbol{\beta}^\top_2} & = \frac{\partial}{\partial \boldsymbol{\beta}^\top_2} \left(\frac{\partial \ell_n(\boldsymbol{\beta}, \boldsymbol{\phi})}{\partial \boldsymbol{\beta}_1} \right) \\
& = \sum^n_{i=1} \textbf{Z}_{i,1} \frac{\frac{\partial \widetilde{g}_3(Y_i)}{\partial \boldsymbol{\beta}^\top_2} \cdot \widetilde{\mathcal{L}}_{ni}(\boldsymbol{\beta}, \boldsymbol{\phi}) - \frac{\partial \widetilde{\mathcal{L}}_{ni}(\boldsymbol{\beta}, \boldsymbol{\phi})}{\partial \boldsymbol{\beta}^\top_2} \cdot \widetilde{g}_3(Y_i)}{[\widetilde{\mathcal{L}}_{ni}(\boldsymbol{\beta}, \boldsymbol{\phi})]^2} 
\end{split}
\end{equation*}
and
\begin{equation*}
\begin{split}
\Ddot{\ell}^{(21)}_n(\boldsymbol{\beta}| \boldsymbol{\phi}) = \frac{\partial^2 \ell_n(\boldsymbol{\beta}, \boldsymbol{\phi})}{\partial \boldsymbol{\beta}_2 \partial \boldsymbol{\beta}^\top_1} & = \frac{\partial}{\partial \boldsymbol{\beta}^\top_1} \left(\frac{\partial \ell_n(\boldsymbol{\beta}, \boldsymbol{\phi})}{\partial \boldsymbol{\beta}_2} \right) \\
& = \sum^n_{i=1} \textbf{Z}_{i,2} \frac{\frac{\partial \widetilde{g}_4(Y_i)}{\partial \boldsymbol{\beta}^\top_1} \cdot \widetilde{\mathcal{L}}_{ni}(\boldsymbol{\beta},\boldsymbol{\phi}) - \frac{\partial \widetilde{\mathcal{L}}_{ni}(\boldsymbol{\beta}, \boldsymbol{\phi})}{\partial \boldsymbol{\beta}^\top_1} \cdot \widetilde{g}_4(Y_i)}{[\widetilde{\mathcal{L}}_{ni}(\boldsymbol{\beta}, \boldsymbol{\phi})]^2},
\end{split}
\end{equation*}
respectively. Using the Leibniz integral rule, it is clear to see that
\begin{equation} \label{dbetatop2}
\begin{split}
  \frac{\partial  \widetilde{g}_3(Y_i)}{\partial \boldsymbol{\beta}^\top_2} &= \textbf{Z}^\top_{i,2} \int^\infty_{-\infty} \widetilde{g}_1(Y_i | u_i) \widetilde{g}_2(Y_i | u_i) \big[n_i - \widetilde{R}_i(Y_i \vert u_i)\big]  \big[\delta_i - \widetilde{H}_i(Y_i \vert u_i) \big] f_\phi(u_i) \; du_i \\
  &= \textbf{Z}^\top_{i,2} \int^{\infty}_{-\infty} m_6(u_i;Y_i) \; du_i
\end{split}
\end{equation}
and
\begin{equation} \label{dbetatop1}
\begin{split}
\frac{\partial  \widetilde{g}_4(Y_i)}{\partial \boldsymbol{\beta}^\top_1} &= \textbf{Z}^\top_{i,1} \int^\infty_{-\infty} \widetilde{g}_1(Y_i | u_i) \widetilde{g}_2(Y_i | u_i) \big[n_i - \widetilde{R}_i(Y_i \vert u_i)\big]  \big[\delta_i - \widetilde{H}_i(Y_i \vert u_i) \big] f_\phi(u_i) \; du_i \\
&= \textbf{Z}^\top_{i,1} \int^{\infty}_{-\infty} m_6(u_i;Y_i) \; du_i.
\end{split}
\end{equation}
Equations \eqref{dbetatop2} and \eqref{dbetatop1} would be equivalent and thus the off-diagonal sub-matrices if the same vector of covariates are used in the recurrent events and terminal event sub-model. Thus, the final expressions of the off-diagonal sub-matrices are
\begin{equation*}
    \Ddot{\ell}^{(12)}_n(\boldsymbol{\beta}| \boldsymbol{\phi}) = \textbf{Z}_{i,1} \textbf{Z}^\top_{i,2} \frac{\int^{\infty}_{-\infty} m_6(u_i;Y_i) \; du_i \cdot \widetilde{\mathcal{L}}_{ni}(\boldsymbol{\beta}, \boldsymbol{\phi}) - \widetilde{g}_3(Y_i)\widetilde{g}_4(Y_i)}{[\widetilde{\mathcal{L}}_{ni}(\boldsymbol{\beta}, \boldsymbol{\phi})]^2} 
\end{equation*}
and
\begin{equation*}
    \Ddot{\ell}^{(21)}_n(\boldsymbol{\beta}| \boldsymbol{\phi}) = \textbf{Z}_{i,2} \textbf{Z}^\top_{i,1} \frac{\int^{\infty}_{-\infty} m_6(u_i;Y_i) \; du_i  \cdot \widetilde{\mathcal{L}}_{ni}(\boldsymbol{\beta}, \boldsymbol{\phi}) - \widetilde{g}_3(Y_i)\widetilde{g}_4(Y_i)}{[\widetilde{\mathcal{L}}_{ni}(\boldsymbol{\beta}, \boldsymbol{\phi})]^2}.
\end{equation*}
This completes the derivations of the gradient vector and Hessian matrix.

\section*{Appendix II: Derivation of the Gauss-Hermite Quadrature}
The numerical integration method we choose to use is the Gauss-Hermite quadrature \citep{liu1994note}. Given an arbitrary function $g(u_i)$, where $g(u_i) > 0$, the Gauss-Hermite quadrature approximate $\int^\infty_{-\infty} g(u_i) \; du_i$ by the following
\begin{equation}\label{GHQuad}
\int^{\infty}_{-\infty} g(u_i) \; du_i \approx \sqrt{2} \widehat{\sigma} \sum^m_{j=1} w_j \exp(x^2_j) g(\widehat{\mu} + \sqrt{2}\widehat{\sigma}x_j),
\end{equation}
where $w_j$ and $x_j$ are the weights and the abscissas, respectively. Tables of ($x_j, w_j$) for $m=1,\ldots,20$ are given by Abramowitz and Stegun (1972, p.924). In \eqref{GHQuad}, $\widehat{\mu}$ is defined as the mode of $g(\cdot)$, and $\widehat{\sigma} = 1/\widehat{k}$, where
$$
\widehat{k} = - \frac{\partial^2}{\partial u_i^2} \log g(u_i) \vert_{u_i=\hat{\mu}}.
$$
To apply the Gauss-Hermite quadrature in our context, we let $g(u_i) = m_j(u_i; Y_i), \; j=1,\ldots,6$, respectively, where
\begin{equation*}
\begin{split}
m_1(u_i;Y_i) &= \widetilde{g}_1(Y_i | u_i) \widetilde{g}_2(Y_i | u_i) f_\phi(u_i), \\
m_2(u_i;Y_i) &= \widetilde{g}_1(Y_i | u_i) \widetilde{g}_2(Y_i | u_i) \left[n_i - \widetilde{R}_i(Y_i | u_i) \right] f_\phi(u_i), \\
m_3(u_i; Y_i) &= \widetilde{g}_1(Y_i | u_i) \widetilde{g}_2(Y_i | u_i) \left[\delta_i - \widetilde{H}_i(Y_i| u_i) \right] f_\phi(u_i), \\
m_4(u_i; Y_i) &= \widetilde{g}_1(Y_i | u_i) \widetilde{g}_2(Y_i | u_i) \Big\{ [n_i - \widetilde{R}_i(Y_i)]^2 - \widetilde{R}_i(Y_i| u_i) \Big\} f_\phi(u_i), \\
m_5(u_i; Y_i) &= \widetilde{g}_1(Y_i | u_i) \widetilde{g}_2(Y_i | u_i) \Big\{ [\delta_i - \widetilde{H}_i(Y_i)]^2 - \widetilde{H}_i(Y_i | u_i) \Big\} 
f_\phi(u_i), \\
m_6(u_i; Y_i) &= \widetilde{g}_1(Y_i | u_i) \widetilde{g}_2(Y_i | u_i) \left[n_i - \widetilde{R}_i(Y_i) \right] \left[\delta_i - \widetilde{H}_i(Y_i| u_i) \right] f_\phi(u_i).
\end{split}
\end{equation*}
Taking the first derivative w.r.t. $u_i$ of $\log m_j(u_i; Y_i), \; j=1,\ldots,6$, gives
\begin{equation*}
    \begin{split}
        \frac{\partial \log m_1(u_i; Y_i)}{\partial u_i} &= -\gamma \widetilde{H}_i(Y_i| u_i) + \delta_i \gamma - \widetilde{R}_i(Y_i| u_i) + n_i - \frac{u_i}{\phi^2}, \\
        \frac{\partial \log m_2(u_i; Y_i)}{\partial u_i} &= \frac{\partial \log m_1(u_i; Y_i)}{\partial u_i} - \frac{\widetilde{R}_i(Y_i| u_i)}{n_i - \widetilde{R}_i(Y_i| u_i)}, \\
        \frac{\partial \log m_3(u_i; Y_i)}{\partial u_i} &= \frac{\partial \log m_1(u_i; Y_i)}{\partial u_i} - \frac{\gamma \widetilde{H}_i(Y_i| u_i)}{\delta_i - \widetilde{H}_i(Y_i| u_i)}, \\
        \frac{\partial \log m_4(u_i; Y_i)}{\partial u_i} &= \frac{\partial \log m_1(u_i; Y_i)}{\partial u_i} - \frac{2[\widetilde{R}_i(Y_i| u_i)]^2 - 2n_i\widetilde{R}_i(Y_i| u_i) - \widetilde{R}_i(Y_i| u_i)}{[n_i - \widetilde{R}_i(Y_i| u_i)]^2 - \widetilde{R}_i(Y_i| u_i)}, \\
        \frac{\partial \log m_5(u_i; Y_i)}{\partial u_i} &= \frac{\partial \log m_1(u_i; Y_i)}{\partial u_i} - \frac{2\gamma[\widetilde{H}_i(Y_i| u_i)]^2 - 2\gamma\delta_i\widetilde{H}_i(Y_i| u_i) - \gamma\widetilde{H}_i(Y_i| u_i)}{[\delta_i - \widetilde{H}_i(Y_i| u_i)]^2 - \widetilde{H}_i(Y_i| u_i)} ,\\
        \frac{\partial \log m_6(u_i; Y_i)}{\partial u_i} &= \frac{\partial \log m_1(u_i; Y_i)}{\partial u_i} - \frac{\widetilde{R}_i(Y_i| u_i)}{n_i - \widetilde{R}_i(Y_i| u_i)} - \frac{\gamma \widetilde{H}_i(Y_i| u_i)}{\delta_i - \widetilde{H}_i(Y_i| u_i)}.
    \end{split}
\end{equation*}
Define $\widehat{u}_{ij}, \; j=1,\ldots,6$, as the mode of $m_j(u_i; Y_i), \; j=1,\ldots,6$, respectively. Then, let $\widehat{k}_j = -\partial^2/\partial u^2_i \log m_j(u_i,Y_i), \; j=1,\ldots,6$, respectively, where
\begin{equation*}
    \begin{split}
        \widehat{k}_1 &= - \left.\frac{\partial^2}{\partial u^2_i} \log m_1(u_i; Y_i) \right|_{u_i = \widehat{u}_{i1}} \\
        &= \gamma^2 \widetilde{H}_0(Y_i) \exp(\boldsymbol{\beta}^\top_2 \textbf{Z}_{i,2} + \gamma \widehat{u}_{i1}) + \widetilde{R}_0(Y_i) \exp(\boldsymbol{\beta}^\top_1 \textbf{Z}_{i,1} + \widehat{u}_{i1}) + \frac{1}{\phi^2} \\
        &= \gamma^2 \widetilde{H}_i(Y_i | \widehat{u}_{i1}) + \widetilde{R}_i(Y_i | \widehat{u}_{i1}) + \frac{1}{\phi^2}, \\
        \widehat{k}_2 &= - \left.\frac{\partial^2}{\partial u^2_i} \log m_2(u_i; Y_i) \right|_{u_i = \widehat{u}_{i2}} \\
        &= -\left.\frac{\partial^2}{\partial u^2_i} \log m_1(u_i; Y_i) \right|_{u_i = \widehat{u}_{i2}} + \frac{n_i\widetilde{R}_i(Y_i | \widehat{u}_{i2})}{[n_i-\widetilde{R}_i(Y_i | \widehat{u}_{i2})]^2}, \\
        \widehat{k}_3 &= - \left.\frac{\partial^2}{\partial u^2_i} \log m_3(u_i; Y_i) \right|_{u_i = \widehat{u}_{i3}} \\
        &= -\left.\frac{\partial^2}{\partial u^2_i} \log m_1(u_i; Y_i) \right|_{u_i = \widehat{u}_{i3}} + \frac{\delta_i \gamma^2 \widetilde{H}_i(Y_i | \widehat{u}_{i3})}{[\delta_i-\widetilde{H}_i(Y_i | \widehat{u}_{i3})]^2}, \\
        \widehat{k}_4 &= -\left.\frac{\partial^2}{\partial u^2_i} \log m_4(u_i; Y_i) \right|_{u_i = \widehat{u}_{i4}} \\
        &= -\left.\frac{\partial^2}{\partial u^2_i} \log m_1(u_i; Y_i) \right|_{u_i = \widehat{u}_{i4}} \\
        &- \frac{4n^2_i [\widetilde{R}_i(Y_i | \widehat{u}_{i4})]^2 - 2n_i[\widetilde{R}_i(Y_i | \widehat{u}_{i4})]^3 -[\widetilde{R}_i(Y_i | \widehat{u}_{i4})]^3 - n^2_i \widetilde{R}_i(Y_i | \widehat{u}_{i4}) - 2n^3_i \widetilde{R}_i(Y_i | \widehat{u}_{i4})}{[(n_i - \widetilde{R}_i(Y_i | \widehat{u}_{i4}))^2 - \widetilde{R}_i(Y_i | \widehat{u}_{i4})]^2}, \\
        \widehat{k}_5 &= -\left.\frac{\partial^2}{\partial u_i} \log m_5(u_i; Y_i) \right|_{u_i = \widehat{u}_{i5}} \\
        &= -\left.\frac{\partial^2}{\partial u_i} \log m_1(u_i; Y_i) \right|_{u_i = \widehat{u}_{i5}} - \frac{\gamma^2[4\delta_i[\widetilde{H}_i(Y_i | \widehat{u}_{i5})]^2 + [\widetilde{H}_i(Y_i | \widehat{u}_{i5})]^3 - 2\delta_i[\widetilde{H}_i(Y_i | \widehat{u}_{i5})]^3 - 3\delta_i\widetilde{H}_i(Y_i | \widehat{u}_{i5})]}{[(\delta_i - \widetilde{H}_i(Y_i | \widehat{u}_{i5}))^2 - \widetilde{H}_i(Y_i | \widehat{u}_{i5})]^2}, \\
        \widehat{k}_6 &= -\left. \frac{\partial^2}{\partial u^2_i} \log m_6(u_i; Y_i) \right|_{u_i = \widehat{u}_{i6}} \\
        &= -\left. \frac{\partial^2}{\partial u^2_i} \log m_1(u_i; Y_i) \right|_{u_i = \widehat{u}_{i6}} + \frac{n_i\widetilde{R}_i(Y_i | \widehat{u}_{i6})}{[n_i-\widetilde{R}_i(Y_i | \widehat{u}_{i6})]^2} + \frac{\delta_i \gamma^2 \widetilde{H}_i(Y_i | \widehat{u}_{i6})}{[\delta_i-\widetilde{H}_i(Y_i | \widehat{u}_{i6})]^2}.
    \end{split}
\end{equation*}
Finally, we are able to obtain $\widehat{\sigma}_{ij} = 1/\widehat{k}_{ij}, \; j=1,\ldots,6$, respectively. This completes the derivations of the Gauss-Hermite quadrature.

\section*{Appendix III: Additional Simulation Studies and Results}

\subsection*{Additional simulation results from Scenario 2}

\begin{table}[H]
\centering
\caption{Summary of the variable selection and estimation results in Scenario 2 when the censoring rate is 40\%. TP: average number of true positives; FP: average number of false positives; SM: similarity measure; TM: frequency of true model selected; MSE: mean squared error.} \label{TS2high}
\scalebox{0.88}{\begin{tabular}{p{3.25cm}|p{2.75cm}p{1.8cm}p{1.8cm}p{1.8cm}p{1.8cm}}
\hline
Method & MSE(SD) & TP & FP & SM & TM \\
\hline
\multicolumn{6}{c}{$n=100, \; p_n = 12 \times 2, \; q_n = 4$} \\
\hline
BAR (*) & 1.261(1.131) & 3.30 & 0.52 & 0.85 & 28\% \\
BAR (**) & 1.269(1.344) & 3.27 & 0.50 & 0.84 & 28\% \\
BAR (***) & 1.253(1.233) & 3.26 & 0.48 & 0.85 & 28\% \\
BAR (****) & 1.309(1.334)& 3.21 & 0.53 & 0.83 & 23\%\\
MIC (*) & 0.719(0.754) & 3.76 & 0.18 & 0.95 & 65\% \\
MIC (**) & 0.767(0.765) & 3.76 & 0.37 & 0.93 & 56\% \\
MIC (***) & 0.890(0.926) & 3.67 & 0.55 & 0.90 & 46\%\\
MIC (****) & 1.255(1.234) & 3.43 & 1.16 & 0.82 & 25\% \\
Oracle & 0.531(0.634) & 4 & 0 & 1 & 100\% \\
\hline
\multicolumn{6}{c}{$n=300, \; p_n = 15 \times 2, \; q_n = 4$} \\
\hline
BAR (*) & 0.312(0.365) & 3.94 & 0.19 & 0.97 & 80\% \\
BAR (**) & 0.328(0.486) & 3.93 & 0.19 & 0.97 & 80\% \\
BAR (***) & 0.336(0.383) & 3.93 & 0.17 & 0.97 & 81\% \\
BAR (****) & 0.307(0.343)& 3.93 & 0.18 & 0.97 & 80\%\\
MIC (*) & 0.172(0.152) & 4 & 0.23 & 0.98 & 81\% \\
MIC (**) & 0.197(0.163) & 3.98 & 0.46 & 0.95 & 66\% \\
MIC (***) & 0.242(0.263) & 3.96 & 0.61 & 0.94 & 60\%\\
MIC (****) & 0.303(0.343)& 3.88 & 0.78 & 0.91 & 48\%\\
Oracle & 0.134(0.111) & 4 & 0 & 1 & 100\% \\
\hline
\multicolumn{6}{c}{$n=500, \; p_n = 17 \times 2, \; q_n = 4$} \\
\hline
BAR (*) & 0.192(0.201) & 3.99 & 0.17 & 0.98 & 83\% \\
BAR (**) & 0.202(0.229) & 3.99 & 0.18 & 0.98 & 83\% \\
BAR (***) & 0.219(0.257) & 3.99 & 0.20 & 0.98 & 82\% \\
BAR (****) & 0.213(0.247)& 3.99 & 0.24 & 0.97 & 78\%\\
MIC (*) & 0.126(0.090) & 4 & 0.26 & 0.97 & 78\% \\
MIC (**) & 0.126(0.107) & 4 & 0.37 & 0.96 & 72\% \\
MIC (***) & 0.176(0.182) & 3.95 & 0.78 & 0.92 & 54\%\\
MIC (****) & 0.182(0.162)& 3.93 & 0.76 & 0.92 & 48\%\\
Oracle & 0.107(0.081) & 4 & 0 & 1 & 100\% \\
\hline
\end{tabular}}
\end{table}

\begin{table}[H]
\centering
\caption{Summary of the estimation results of $\gamma$ and $\phi$ in Scenario 2 when the censoring rate is 40\%. $Ave()$: the sample mean of the 200 parameter estimates; $SD()$: the sample standard deviation of the 200 parameter estimates. The true values are $\gamma = 1$ and $\phi = 1$. } \label{gammaphiS2high}
\begin{tabular}{p{3cm}|p{2cm}p{2cm}p{2cm}p{2cm}}
\hline
Method & $Ave(\widehat{\gamma})$ & $Ave(\widehat{\phi})$ & $SD(\widehat{\gamma})$ & $SD(\widehat{\phi})$ \\
\hline
\multicolumn{5}{c}{$n=100, \; p_n = 12\times2, \; q_n = 4$} \\
\hline
BAR (*) & 1.131 & 1.428 & 0.378 & 0.543\\
BAR (**) & 1.115 & 1.428 & 0.387 & 0.555\\
BAR (***) & 1.080 & 1.437 & 0.367 & 0.572\\
BAR (****) & 1.065 & 1.423 & 0.354 & 0.532\\
MIC (*) & 1.129 & 1.383 & 0.388 & 0.494 \\
MIC (**) & 1.081 & 1.376 & 0.377 & 0.500\\
MIC (***) & 1.042 & 1.386 & 0.397 & 0.513\\
MIC (****) & 1.025 & 1.338 & 0.428 & 0.511\\
Oracle & 1.109 & 1.359 & 0.366 & 0.435\\
\hline
\multicolumn{5}{c}{$n=300, \; p_n = 15\times2, \; q_n = 4$} \\
\hline
BAR (*) & 0.945 & 1.446 & 0.163 & 0.266 \\
BAR (**) & 0.923 & 1.441 & 0.166 & 0.271 \\
BAR (***) & 0.891 & 1.442 & 0.161 & 0.286 \\
BAR (****) & 0.910 & 1.439 & 0.153 & 0.267 \\
MIC (*) & 0.954 & 1.390 & 0.172 & 0.214\\
MIC (**) & 0.904 & 1.382 & 0.169 & 0.224\\
MIC (***) & 0.838 & 1.384 & 0.180 & 0.236\\
MIC (****) & 0.896 & 1.359 & 0.162 & 0.215\\
Oracle & 0.94 & 1.381 & 0.164 & 0.221 \\
\hline
\multicolumn{5}{c}{$n=500, \; p_n = 17\times2, \; q_n = 4$} \\
\hline
BAR (*) & 0.945 & 1.423 & 0.136 & 0.201\\
BAR (**) & 0.920 & 1.418 & 0.126 & 0.206\\
BAR (***) & 0.888 & 1.413 & 0.120 &  0.209\\
BAR (****) & 0.923 & 1.421 & 0.127 & 0.207\\
MIC (*) & 0.953 & 1.382 & 0.124 & 0.176\\
MIC (**) & 0.895 & 1.372 & 0.122 & 0.179\\
MIC (***) & 0.828 & 1.374 & 0.132 & 0.181\\
MIC (****) & 0.907 & 1.357 & 0.120 & 0.179\\
Oracle & 0.947 & 1.376 & 0.123 & 0.17\\
\hline
\end{tabular}
\end{table}

\subsection{Scenario 3: Grouped variables}

In situations where there exist groups of highly correlated covariates, it is desirable that all important covariates in a group are simultaneously selected. For Scenario 2, we only consider continuous covariates, and we generate the covariates $\textbf{Z}$ using the same way as described in Scenario 1. We set $n=500$, and $d_1 = d_2 = 10,$. Here, the covariates are placed into four groups as $(\textbf{Z}_1, \textbf{Z}_2), (\textbf{Z}_3,\textbf{Z}_4,\textbf{Z}_5), (\textbf{Z}_6,\textbf{Z}_7,\textbf{Z}_8), (\textbf{Z}_9, \textbf{Z}_{10})$. We only consider the first group and last group to have non-zero effects, meaning the true values of $\boldsymbol{\beta}_1$ and $\boldsymbol{\beta}_2$ are
\begin{align*}
  \boldsymbol{\beta}_{01} & = (0.8,0.8,0,\ldots,0,-0.8,0.8)^\top, \\
  \boldsymbol{\beta}_{02} &= (0.95,0.95,0,\ldots,0,-0.75,-0.75)^\top.
\end{align*}
We generate all four groups from the multivariate normal distribution with mean 0 and $\text{Cov}(\textbf{Z}_i, \textbf{Z}_j) = \rho^{|i-j|}$, where $i,j \in (1,2) \; \text{or} \; (3,4,5) \; \text{or} \; (6,7,8) \; \text{or} \; (9,10)$, and $\rho = 0.75, 0.8, \; \text{or} \; 0.85$, for $i \neq j$. In addition to TP and FP, we also calculate the following statistic $G$, which has the equation
\begin{equation*}
\begin{split}
G = & 0.1 \times G_{11} + 0.15 \times G_{12} +  0.15 \times G_{13} + 0.1 \times G_{14} \\
& + 0.1 \times G_{21} + 0.15 \times G_{22} +  0.15 \times G_{23} + 0.1 \times G_{24},
\end{split}
\end{equation*}
which measures the grouping effect. In the above equation, $G_{k1} \; \text{and} \; G_{k4}$ represent the percentages of the first and last groups for the estimated regression coefficients being \textit{both non-zero}, respectively, for $k=1,2$. Additionally, $G_{k2} \; \text{and} \; G_{k3}$ represent the percentages of the second and third groups for the estimated regression coefficients being \textit{all zeros}, respectively, for $k=1,2$. The coefficients of $G_{kj}, \; k=1,2, \; j=1,2,3,4$, represent the percentage weights of the groups. We use the same initial values, as described in \eqref{init1}, \eqref{init2}, and \eqref{init3} in Section \ref{S1inSS1}. We use 200 replications for each value of $\rho$.

\begin{table}[H]
\centering
\caption{Summary of variable selection and estimation results in Scenario 3. TP: average number of true positives; FP: average number of false positives; G: grouping effect; MSE: mean squared error.} \label{TS3}
\scalebox{0.975}{\begin{tabular}{p{3.75cm}|p{3cm}p{2cm}p{2cm}p{2cm}}
\hline
Method & MSE(SD) & TP & FP & G \\
\hline
\multicolumn{5}{c}{$\rho=0.75, \; q_n = 8$} \\
\hline
BAR (*) & 0.518(0.298) & 8 & 0.01 & 0.999\\
BAR (**) & 0.592(0.314) & 7.99 & 0.015 & 0.997\\
BAR (***) & 0.482(0.296) & 8 & 0.005 & 0.999\\
MIC (*) & 0.643(0.549) & 7.95 & 0.515 & 0.959\\
MIC (**) & 0.561(0.448) & 7.94 & 0.230 & 0.975\\
MIC (***) & 0.578(0.466) & 7.915 & 0.49 & 0.948 \\
Oracle & 0.477(0.286) & 8 & 0 & 1 \\
\hline
\multicolumn{5}{c}{$\rho=0.80, \; q_n = 8$} \\
\hline
BAR (*) & 0.616(0.431) & 7.98 & 0.005 & 0.997 \\
BAR (**) & 0.596(0.392) & 7.98 & 0.01 & 0.997 \\
BAR (***) & 0.584(0.582) & 7.965 & 0.015 & 0.994 \\
MIC (*) & 0.863(0.734) & 7.82 & 0.93 & 0.924\\
MIC (**) & 0.790(0.645) & 7.845 & 0.935 & 0.924 \\
MIC (***) & 0.777(0.682) & 7.835 & 0.985 & 0.912 \\
Oracle & 0.495(0.276) & 8 & 0 & 1 \\
\hline
\multicolumn{5}{c}{$\rho=0.85, \; q_n = 8$} \\
\hline
BAR (*) & 0.908(0.834) & 7.88 & 0 & 0.988\\
BAR (**) & 0.919(0.906) & 7.81 & 0 & 0.985 \\
BAR (***) & 0.900(0.900) & 7.78 & 0 & 0.984 \\
MIC (*) & 1.143(0.908) & 7.80 & 2.145 & 0.854  \\
MIC (**) & 0.930(0.778) & 7.845 & 1.60 & 0.890 \\
MIC (***) & 0.970(0.848) & 7.76 & 1.62 & 0.872 \\
Oracle & 0.554(0.314) & 8 & 0 & 1 \\
\hline
\end{tabular}}
\end{table}

From the simulation results of Scenario 2 summarized in Table \ref{TS3}, we observe that the overall misclassification rate of the BAR method is lower than the MIC method, where the average number of FP is higher for the MIC method. Additionally, the $G$ statistic is higher for the BAR method than the MIC method for $\rho = 0.75,0.8,0.85$, implying a higher frequency in the number of groups of correlated variables that are wholly selected by the BAR method. One can notice the overall selection and estimation error increases for both methods, as the within-group correlation $\rho$ increases. We also report the estimation results of $\gamma$ and $\phi$ in Table \ref{gammaphiS3} in the Appendix.

\begin{table}[H]
    \centering
    \caption{Summary of the biases and the sample standard deviation of the estimates of $\gamma$ and $\phi$ in Scenario 3.} \label{gammaphiS3}
    \scalebox{0.95}{\begin{tabular}{p{4cm}|p{2.5cm}p{2.5cm}p{2.5cm}p{2.5cm}}
    \hline
     Method & $\vert \text{Bias}(\hat{\gamma}) \vert$ & $\vert \text{Bias}(\hat{\phi}) \vert$  & $\text{SD}(\hat{\gamma})$ & $\text{SD}(\hat{\phi})$ \\
     \hline
    \multicolumn{5}{c}{$\rho = 0.75$} \\
    \hline
      BAR(*)  & 0.1111 & 0.4404 & 0.0960 & 0.1745 \\
      BAR(**)  & 0.1243 &  0.4417 & 0.0957 & 0.1961\\
      BAR(***)  & 0.1509 & 0.4256 & 0.0990 & 0.1872 \\
      MIC(*)  & 0.1006 & 0.4163 & 0.1030 & 0.1669 \\
      MIC(**)  & 0.1247 & 0.4121 & 0.1003 & 0.1780 \\
      MIC(***)  & 0.1721 & 0.4110 & 0.1104 & 0.1857 \\
      Oracle & 0.1088 & 0.4142 & 0.0995 & 0.1737 \\
    \hline
    \multicolumn{5}{c}{$\rho = 0.80$} \\
    \hline
      BAR(*)  & 0.1220 & 0.4442 & 0.0894 & 0.1685 \\
      BAR(**)  & 0.1359 & 0.4380 & 0.0878 & 0.1712\\
      BAR(***)  & 0.1598 & 0.4256 & 0.0896 & 0.1828\\
      MIC(*)  & 0.1116 & 0.4125 & 0.0974 & 0.1600 \\
      MIC(**)  & 0.1420 & 0.4129 & 0.0928 & 0.1656\\
      MIC(***)  & 0.1924 & 0.417 & 0.1030 & 0.1756\\
      Oracle & 0.1163 & 0.4197 & 0.0917 & 0.1606\\
    \hline
    \multicolumn{5}{c}{$\rho = 0.85$} \\
    \hline
     BAR(*)  & 0.1203 & 0.4524 & 0.104 & 0.1825 \\
      BAR(**)  & 0.1397 & 0.4389  & 0.120 & 0.2122\\
      BAR(***)  & 0.1740 & 0.4189 & 0.1468 & 0.2612\\
      MIC(*)  & 0.1207 & 0.3743 & 0.1156 & 0.1816 \\
      MIC(**)  & 0.1483 & 0.3931 & 0.1025 & 0.1733\\
      MIC(***)  & 0.1945 & 0.3850 & 0.1329 & 0.2169\\
      Oracle & 0.1176 & 0.4221 & 0.1039 & 0.1751 \\
    \hline
    \end{tabular}}
\end{table}

\section*{Appendix IV: Proof of Oracle Properties of the BAR Estimator}
Let $\ell_n(\boldsymbol{\beta}, \boldsymbol{\phi}) = \log \widetilde{\mathcal{L}}_n(\boldsymbol{\beta}, \boldsymbol{\phi})$ be the log-likelihood function defined in \eqref{Approx2}, and let $(\widetilde{\boldsymbol{\beta}}, \widetilde{\boldsymbol{\phi}})$ be the un-penalized estimates of $(\boldsymbol{\beta}, \boldsymbol{\phi})$. 

We consider the total number of non-zero components and zero components in $\boldsymbol{\beta}$ to be $p_n$, and $p_n$ is diverging, i.e., $p_n \longrightarrow \infty$ and $q_n \longrightarrow \infty$ when $n \longrightarrow \infty$. However, $p_n$ and $q_n$ need to satisfy condition \textbf{C6}.

Let $\boldsymbol{\beta} = (\boldsymbol{\beta}^\top_{s1}, \boldsymbol{\beta}^\top_{s2})^\top$, where
$$
\boldsymbol{\beta}_{s1} = (\beta_{s1,1},\ldots,\beta_{s1,q_n})^\top
$$
is the $q_n$-dimensional vector that consists of the non-zero regression coefficients for the joint frailty model of the recurrent and terminal events. And,
$$
\boldsymbol{\beta}_{s2} = (\beta_{s2,q_n+1},\ldots,\beta_{s2,p_n})^\top
$$
is the $(p_n - q_n)$-dimensional vector that consists of the zero regression coefficients for the model. To implement our novel simultaneous variable selection and estimation method, we consider the following penalized likelihood
\begin{equation}\label{pl}
\begin{split}
\ell_{pp}(\boldsymbol{\beta} | \check{\boldsymbol{\beta}}) &= -2\ell_p(\boldsymbol{\beta}) + \lambda_n \sum^2_{j=1} \sum^{d_j}_{k=1} \frac{\beta^2_{j,k}}{(\check{\beta}_{j,k})^2} \\
& = -2\ell_p(\boldsymbol{\beta}) + \lambda_n \sum^{p_n}_{j=1} \frac{\beta^2_{j}}{(\check{\beta}_{j})^2}.
\end{split}
\end{equation}
To establish the oracle properties, we show that minimizing \eqref{pl} is asymptotically equivalent to minimizing the following penalized least-squares function
$$
\frac{1}{2} ||\textbf{Y}(\boldsymbol{\beta}) - \textbf{X}(\boldsymbol{\beta})\boldsymbol{\beta}||^2 + \lambda_n \sum^{p_n}_{j=1} \frac{\beta^2_{j}}{(\check{\beta}_{j})^2},
$$
by using Cholesky decomposition. Since $(\widetilde{\boldsymbol{\beta}}, \widetilde{\boldsymbol{\phi}}) = \argmax_{(\boldsymbol{\beta}, \boldsymbol{\phi})} \ell_n(\boldsymbol{\beta}, \boldsymbol{\phi})$, then
$$
\widetilde{\boldsymbol{\beta}} = \argmax_{\boldsymbol{\beta}} \ell_n(\boldsymbol{\beta}, \widetilde{\boldsymbol{\phi}}) = \argmax_{\boldsymbol{\beta}} \ell_n(\boldsymbol{\beta}| \widetilde{\boldsymbol{\phi}}),
$$
where $\ell_n(\boldsymbol{\beta}| \widetilde{\boldsymbol{\phi}}) = \log \widetilde{\mathcal{L}}_n(\boldsymbol{\beta}, \widetilde{\boldsymbol{\phi}})$ and $\ell_n(\boldsymbol{\beta} | \boldsymbol{\phi}) = \ell_n(\boldsymbol{\beta}, \boldsymbol{\phi})$.

Let $\Dot{\ell}_n(\boldsymbol{\beta} | \boldsymbol{\phi}) = \partial \ell_n(\boldsymbol{\beta}, \boldsymbol{\phi}) / \partial \boldsymbol{\beta}$ and $\Ddot{\ell}_n(\boldsymbol{\beta} | \boldsymbol{\phi}) = \partial^2 \ell_n(\boldsymbol{\beta}, \boldsymbol{\phi}) / \partial \boldsymbol{\beta} \partial \boldsymbol{\beta}^\top$. Then, $(\widetilde{\boldsymbol{\beta}}, \widetilde{\boldsymbol{\phi}})$ satisfies $\Dot{\ell}_n(\widetilde{\boldsymbol{\beta}} | \widetilde{\boldsymbol{\phi}}) = \boldsymbol{0}$. By the first-order Taylor expansion of $\Dot{\ell}_n(\widetilde{\boldsymbol{\beta}} | \widetilde{\boldsymbol{\phi}})$ around $\boldsymbol{\beta}$, we have 
$$
\boldsymbol{0} = \Dot{\ell}_n(\widetilde{\boldsymbol{\beta}} | \widetilde{\boldsymbol{\phi}}) \approx \Dot{\ell}_n(\boldsymbol{\beta} | \widetilde{\boldsymbol{\phi}}) + \Ddot{\ell}_n(\boldsymbol{\beta} | \widetilde{\boldsymbol{\phi}}) (\widetilde{\boldsymbol{\beta}} - \boldsymbol{\beta}),
$$
which yields
$$
\widetilde{\boldsymbol{\beta}} - \boldsymbol{\beta} \approx -[\Ddot{\ell}_n(\boldsymbol{\beta} | \widetilde{\boldsymbol{\phi}})]^{-1} \Dot{\ell}_n(\boldsymbol{\beta} | \widetilde{\boldsymbol{\phi}}).
$$
On the other hand, by the second-order Taylor expansion of $\ell_n(\widetilde{\boldsymbol{\beta}} | \widetilde{\boldsymbol{\phi}})$ around $\boldsymbol{\beta}$ yields
$$
\ell_n(\widetilde{\boldsymbol{\beta}} | \widetilde{\boldsymbol{\phi}}) \approx \ell_n(\boldsymbol{\beta} | \widetilde{\boldsymbol{\phi}}) + (\widetilde{\boldsymbol{\beta}} - \boldsymbol{\beta})^\top \Dot{\ell}_n(\boldsymbol{\beta} | \widetilde{\boldsymbol{\phi}}) + (\widetilde{\boldsymbol{\beta}} - \boldsymbol{\beta})^\top \frac{\Ddot{\ell}_n(\boldsymbol{\beta} | \widetilde{\boldsymbol{\phi}})}{2} (\widetilde{\boldsymbol{\beta}} - \boldsymbol{\beta}).
$$
Thus we have
\begin{equation*}
\begin{split}
\ell_p(\boldsymbol{\beta}) = \ell_n(\boldsymbol{\beta} | \widetilde{\boldsymbol{\phi}}) &\approx \ell_n(\widetilde{\boldsymbol{\beta}} | \widetilde{\boldsymbol{\phi}}) + [\Dot{\ell}_n(\boldsymbol{\beta} | \widetilde{\boldsymbol{\phi}})]^\top [\Ddot{\ell}_n(\boldsymbol{\beta} | \widetilde{\boldsymbol{\phi}})]^{-1} \Dot{\ell}_n(\boldsymbol{\beta} | \widetilde{\boldsymbol{\phi}}) \\
&- \frac{1}{2} [\Dot{\ell}_n(\boldsymbol{\beta} | \widetilde{\boldsymbol{\phi}})]^\top [\Ddot{\ell}_n(\boldsymbol{\beta} | \widetilde{\boldsymbol{\phi}})]^{-1} [\Ddot{\ell}_n(\boldsymbol{\beta} | \widetilde{\boldsymbol{\phi}})] [\Ddot{\ell}_n(\boldsymbol{\beta} | \widetilde{\boldsymbol{\phi}})]^{-1} \Dot{\ell}_n(\boldsymbol{\beta} | \widetilde{\boldsymbol{\phi}}).
\end{split}
\end{equation*}
Hence,
\begin{equation*}
\ell_p(\boldsymbol{\beta}) =  \frac{1}{2} [\Dot{\ell}_n(\boldsymbol{\beta} | \widetilde{\boldsymbol{\phi}})]^\top [\Ddot{\ell}_n(\boldsymbol{\beta} | \widetilde{\boldsymbol{\phi}})]^{-1} \Dot{\ell}_n(\boldsymbol{\beta} | \widetilde{\boldsymbol{\phi}}) + c_2,
\end{equation*}
where $c_2 = \ell_n(\widetilde{\boldsymbol{\beta}} | \widetilde{\boldsymbol{\phi}})$ is a constant independent of $\boldsymbol{\beta}$. Hence, maximizing $\ell_p(\boldsymbol{\beta})$ is equivalent to minimizing 
\begin{equation*}
-\ell_p(\boldsymbol{\beta}) = -\frac{1}{2} [\Dot{\ell}_n(\boldsymbol{\beta} | \widetilde{\boldsymbol{\phi}})]^\top [\Ddot{\ell}_n(\boldsymbol{\beta} | \widetilde{\boldsymbol{\phi}})]^{-1} \Dot{\ell}_n(\boldsymbol{\beta} | \widetilde{\boldsymbol{\phi}}).
\end{equation*}
Next, we show that $-\ell_p(\boldsymbol{\beta}) = \frac{1}{2} ||\textbf{Y}(\boldsymbol{\beta}) - \textbf{X}(\boldsymbol{\beta})\boldsymbol{\beta} ||^2$ by the Cholesky decomposition. 

Let $\textbf{X}(\boldsymbol{\beta})$ be the Cholesky decomposition of $-\Ddot{\ell}_n(\boldsymbol{\beta} | \widetilde{\boldsymbol{\phi}})$ as $-\Ddot{\ell}_n(\boldsymbol{\beta} | \widetilde{\boldsymbol{\phi}}) = \textbf{X}^\top(\boldsymbol{\beta}) \textbf{X}(\boldsymbol{\beta})$, and let $\textbf{Y}(\boldsymbol{\beta}) = [\textbf{X}^\top(\boldsymbol{\beta})]^{-1} [\Dot{\ell}_n(\boldsymbol{\beta} | \widetilde{\boldsymbol{\phi}}) - \Ddot{\ell}_n(\boldsymbol{\beta} | \widetilde{\boldsymbol{\phi}})\boldsymbol{\beta}]$ be the pseudo-response vector. Then, we have
$$
\frac{1}{2} ||\textbf{Y}(\boldsymbol{\beta}) - \textbf{X}(\boldsymbol{\beta})\boldsymbol{\beta}||^2 = -\frac{1}{2} [\Dot{\ell}_n(\boldsymbol{\beta} | \widetilde{\boldsymbol{\phi}})]^\top [\Ddot{\ell}_n(\boldsymbol{\beta} | \widetilde{\boldsymbol{\phi}})]^{-1} \Dot{\ell}_n(\boldsymbol{\beta} | \widetilde{\boldsymbol{\phi}}).
$$
Unlike \citet{zhao2019simultaneous}, here we write $\textbf{Y}(\boldsymbol{\beta})$ and $\textbf{X}(\boldsymbol{\beta})$ to emphasize the dependence of \textbf{X} and \textbf{Y} on $\boldsymbol{\beta}$. Note that in terms of notation, we consider $\textbf{X}(\boldsymbol{\beta}) = \textbf{X}(\boldsymbol{\beta} | \widetilde{\boldsymbol{\phi}})$, and $\textbf{Y}(\boldsymbol{\beta}) = \textbf{Y}(\boldsymbol{\beta} | \widetilde{\boldsymbol{\phi}})$.

To prove Theorem 1, first we introduce the following notations. Define
\begin{equation} \label{A1}
\begin{pmatrix} \boldsymbol{\alpha}^*(\boldsymbol{\beta}) \\ \boldsymbol{\gamma}^*(\boldsymbol{\beta})\end{pmatrix} \equiv g(\boldsymbol{\beta}) = \{ \boldsymbol{\Omega}_n(\boldsymbol{\beta}) + \lambda_n \textbf{D}(\boldsymbol{\beta}) \}^{-1} \textbf{v}_n(\boldsymbol{\beta})
\end{equation}
and partition the matrix $\{ n^{-1} \boldsymbol{\Omega}_n(\boldsymbol{\beta})\}^{-1}$ into
$$
\{ n^{-1} \boldsymbol{\Omega}_n(\boldsymbol{\beta})\}^{-1} = \begin{pmatrix} \textbf{A}(\boldsymbol{\beta}) &  \textbf{B}(\boldsymbol{\beta}) \\ \textbf{B}^\top(\boldsymbol{\beta}) & \textbf{G}(\boldsymbol{\beta}) \end{pmatrix},
$$
where $\textbf{A}(\boldsymbol{\beta}), \textbf{B}(\boldsymbol{\beta})$ and $\textbf{G}(\boldsymbol{\beta})$ are $q_n \times q_n, q_n \times (p_n - q_n)$ and $(p_n - q_n) \times (p_n - q_n)$ matrices, respectively. Here, we use $\boldsymbol{\Omega}_n(\boldsymbol{\beta})$ and $\textbf{v}_n(\boldsymbol{\beta})$ instead of $\boldsymbol{\Omega}_n$ and $\textbf{v}_n$, respectively, to emphasize the dependence of $\boldsymbol{\Omega}_n$ and $\textbf{v}_n$ on $\boldsymbol{\beta}$. This is important in the subsequent proofs, especially in Lemma 2.

Multiplying $\boldsymbol{\Omega}^{-1}_n(\boldsymbol{\beta})(\boldsymbol{\Omega}_n(\boldsymbol{\beta}) + \lambda_n \textbf{D}(\boldsymbol{\beta}))$ and substituting $\boldsymbol{\beta}_{s0} = (\boldsymbol{\beta}^\top_{s01}, \boldsymbol{\beta}^\top_{s02})^\top$ on both sides of 
\begin{equation} \label{A2}
\begin{pmatrix} \boldsymbol{\alpha}^*(\boldsymbol{\beta}) - \boldsymbol{\beta}_{s01} \\ \boldsymbol{\gamma}^*(\boldsymbol{\beta}) \end{pmatrix} + \frac{\lambda_n}{n} \begin{pmatrix} \textbf{A}(\boldsymbol{\beta}) \textbf{D}_1(\boldsymbol{\beta}_{s1}) \boldsymbol{\alpha}^*(\boldsymbol{\beta})  + \textbf{B}(\boldsymbol{\beta}) \textbf{D}_2(\boldsymbol{\beta}_{s2}) \boldsymbol{\gamma}^*(\boldsymbol{\beta}) \\ \textbf{B}^\top(\boldsymbol{\beta})\textbf{D}_1(\boldsymbol{\beta}_{s1})\boldsymbol{\alpha}^*(\boldsymbol{\beta}) + \textbf{G}(\boldsymbol{\beta}) \textbf{D}_2(\boldsymbol{\beta}_{s2}) \boldsymbol{\gamma}^*(\boldsymbol{\beta})\end{pmatrix} = \widehat{\textbf{b}}(\boldsymbol{\beta}) - \boldsymbol{\beta}_{s0},
\end{equation}
where $\widehat{\textbf{b}}(\boldsymbol{\beta}) = \boldsymbol{\Omega}^{-1}_n(\boldsymbol{\beta}) \textbf{v}_n(\boldsymbol{\beta}), \; \textbf{D}_1(\boldsymbol{\beta}_{s1}) = \text{diag}(\beta^{-2}_{s1,1},\ldots,\beta^{-2}_{s1,q_n})$ and $\textbf{D}_2(\boldsymbol{\beta}_{s2}) = \text{diag}(\beta^{-2}_{s2,q_n +1},\ldots,\beta^{-2}_{s2,p_n})$. 

The following three Lemmas (Lemmas 1 - 3) are needed to prove Theorem 1.
\vspace*{20px}

\noindent
\textbf{Lemma 1}: \textit{Let $\delta$ be a large positive constant. Define $H_{n1} = \{\boldsymbol{\beta}_{s1} : ||\boldsymbol{\beta}_{s1} - \boldsymbol{\beta}_{s01}|| \leq \delta \sqrt{p_n / n} \}$ and $H_{n2} = \{\boldsymbol{\beta}_{s2} : ||\boldsymbol{\beta}_{s2} - \boldsymbol{\beta}_{s02}|| \leq \delta \sqrt{p_n / n} \}, H_n = H_{n1} \otimes H_{n2}$. Then, under conditions \textbf{C1} - \textbf{C9}, with probability tending to one, we have}
\vspace*{7px}

\noindent
(i) $\sup_{\boldsymbol{\beta} \in H_n} ||\widehat{\textbf{b}}(\boldsymbol{\beta}) - \boldsymbol{\beta}_{s0}|| = O_p(\sqrt{p_n / n})$, \\
(ii) $\sup_{\boldsymbol{\beta} \in H_n} \frac{||\boldsymbol{\gamma}^*(\boldsymbol{\beta})||}{||\boldsymbol{\beta}_{s2}||} < \frac{1}{c_0}$ for some constant $c_0 > 1$, \\
(iii) $g(\cdot)$ is mapping from $H_n$ to itself. \\

\noindent 
\textbf{Proof of Lemma 1}. \\
We want to show 
$$
(i) \sup_{\boldsymbol{\beta} \in H_n} ||\widehat{\textbf{b}}(\boldsymbol{\beta}) - \boldsymbol{\beta}_{s0}|| = O_p(\sqrt{p_n / n}).
$$
Let $(\widetilde{\boldsymbol{\beta}}, \widetilde{\boldsymbol{\phi}})$ be the values that satisfy $\Dot{\ell}_n(\widetilde{\boldsymbol{\beta}} | \widetilde{\boldsymbol{\phi}}) = \boldsymbol{0}$. Since $\boldsymbol{\Omega}_n(\boldsymbol{\beta}) = -\Ddot{\ell}_n (\boldsymbol{\beta} | \widetilde{\boldsymbol{\phi}}), \; \textbf{v}_n(\boldsymbol{\beta}) = \Dot{\ell}_n(\boldsymbol{\beta} | \widetilde{\boldsymbol{\phi}}) - \Ddot{\ell}_n (\boldsymbol{\beta} | \widetilde{\boldsymbol{\phi}})\boldsymbol{\beta}$, we have
\begin{equation*}
\begin{split}
\widehat{\textbf{b}}(\boldsymbol{\beta}) & = \boldsymbol{\Omega}^{-1}_n(\boldsymbol{\beta}) \textbf{v}_n(\boldsymbol{\beta}) \\
& = [-\Ddot{\ell}_n(\boldsymbol{\beta} | \widetilde{\boldsymbol{\phi}})]^{-1} [\Dot{\ell}_n(\boldsymbol{\beta} | \widetilde{\boldsymbol{\phi}}) - \Ddot{\ell}_n (\boldsymbol{\beta} | \widetilde{\boldsymbol{\phi}})\boldsymbol{\beta}] \\
& = \boldsymbol{\beta} - [\Ddot{\ell}_n(\boldsymbol{\beta} | \widetilde{\boldsymbol{\phi}})]^{-1}[\Dot{\ell}_n(\boldsymbol{\beta} | \widetilde{\boldsymbol{\phi}})].
\end{split}
\end{equation*}
Using first-order Taylor expansion of $\Dot{\ell}_n(\boldsymbol{\beta} | \widetilde{\boldsymbol{\phi}})$ at $\widetilde{\boldsymbol{\beta}}$, we obtain
\begin{equation*}
\begin{split}
\Dot{\ell}_n(\boldsymbol{\beta} | \widetilde{\boldsymbol{\phi}}) &= \Dot{\ell}_n(\widetilde{\boldsymbol{\beta}} | \widetilde{\boldsymbol{\phi}}) + \Ddot{\ell}_n( \widetilde{\boldsymbol{\beta}}^{*} | \widetilde{\boldsymbol{\phi}})(\boldsymbol{\beta} - \widetilde{\boldsymbol{\beta}}) \\
&= \boldsymbol{0} + \Ddot{\ell}_n( \widetilde{\boldsymbol{\beta}}^{*} | \widetilde{\boldsymbol{\phi}})(\boldsymbol{\beta} - \widetilde{\boldsymbol{\beta}}),
\end{split}
\end{equation*}
where $\widetilde{\boldsymbol{\beta}}^{*}$ is between $\widetilde{\boldsymbol{\beta}}$ and $\boldsymbol{\beta}$. Then,
\begin{equation*}
\begin{split}
\widehat{\textbf{b}}(\boldsymbol{\beta}) &= \boldsymbol{\beta} - [\Ddot{\ell}_n(\boldsymbol{\beta} | \widetilde{\boldsymbol{\phi}})]^{-1} [\Ddot{\ell}_n( \widetilde{\boldsymbol{\beta}}^{*} | \widetilde{\boldsymbol{\phi}})](\boldsymbol{\beta} - \widetilde{\boldsymbol{\beta}}) \\
&= \boldsymbol{\beta} - [\boldsymbol{\Omega}_n(\boldsymbol{\beta})]^{-1} [\boldsymbol{\Omega}_n(\widetilde{\boldsymbol{\beta}}^*)] (\boldsymbol{\beta} - \widetilde{\boldsymbol{\beta}}).
\end{split}
\end{equation*}
Since $||\widetilde{\boldsymbol{\beta}} - \boldsymbol{\beta}_{s0}|| = O_p(\sqrt{p_n / n}) = o_p(1)$, if $\boldsymbol{\beta} \in H_n$, then $\sup_{\boldsymbol{\beta} \in H_n} ||\boldsymbol{\beta} - \boldsymbol{\beta}_{s0}|| \leq \sqrt{2}\delta\sqrt{p_n / n} = O_p(1)$, and $||\widetilde{\boldsymbol{\beta}}^{*} - \boldsymbol{\beta}_{s0}|| = o_p(1)$. By Condition \textbf{C4}, we have
$$
n^{-1} \boldsymbol{\Omega}_n(\boldsymbol{\beta}) = \textbf{I}(\boldsymbol{\beta}_{s0}) + o_p(1) \;\; \text{and} \;\; n^{-1} \boldsymbol{\Omega}_n(\widetilde{\boldsymbol{\beta}}^{*}) = \textbf{I}(\boldsymbol{\beta}_{s0}) + o_p(1)
$$
uniformly for $\boldsymbol{\beta} \in H_n$. Therefore, 
$$
[n^{-1} \boldsymbol{\Omega}_n(\boldsymbol{\beta})]^{-1} = \textbf{I}(\boldsymbol{\beta}_{s0})^{-1} + o_p(1) \;\; \text{and} \;\; [n^{-1} \boldsymbol{\Omega}_n(\boldsymbol{\beta})]^{-1} [n^{-1} \boldsymbol{\Omega}_n(\widetilde{\boldsymbol{\beta}}^*)] = \textbf{I}_{p_n} + o_p(1),
$$
and
\begin{equation*}
\begin{split}
\widehat{\textbf{b}}(\boldsymbol{\beta}) &= \boldsymbol{\beta} - (\textbf{I}_{p_n} + o_p(1))(\boldsymbol{\beta} - \widetilde{\boldsymbol{\beta}}) \\
& = \boldsymbol{\beta} - (\boldsymbol{\beta} - \widetilde{\boldsymbol{\beta}}) + o_p(1) (\boldsymbol{\beta} - \widetilde{\boldsymbol{\beta}}) \\
& = \widetilde{\boldsymbol{\beta}} + o_p(1) (\boldsymbol{\beta} - \widetilde{\boldsymbol{\beta}}), \\
\widehat{\textbf{b}}(\boldsymbol{\beta}) - \boldsymbol{\beta}_{s0} &=  \widetilde{\boldsymbol{\beta}} - \boldsymbol{\beta}_{s0} + o_p(1) (\boldsymbol{\beta} - \boldsymbol{\beta}_{s0} - (\widetilde{\boldsymbol{\beta}} - \boldsymbol{\beta}_{s0})) \\
& = \widetilde{\boldsymbol{\beta}} - \boldsymbol{\beta}_{s0} + o_p(1)(\boldsymbol{\beta} - \boldsymbol{\beta}_{s0}) + o_p(1)(\widetilde{\boldsymbol{\beta}} - \boldsymbol{\beta}_{s0}),
\end{split}
\end{equation*}
therefore
\begin{equation*}
||\widehat{\textbf{b}}(\boldsymbol{\beta}) - \boldsymbol{\beta}_{s0}|| \leq ||\widetilde{\boldsymbol{\beta}} - \boldsymbol{\beta}_{s0}|| + o_p(1) ||\boldsymbol{\beta} - \boldsymbol{\beta}_{s0}|| + o_p(1) ||\widetilde{\boldsymbol{\beta}} - \boldsymbol{\beta}_{s0}||.
\end{equation*}
Thus, 
\begin{equation*}
\begin{split}
\sup_{\boldsymbol{\beta} \in H_n}||\widehat{\textbf{b}}(\boldsymbol{\beta}) - \boldsymbol{\beta}_{s0}|| &\leq ||\widetilde{\boldsymbol{\beta}} - \boldsymbol{\beta}_{s0}|| + o_p(1) \sup_{\boldsymbol{\beta} \in H_n} ||\boldsymbol{\beta} - \boldsymbol{\beta}_{s0}|| + o_p(1) ||\widetilde{\boldsymbol{\beta}} - \boldsymbol{\beta}_{s0}|| \\
& \leq O_p(\sqrt{p_n/n}) + o_p(1) (\sqrt{2}\delta \sqrt{p_n / n}) + o_p(1) O_p(\sqrt{p_n /n}) \\
& =  O_p(\sqrt{p_n/n}) + o_p(1) (\sqrt{2}\delta \sqrt{p_n / n}) + o_p(\sqrt{p_n / n}) \\
& = O_p(\sqrt{p_n/n}), 
\end{split}
\end{equation*}
and \textbf{\textit{Lemma 1(i)}} is proven. Since we have proved (i), $\sup_{\boldsymbol{\beta} \in H_n} \big\lVert \widehat{\textbf{b}}(\boldsymbol{\beta}) - \boldsymbol{\beta}_{s0} \big\rVert = O_p(\sqrt{p_n /n})$, it follows from \eqref{A2} that 
\begin{equation*}
\sup_{\boldsymbol{\beta} \in H_n} \left\lVert \boldsymbol{\gamma}^*(\boldsymbol{\beta}) + \frac{\lambda_n}{n} \textbf{B}^\top(\boldsymbol{\beta}) \textbf{D}_1(\boldsymbol{\beta}_{s1})\boldsymbol{\alpha}^*(\boldsymbol{\beta}) + \frac{\lambda_n}{n} \textbf{G}(\boldsymbol{\beta})\textbf{D}_2(\boldsymbol{\beta}_{s2}) \boldsymbol{\gamma}^*(\boldsymbol{\beta})  \right\rVert = O_p(\sqrt{p_n/n}).
\end{equation*}
In sequel, for a matrix $\textbf{A}$, $\left\lVert \textbf{A} \right\rVert$ represents the induced 2-norm. Then, using the properties of the matrix 2-norm, we have
\begin{equation*}
\begin{split}
\left\lVert \textbf{B}(\boldsymbol{\beta}) \right\rVert &\leq \left\lVert (n^{-1} \boldsymbol{\Omega}_n(\boldsymbol{\beta}))^{-1} \right\rVert  = \lambda_{\max}\{ (n^{-1}\boldsymbol{\Omega}_n(\boldsymbol{\beta}))^{-1} \} = [\lambda_{\min}(n^{-1}\boldsymbol{\Omega}_n(\boldsymbol{\beta}))]^{-1} \\
& \leq (1/c)^{-1} = c, \text{ where $c$ is given in condition \textbf{C5}.}
\end{split}
\end{equation*}
i.e., $\sup_{\boldsymbol{\beta} \in H_n} \left\lVert \textbf{B}(\boldsymbol{\beta}) \right\rVert \leq c$. Similarly, we have $\sup_{\boldsymbol{\beta} \in H_n} \left\lVert \textbf{B}^\top(\boldsymbol{\beta}) \right\rVert \leq c$. Thus,
\begin{equation}\label{A3}
\sup_{\boldsymbol{\beta} \in H_n} \left\lVert \frac{\lambda_n}{n} \textbf{B}^\top (\boldsymbol{\beta}) \textbf{D}_1(\boldsymbol{\beta}_{s1}) \boldsymbol{\alpha}^*(\boldsymbol{\beta}) \right\rVert \leq \bigg(\frac{\lambda_n}{\sqrt{n}} \bigg) O_p(\sqrt{p_n /n}) = o_p(\sqrt{p_n/n}).
\end{equation}
Since $g(\boldsymbol{\beta}) = \begin{pmatrix} \boldsymbol{\alpha}^*(\boldsymbol{\beta}) \\ \boldsymbol{\gamma}^*(\boldsymbol{\beta}) \end{pmatrix}$,
\begin{equation*}
    \begin{split}
        g(\boldsymbol{\beta}) &= \{ \boldsymbol{\Omega}_n(\boldsymbol{\beta}) + \lambda_n \textbf{D}_n(\boldsymbol{\beta})\}^{-1} \textbf{v}_n(\boldsymbol{\beta}) \\
        &= [(\boldsymbol{\Omega}_n(\boldsymbol{\beta}) + \lambda_n \textbf{D}_n(\boldsymbol{\beta}))^{-1} \boldsymbol{\Omega}_n(\boldsymbol{\beta})][(\boldsymbol{\Omega}_n(\boldsymbol{\beta}))^{-1}\textbf{v}_n(\boldsymbol{\beta})] \\
        &= [(\boldsymbol{\Omega}_n(\boldsymbol{\beta}) + \lambda_n \textbf{D}_n(\boldsymbol{\beta}))^{-1} \boldsymbol{\Omega}_n(\boldsymbol{\beta})] \widehat{\textbf{b}}(\boldsymbol{\beta}),
    \end{split}
\end{equation*}
by condition \textbf{C5}, there exists a constant $M > 0$, such that
$$
\sup_{\boldsymbol{\beta} \in H_n} \left\lVert (\boldsymbol{\Omega}_n(\boldsymbol{\beta}) + \lambda_n \textbf{D}_n(\boldsymbol{\beta}))^{-1} \boldsymbol{\Omega}_n(\boldsymbol{\beta}) \right\rVert \leq M.
$$
Then, we have
$$
\left\lVert g(\boldsymbol{\beta}) \right\rVert \leq \left\lVert (\boldsymbol{\Omega}_n(\boldsymbol{\beta}) + \lambda_n \textbf{D}_n(\boldsymbol{\beta}))^{-1} \boldsymbol{\Omega}_n(\boldsymbol{\beta}) \right\rVert \cdot \left\lVert \widehat{\textbf{b}}(\boldsymbol{\beta}) \right\rVert \leq M \left\lVert \widehat{\textbf{b}}(\boldsymbol{\beta})\right\rVert.
$$
On the other hand, $\left\lVert g(\boldsymbol{\beta}) \right\rVert^2 = \left\lVert \boldsymbol{\alpha}^*(\boldsymbol{\beta}) \right\rVert^2 + \left\lVert \boldsymbol{\gamma}^*(\boldsymbol{\beta}) \right\rVert^2$, and
\begin{equation*}
\begin{split}
    \left\lVert \widehat{\textbf{b}}(\boldsymbol{\beta}) \right\rVert=\left\lVert \widehat{\textbf{b}}(\boldsymbol{\beta})-\boldsymbol{\beta}_{s0} \right\rVert &\leq o_p\left(\sqrt{p_n/n} \right)+a_1\sqrt{q_n}\\
    &=O_p(q_n).
\end{split}
\end{equation*}
i.e.,
\begin{eqnarray}\label{B1}
    \sup_{\boldsymbol{\beta}\in H_n}\left\lVert \widehat{\textbf{b}}(\boldsymbol{\beta})\right\rVert=O_p(\sqrt{q_n}).
\end{eqnarray}
By \textbf{\textit{Lemma 1(i)}} and condition \textbf{C7}, we also have
$$
\left\lVert \boldsymbol{\alpha}^*(\boldsymbol{\beta}) \right\rVert \leq \left\lVert g(\boldsymbol{\beta}) \right\rVert \leq M || \widehat{\textbf{b}}(\boldsymbol{\beta}) ||.
$$
Then
\begin{eqnarray}\label{B2}
\sup_{\boldsymbol{\beta} \in H_n} \left\lVert \boldsymbol{\alpha}^*(\boldsymbol{\beta})\right\rVert \leq M \sup_{\boldsymbol{\beta} \in H_n} \left\lVert \widehat{\textbf{b}}(\boldsymbol{\beta})\right\rVert = O_p(\sqrt{q_n}).
\end{eqnarray}
Now, we consider $\left\lVert \textbf{D}_1(\boldsymbol{\beta}_{s1})\right\rVert$. Since
$$
\left\lVert \textbf{D}_1(\boldsymbol{\beta}_{s1})\right\rVert = \lambda_{\max}\{ \textbf{D}_1(\boldsymbol{\beta}_{s1})\} = \max_{1 \leq j \leq q_n} \bigg\{\frac{1}{\beta^2_{s1,j}} \bigg\} = \frac{1}{\min_{1 \leq j \leq q_n} \{1/\beta^2_{s1,j} \}},
$$
when $\boldsymbol{\beta} \in H_n$, we have $\left\lVert \boldsymbol{\beta} - \boldsymbol{\beta}_{s0} \right\rVert \leq \sqrt{2} \delta \sqrt{p_n/n}$, then $\left\lVert \beta_{s1,j} - \beta_{s0,j}\right\rVert = | \beta_{s1,j} - \beta_{s0,j}| \leq \delta \sqrt{p_n /n}$, i.e.,
$$
|\beta_{s0,j}| - \delta\sqrt{\frac{p_n}{n}} \leq |\beta_{s1,j}| \leq |\beta_{s0,j}| + \delta \sqrt{\frac{p_n}{n}}, \; 1 \leq j \leq q_n.
$$
By Condition \textbf{C7}, when $n$ is sufficiently large, we have
$$
\frac{a_0}{2} \leq |\beta_{s1,j}| \leq 2a_1, \text{ since } \delta \sqrt{\frac{p_n}{n}} \rightarrow 0, \text{ as } n \rightarrow \infty.
$$
Then, 
$$
\big( \frac{a_0}{2} \big)^2 \leq \min_{1 \leq j \leq q_n} \{\beta^2_{s1,j}\} \leq \max_{1 \leq j \leq q_n} \{\beta^2_{s1,j}\} \leq (2a_1)^2,
$$
$$
\left\lVert\textbf{D}_1(\boldsymbol{\beta}_{s1})\right\rVert = \frac{1}{\min_{1 \leq j \leq q_n} \{\beta^2_{s1,j}\}} \leq \frac{1}{(a_0/2)^2} = \frac{4}{a^2_0}.
$$
This implies
\begin{eqnarray}\label{B3}
\sup_{\boldsymbol{\beta} \in H_n} \left\lVert \textbf{D}_1(\boldsymbol{\beta}_{s1})\right\rVert \leq 4/a^2_0. 
\end{eqnarray}
Therefore, by \eqref{B1}, \eqref{B2}, and \eqref{B3}, we have
\begin{equation*}
    \begin{split}
        \sup_{\boldsymbol{\beta} \in H_n}  \left\lVert \frac{\lambda_n}{n} \textbf{B}^\top (\boldsymbol{\beta}) \textbf{D}_1(\boldsymbol{\beta}_{s1}) \boldsymbol{\alpha}^*(\boldsymbol{\beta}) \right\rVert & \leq \frac{\lambda_n}{n} \cdot \sup_{\boldsymbol{\beta} \in H_n} \vert\vert \textbf{B}^\top (\boldsymbol{\beta}) \vert\vert \cdot \sup_{\boldsymbol{\beta} \in H_n} \left\lVert\textbf{D}_1(\boldsymbol{\beta}_{s1}) \right\rVert \cdot \sup_{\boldsymbol{\beta} \in H_n} \left\lVert \boldsymbol{\alpha}^*(\boldsymbol{\beta})\right\rVert \\
        & \leq \frac{\lambda_n}{n} \cdot c \cdot \frac{4}{a^2_0} \cdot O_p(\sqrt{q_n}) \\
        & = \frac{4c\lambda_n}{a^2_0 \sqrt{n}}  O_p(\sqrt{q_n / n}) .
    \end{split}
\end{equation*}
Since Condition \textbf{C6} states that $\lambda_n / \sqrt{n} \longrightarrow 0$ as $n \longrightarrow \infty$,
$$
 \sup_{\boldsymbol{\beta} \in H_n} \left\lVert \frac{\lambda_n}{n} \textbf{B}^\top (\boldsymbol{\beta}) \textbf{D}_1(\boldsymbol{\beta}_{s1}) \boldsymbol{\alpha}^*(\boldsymbol{\beta}) \right\rVert = o(1) \cdot O_p(\sqrt{q_n / n}) = o_p(\sqrt{q_n / n}) = o_p(\sqrt{p_n / n}).
$$
The proof of \eqref{A3} is completed. Next, we prove \eqref{A4}
\begin{equation} \label{A4}
    c^{-1} \left\lVert\frac{\lambda_n}{n} \textbf{D}_2(\boldsymbol{\beta}_{s2}) \boldsymbol{\gamma}^*(\boldsymbol{\beta}) \right\rVert - \left\lVert \boldsymbol{\gamma}^*(\boldsymbol{\beta}) \right\rVert \leq o_p(\delta \sqrt{p_n / n}).
\end{equation}
From \eqref{A2}, we obtain
$$
\left\lVert\boldsymbol{\gamma}^*(\boldsymbol{\beta}) + \frac{\lambda_n}{n} \{\textbf{B}^\top (\boldsymbol{\beta}) \textbf{D}_1(\boldsymbol{\beta}_{s1}) \boldsymbol{\alpha}^*(\boldsymbol{\beta}) + \textbf{G}(\boldsymbol{\beta})\textbf{D}_2(\boldsymbol{\beta}_{s2})\boldsymbol{\gamma}^*(\boldsymbol{\beta}) \} \right\rVert \leq \left\lVert \widehat{\textbf{b}}(\boldsymbol{\beta}) - \boldsymbol{\beta}_{s0} \right\rVert,
$$
it implies 
\begin{eqnarray}
\label{E1}
\left\lVert\frac{\lambda_n}{n} \textbf{G}(\boldsymbol{\beta})\textbf{D}_2(\boldsymbol{\beta}_{s2})\boldsymbol{\gamma}^*(\boldsymbol{\beta})\right\rVert - \left\lVert\boldsymbol{\gamma}^*(\boldsymbol{\beta})\right\rVert - \left\Vert\frac{\lambda_n}{n}\textbf{B}^\top (\boldsymbol{\beta}) \textbf{D}_1(\boldsymbol{\beta}_{s1}) \boldsymbol{\alpha}^*(\boldsymbol{\beta})\right\rVert \leq  \left\lVert \widehat{\textbf{b}}(\boldsymbol{\beta}) - \boldsymbol{\beta}_{s0} \right\rVert.
\end{eqnarray}
Now consider 
\begin{equation*}
  \begin{split}
    \left\lVert \frac{\lambda_n}{n} \textbf{D}_2(\boldsymbol{\beta}_{s2}) \boldsymbol{\gamma}^*(\boldsymbol{\beta}) \right\rVert & = \left\lVert \frac{\lambda_n}{n} \textbf{G}^{-1}(\boldsymbol{\beta})\textbf{G}(\boldsymbol{\beta})\textbf{D}_2(\boldsymbol{\beta}_{s2})\boldsymbol{\gamma}^*(\boldsymbol{\beta}) \right\rVert \\
    & \leq \left\lVert \textbf{G}^{-1}(\boldsymbol{\beta}) \right\rVert \cdot \left\lVert\frac{\lambda_n}{n} \textbf{G}(\boldsymbol{\beta})\textbf{D}_2(\boldsymbol{\beta}_{s2})\boldsymbol{\gamma}^*(\boldsymbol{\beta}) \right\rVert,
  \end{split}
\end{equation*}
it yields
$$
\left\lVert \frac{\lambda_n}{n} \textbf{G}(\boldsymbol{\beta})\textbf{D}_2(\boldsymbol{\beta}_{s2})\boldsymbol{\gamma}^*(\boldsymbol{\beta})\right\rVert \geq \{ 1/ ||\textbf{G}^{-1}(\boldsymbol{\beta})|| \} \cdot \left\lVert\frac{\lambda_n}{n} \textbf{D}_2(\boldsymbol{\beta}_{s2})\boldsymbol{\gamma}^*(\boldsymbol{\beta}) \right\rVert.
$$
Since
\begin{equation*}
    \begin{split}
        \left\lVert\textbf{G}^{-1}(\boldsymbol{\beta})\right\rVert  = \lambda_{\max}\{\textbf{G}^{-1}(\boldsymbol{\beta}) \} &= \frac{1}{\lambda_{\min}\{\textbf{G}(\boldsymbol{\beta})\}} \\
    &\leq \frac{1}{\inf_{\boldsymbol{\beta} \in H_n} \lambda_{\min} \{ \textbf{G}(\boldsymbol{\beta})\}} \\
        & \leq 1/(1/c) = c,
    \end{split}
\end{equation*}
then $1/||\textbf{G}^{-1}(\boldsymbol{\beta})|| \leq 1/c$ and $\inf_{\boldsymbol{\beta} \in H_n} \{1 /||\textbf{G}^{-1}(\boldsymbol{\beta})|| \} \geq 1/c$, and
\begin{eqnarray} \label{E2}
\left\lVert \frac{\lambda_n}{n} \textbf{G}(\boldsymbol{\beta}) \textbf{D}_2(\boldsymbol{\beta}_{s2}) \boldsymbol{\gamma}^*(\boldsymbol{\beta}) \right\rVert \geq \frac{1}{c} \left\lVert \frac{\lambda_n}{n} \textbf{D}_2(\boldsymbol{\beta}_{s2}) \boldsymbol{\gamma}^*(\boldsymbol{\beta}) \right\rVert   
\end{eqnarray}
Then \eqref{E1}, \eqref{E2}, and \eqref{A3} and \textbf{\textit{Lemma 1(i)}} imply \eqref{A4}. 
\vspace*{15px}

\noindent
Here we explain why $\inf_{\boldsymbol{\beta} \in H_n} \{\lambda_{\min} \{\textbf{G}(\boldsymbol{\beta}) \} \} \geq 1/c$.
\begin{equation*}
    \begin{split}
        \text{Since } \lambda_{\min} \{\textbf{G}(\boldsymbol{\beta}) \} & \geq \lambda_{\min} \{ (n^{-1} \boldsymbol{\Omega}_n(\boldsymbol{\beta}))^{-1} \} \\
        & = \lambda_{\max} \{n^{-1} \boldsymbol{\Omega}_n(\boldsymbol{\beta}) \} \\
        & \geq \lambda_{\min} \{n^{-1} \boldsymbol{\Omega}_n(\boldsymbol{\beta}) \} \\
        & \geq \inf_{\boldsymbol{\beta} \in H_n} \{ \lambda_{\min} \{n^{-1} \boldsymbol{\Omega}_n(\boldsymbol{\beta}) \} \}, \text{ by condition \textbf{C5}.}
    \end{split}
\end{equation*}
Let $\frac{m_{\boldsymbol{\gamma}^*(\boldsymbol{\beta})}}{\boldsymbol{\beta}_{s2}} = (\boldsymbol{\gamma}^*(\boldsymbol{\beta})/\beta_{s2, q_n+1},\ldots,\boldsymbol{\gamma}^*(\boldsymbol{\beta}) /\beta_{s2, p_n})^\top$, then $\frac{m_{\boldsymbol{\gamma}^*(\boldsymbol{\beta})}}{\boldsymbol{\beta}_{s2}} = \text{diag}(\boldsymbol{\beta}_{s2}) \textbf{D}_2(\boldsymbol{\beta}_{s2}) \boldsymbol{\gamma}^*(\boldsymbol{\beta}) $, then
\begin{equation}
\label{E3}
\begin{split}
\left\lVert \frac{m_{\boldsymbol{\gamma}^*(\boldsymbol{\beta})}}{\boldsymbol{\beta}_{s2}} \right\rVert & \leq \left\lVert \text{diag}(\boldsymbol{\beta}_{s2}) \right\rVert \cdot \left\lVert \textbf{D}_2(\boldsymbol{\beta}_{s2}) \boldsymbol{\gamma}^*(\boldsymbol{\beta}) \right\rVert \\
& = \sqrt{\left\lVert \text{diag}(\boldsymbol{\beta}_{s2}) \right\rVert^2} \cdot \left\lVert\textbf{D}_2(\boldsymbol{\beta}_{s2}) \boldsymbol{\gamma}^*(\boldsymbol{\beta})\right\rVert \\
& = \sqrt{\max_{q_n + 1 \leq j \leq p_n} \beta^2_{s2,j}} \cdot \left\lVert\textbf{D}_2(\boldsymbol{\beta}_{s2}) \boldsymbol{\gamma}^*(\boldsymbol{\beta})\right\rVert \\
& \leq \left\lVert\boldsymbol{\beta}_{s2} \right\rVert \cdot \left\lVert\textbf{D}_2(\boldsymbol{\beta}_{s2}) \boldsymbol{\gamma}^*(\boldsymbol{\beta})\right\rVert \\
&\leq \delta \sqrt{p_n/n} \left\lVert\textbf{D}_2(\boldsymbol{\beta}_{s2}) \boldsymbol{\gamma}^*(\boldsymbol{\beta})\right\rVert.
\end{split}
\end{equation}
Write $\boldsymbol{\gamma}^*(\boldsymbol{\beta})=\text{diag}(\boldsymbol{\beta}_{s2})\frac{\boldsymbol{m}_{\boldsymbol{\gamma}^*(\boldsymbol{\beta})}}{\boldsymbol{\beta}_{s2}}$. Then 
\begin{eqnarray}
\label{E4}
    \left\lVert \boldsymbol{\gamma}^*(\boldsymbol{\beta})\right\rVert\leq \left\lVert \text{diag}(\boldsymbol{\beta}_{s2})\right\rVert\left\lVert\frac{\boldsymbol{m}_{\boldsymbol{\gamma}^*(\boldsymbol{\beta})}}{\boldsymbol{\beta}_{s2}}\right\rVert&\leq& \left\lVert \boldsymbol{\beta}_{s2}\right\rVert\left\lVert\frac{\boldsymbol{m}_{\boldsymbol{\gamma}^*(\boldsymbol{\beta})}}{\boldsymbol{\beta}_{s2}}\right\rVert\leq\delta\sqrt{p_n/n}\left\lVert\frac{\boldsymbol{m}_{\boldsymbol{\gamma}^*(\boldsymbol{\beta})}}{\boldsymbol{\beta}_{s2}}\right\rVert
\end{eqnarray}
\eqref{E2} and \eqref{E3} imply 
\begin{equation}\label{E5}
\begin{split}
    \left\lVert \frac{\lambda_n}{n}\textbf{G}(\boldsymbol{\beta})\textbf{D}_2(\boldsymbol{\beta}_{s2})\boldsymbol{\gamma}^*(\boldsymbol{\beta})) \right\rVert &\geq \frac{c\lambda_n}{n} \left\lVert \textbf{D}_2(\boldsymbol{\beta})\boldsymbol{\gamma}^*(\boldsymbol{\beta})) \right\rVert \nonumber\\
    &\geq \frac{c\lambda_n}{n} \left(\frac{\sqrt{n}}{\delta\sqrt{p_n}} \right)\left\lVert\frac{\boldsymbol{m}_{\boldsymbol{\gamma}^*(\boldsymbol{\beta})}}{\boldsymbol{\beta}_{s2}}\right\rVert.
\end{split}
\end{equation}
By \eqref{E1}, \eqref{E4}, \eqref{E5}, and \textbf{Lemma 1(i)}, we can conclude that
\begin{eqnarray*}
    (1/c)(\lambda_n/n)\left(\frac{\sqrt{n}}{\delta \sqrt{p_n}} \right)\left\lVert\frac{\boldsymbol{m}_{\boldsymbol{\gamma}^*(\boldsymbol{\beta})}}{\boldsymbol{\beta}_{s2}}\right\rVert-\delta \sqrt{p_n/n}\left\lVert\frac{\boldsymbol{m}_{\boldsymbol{\gamma}^*(\boldsymbol{\beta})}}{\boldsymbol{\beta}_{s2}}\right\rVert\leq o_p\left(\delta \sqrt{p_n/n} \right).
\end{eqnarray*}
Therefore, 
\begin{eqnarray*}
    \left[\frac{\lambda_n}{cn}\left(\frac{\sqrt{n}}{\delta \sqrt{p_n}} \right)^2-1 \right]\left\lVert\frac{\boldsymbol{m}_{\boldsymbol{\gamma}^*(\boldsymbol{\beta})}}{\boldsymbol{\beta}_{s2}}\right\rVert\leq o_p(1),
\end{eqnarray*}
and since $\lambda_n/(p_n\delta^2)\longrightarrow 0$,
\begin{eqnarray*}
    \left\lVert\frac{\boldsymbol{m}_{\boldsymbol{\gamma}^*(\boldsymbol{\beta})}}{\boldsymbol{\beta}_{s2}}\right\rVert\leq \frac{1}{{\frac{\lambda_n}{c K_c^2 p_n\delta^2 }-1}}o_p(1)=o_p(1).
\end{eqnarray*}
It implies \begin{eqnarray}
  \label{A6}\sup_{\boldsymbol{\beta}\in H_{n}} \left\lVert\frac{\boldsymbol{m}_{\boldsymbol{\gamma}^*(\boldsymbol{\beta})}}{\boldsymbol{\beta}_{s2}}\right\rVert=o_p(1).\end{eqnarray}
It follows from \eqref{E4} and \eqref{A6} that
\begin{eqnarray}
    \label{A7}
    \left\lVert 
    \boldsymbol{\gamma}^*(\boldsymbol{\beta})\right\rVert\leq \left\lVert 
    \boldsymbol{\beta}_{s2}\right\rVert\left\lVert\frac{\boldsymbol{m}_{\boldsymbol{\gamma}^*(\boldsymbol{\beta})}}{\boldsymbol{\beta}_{s2}}\right\rVert\leq \delta\sqrt{p_n/n} \cdot o_p(1).
\end{eqnarray}
Hence, 
\begin{eqnarray*}
    \sup_{\boldsymbol{\beta}\in H_{n}} \left\{\frac{\left\lVert 
    \boldsymbol{\gamma}^*(\boldsymbol{\beta})\right\rVert}{\left\lVert 
    \boldsymbol{\beta}_{s2}\right\rVert} \right\}\leq \sup_{\boldsymbol{\beta}\in H_{n}} \left\lVert\frac{\boldsymbol{m}_{\boldsymbol{\gamma}^*(\boldsymbol{\beta})}}{\boldsymbol{\beta}_{s2}}\right\rVert=o_p(1),
\end{eqnarray*}
which implies that \textbf{\textit{Lemma 1(ii)}} holds. To prove \textbf{\textit{Lemma 1(iii)}}, from \eqref{E4} and \eqref{A6}, we have already shown that with probability tending to 1, 
\begin{eqnarray*}
\left\lVert\boldsymbol{\gamma}^*(\boldsymbol{\beta})\right\rVert \leq o_p(1)\delta\sqrt{p_n/n}\leq \delta\sqrt{p_n/n}.  
\end{eqnarray*}
Therefore, we are left to show that
\begin{eqnarray*}
    \left\lVert \boldsymbol{\alpha}^*(\boldsymbol{\beta})-\boldsymbol{\beta}_{s01}\right\rVert\leq \delta \sqrt{p_n/n}
\end{eqnarray*}
with probability tending to 1. Similar to the proof of \eqref{A3}, we have 
\begin{eqnarray*}
    \sup_{\boldsymbol{\beta}\in H_{n}}\left\lVert \frac{\lambda_n}{n}\textbf{A}(\boldsymbol{\beta})\textbf{D}_1(\boldsymbol{\beta}_{s1}) \boldsymbol{\alpha}^*(\boldsymbol{\beta})\right\rVert=o_p(\sqrt{p_n/n})=o_p(\delta\sqrt{p_n/n}).
\end{eqnarray*}
Subsequently, from \eqref{A2}, we have 
\begin{eqnarray}\label{A8}
\sup_{\boldsymbol{\beta}\in H_{n}}\left\lVert \boldsymbol{\alpha}^*(\boldsymbol{\beta})-\boldsymbol{\beta}_{s01}+\frac{\lambda_n}{n} \textbf{B}(\boldsymbol{\beta})\textbf{D}_2(\boldsymbol{\beta}_{s2})\boldsymbol{\gamma}^*(\boldsymbol{\beta})\right\rVert=o_p\left( \delta \sqrt{p_n/n}\right).
\end{eqnarray}
According to \eqref{A4} and \eqref{A7}, we have 
\begin{equation*}
\begin{split}
    c^{-1}\left\lVert \frac{\lambda_n}{n}\textbf{D}_2(\boldsymbol{\beta_{s2}})\boldsymbol{\gamma}^*(\boldsymbol{\beta}) \right\rVert &\leq \left\lVert \boldsymbol{\gamma}^*(\boldsymbol{\beta}) \right\rVert + o_p\left(\delta \sqrt{p_n/n} \right)\\
    & \leq o_p\left(\delta \sqrt{p_n/n} \right)+o_p\left(\delta \sqrt{p_n/n} \right)\\
    &=o_p\left(\delta \sqrt{p_n/n} \right).
\end{split}
\end{equation*}
Then $\left\lVert \frac{\lambda_n}{n}\textbf{D}_2(\boldsymbol{\beta_{s2}})\boldsymbol{\gamma}^*(\boldsymbol{\beta}) \right\rVert \leq c \cdot o_p\left(\delta \sqrt{p_n/n}\right)=o_p\left(\delta \sqrt{p_n/n} \right)$, and therefore, 
\begin{equation}\label{A9}
\begin{split}
    \sup_{\boldsymbol{\beta}\in H_n}\left\lVert \frac{\lambda_n}{n}\textbf{B}(\boldsymbol{\beta})\textbf{D}_2(\boldsymbol{\beta_{s2}})\boldsymbol{\gamma}^*(\boldsymbol{\beta}) \right\rVert &\leq \left\lVert \textbf{B}(\boldsymbol{\beta}) \right\rVert \left\lVert \frac{\lambda_n}{n}\textbf{D}_2(\boldsymbol{\beta_{s2}})\boldsymbol{\gamma}^*(\boldsymbol{\beta}) \right\rVert \nonumber\\
    & \leq c \cdot o_p\left(\delta \sqrt{p_n/n} \right) \nonumber\\
    & = o_p \left( \delta \sqrt{p_n/n} \right).
\end{split}
\end{equation}
Note: here in \eqref{A9}, we have improved the result in \cite{zhao2019simultaneous} as they have reached $O_p\left(\delta \sqrt{p_n/n} \right)$ only, and we have obtained $o_p\left(\delta \sqrt{p_n/n} \right)$.

Thus, \eqref{A8} and \eqref{A9} yields 
\begin{eqnarray}
    \label{extra2}
    \sup_{\boldsymbol{\beta}\in H_n}\left\lVert \boldsymbol{\alpha}^*(\boldsymbol{\beta})-\boldsymbol{\beta}_{s01}\right\rVert\leq o_p\left(\delta \sqrt{p_n/n}\right).
\end{eqnarray}
The inequality of \eqref{extra2} implies that, with probability tending to 1, $\forall \boldsymbol{\beta} \in H_n$, we have
\begin{eqnarray*}
    \left\lVert \boldsymbol{\alpha}^*(\boldsymbol{\beta})-\boldsymbol{\beta}_{s01}\right\rVert\leq \delta \sqrt{p_n/n},
\end{eqnarray*}
for large $n$, and hence, \textit{\textbf{Lemma 1 (iii)}} holds.
\vspace*{20px}

\noindent
Note: Let $\boldsymbol{\beta}_{s1}=\boldsymbol{\alpha}$ and $\boldsymbol{\beta}_{s2}=\boldsymbol{0}$ in $\boldsymbol{\Omega}_n(\boldsymbol{\beta})$ and $\textbf{v}_n(\boldsymbol{\beta})$,  Then, we define $\boldsymbol{\Omega}_n(\boldsymbol{\alpha})=\boldsymbol{\Omega}_n(\boldsymbol{\beta})$. Similarly, $\textbf{v}_n(\boldsymbol{\alpha})=\boldsymbol{v}_n(\boldsymbol{\beta})$. The same applies to $\boldsymbol{\Omega}_n^{(1)}(\boldsymbol{\alpha})$ and $\textbf{v}_n^{(1)}(\boldsymbol{\alpha})$.
\vspace*{20px}

\noindent
\textbf{Lemma 2:} \textit{A matrix calculus identity: Assume a vector $\boldsymbol{\alpha} \in \mathbb{R}^{q_n}$, $q_n\geq 1$, $f$ is a function mapping from $\mathbb{R}^{q_n}$ to $\mathbb{R}^{q_n}$ defined by $f(\boldsymbol{\alpha})=(f_1(\boldsymbol{\alpha}),\ldots,f_{q_n}(\boldsymbol{\alpha}))^\top$, and $f$ is differentiable. Also, $\boldsymbol{\omega}(\boldsymbol{\alpha})$ is a $q_n \times q_n$ matrix, mapping from $\mathbb{R}^{q_n}$ to $\mathbb{R}^{q_n\times q_n}$, and differentiable. Then 
\begin{eqnarray*}
    \frac{\partial \left[\boldsymbol{w}(\boldsymbol{\alpha})f(\boldsymbol{\alpha})\right]}{\partial \boldsymbol{\alpha}^\top}=\boldsymbol{w}(\boldsymbol{\alpha})\frac{\partial f(\boldsymbol{\alpha})}{\partial \boldsymbol{\alpha}^\top}+\begin{pmatrix}
        f^\top(\boldsymbol{\alpha})&\ldots&0\\
        \vdots&\ddots&\vdots\\
        0&\ldots&f^\top(\boldsymbol{\alpha})
    \end{pmatrix}
    \begin{pmatrix}
        \left( \frac{\partial \boldsymbol{w}_1^\top (\boldsymbol{\alpha})}{\partial \boldsymbol{\alpha}}\right)^\top\\
        \vdots\\
        \left( \frac{\partial \boldsymbol{w}_{q_n}^\top (\boldsymbol{\alpha})}{\partial \boldsymbol{\alpha}}\right)^\top
    \end{pmatrix},
\end{eqnarray*}
where the two matrices in the last term of the above equation are block matrices, $\boldsymbol{w}_j^\top (\boldsymbol{\alpha})$ is the $j$th row of $\boldsymbol{W}(\boldsymbol{\alpha})$ and $\partial \boldsymbol{w}_j^\top (\boldsymbol{\alpha})/\partial \boldsymbol{\alpha}$ is a $q_n\times q_n$ matrix, $1\leq j \leq q_n$.} 
\vspace*{20px}

\noindent
\textbf{Lemma 3:} \textit{Under the conditions \textbf{C1}-\textbf{C9}, with probability tending to one, the equation $\boldsymbol{\alpha}=(\boldsymbol{\Omega}_n^{(1)}(\boldsymbol{\alpha})+\lambda_n \textbf{D}_1(\boldsymbol{\alpha}))^{-1}\textbf{v}_n^{(1)}(\boldsymbol{\alpha})$ has a unique fixed-point $\widehat{\boldsymbol{\alpha}}^{*}$ in the domain $H_{n_1}$.}
\vspace*{20px}

\noindent
Our Lemma 3 is different from that in \cite{zhao2019simultaneous}. Our proofs can be different in three separate ways:
\begin{enumerate}
    \item In \cite{zhao2019simultaneous}, $\boldsymbol{\Omega}_n^{(1)}$ and $\textbf{v}_n^{(1)}$ are treated as constants while here, we have $\boldsymbol{\Omega}_n^{(1)}=\boldsymbol{\Omega}_n^{(1)}(\boldsymbol{\alpha})$ and $\textbf{v}_n^{(1)}=\textbf{v}_n^{(1)}(\boldsymbol{\alpha})$.
    \item The domain $H_{n_1}$ is defined to have a different form from $[1/K , K_0]^{q_n}$.
    \item The proofs in the following are different due to the dependence of $\boldsymbol{\Omega}_n^{(1)}(\boldsymbol{\alpha})$ and $\textbf{v}_n^{(1)}(\boldsymbol{\alpha})$ on $\boldsymbol{\alpha}$.
\end{enumerate}
\noindent
\textbf{Proof of Lemma 3.} \\
Define 
\begin{equation} \label{A10}
    f(\boldsymbol{\alpha})=(f_1(\boldsymbol{\alpha}),\ldots,f_{q_n}(\boldsymbol{\alpha})) \equiv (\boldsymbol{\Omega}_n^{(1)}(\boldsymbol{\alpha})+\lambda_n \textbf{D}_1(\boldsymbol{\alpha}))^{-1}\textbf{v}_n^{(1)}(\boldsymbol{\alpha}),
\end{equation}
where $\boldsymbol{\alpha}=(\alpha_1,\ldots,\alpha_{q_n})^\top$. By multiplying $(\boldsymbol{\Omega}_n^{(1)}(\boldsymbol{\alpha}))^{-1}(\boldsymbol{\Omega}_n^{(1)}(\boldsymbol{\alpha})+\lambda_n \textbf{D}_1(\boldsymbol{\alpha}))$ and subtracting $\boldsymbol{\beta}_{s01}$ on both sides of \eqref{A10}, we have 
\begin{eqnarray} \label{F1}
    f(\boldsymbol{\alpha})-\boldsymbol{\beta}_{s01}+\lambda_n(\boldsymbol{\Omega}_n^{(1)}(\boldsymbol{\alpha}))^{-1}\textbf{D}_1(\boldsymbol{\alpha})f(\boldsymbol{\alpha})=(\boldsymbol{\Omega}^{(1)}_n(\boldsymbol{\alpha}))^{-1}\textbf{v}_n^{(1)}-\boldsymbol{\beta}_{s01}.
\end{eqnarray}
Since $\boldsymbol{\Omega}_n(\boldsymbol{\alpha})=\textbf{X}^\top(\boldsymbol{\alpha})\textbf{X}(\boldsymbol{\alpha})$, $\textbf{v}_n(\boldsymbol{\alpha})=\textbf{X}^\top (\boldsymbol{\alpha}) \textbf{Y}(\boldsymbol{\alpha})$  by Cholesky decomposition. Let $\textbf{X}(\boldsymbol{\alpha})=(\textbf{X}_1(\boldsymbol{\alpha}),\textbf{X}_2(\boldsymbol{\alpha}))$, $\textbf{X}_1(\boldsymbol{\alpha})$ is a $p_n \times q_n$ matrix and $\textbf{X}_2(\boldsymbol{\alpha})$ is a $p_n\times (p_n - q_n)$ matrix. Then 
\begin{eqnarray*}
    \textbf{X}^\top(\boldsymbol{\alpha})=\begin{pmatrix}
      \textbf{X}_1^\top(\boldsymbol{\alpha}) \\
      \textbf{X}_2^\top(\boldsymbol{\alpha})
    \end{pmatrix},
\end{eqnarray*}
and
\begin{equation*}
\begin{split}
\boldsymbol{\Omega}_n(\boldsymbol{\alpha})=\textbf{X}^\top(\boldsymbol{\alpha})\textbf{X}(\boldsymbol{\alpha})& = \begin{pmatrix}
      \textbf{X}_1^\top(\boldsymbol{\alpha}) \\
      \textbf{X}_2^\top(\boldsymbol{\alpha})
    \end{pmatrix} \begin{pmatrix} \textbf{X}_1(\boldsymbol{\alpha}) & \textbf{X}_2(\boldsymbol{\alpha}) \end{pmatrix}\\
    &=\begin{pmatrix}
     \textbf{X}_1^\top(\boldsymbol{\alpha}) \textbf{X}_1(\boldsymbol{\alpha}) & \textbf{X}_1^\top(\boldsymbol{\alpha})\textbf{X}_2(\boldsymbol{\alpha}) \\
     \textbf{X}_2^\top(\boldsymbol{\alpha}) \textbf{X}_1(\boldsymbol{\alpha}) & \textbf{X}_2^\top(\boldsymbol{\alpha}) \textbf{X}_2(\boldsymbol{\alpha})
    \end{pmatrix}.
\end{split}
\end{equation*}
We obtain $\boldsymbol{\Omega}_n^{(1)}(\boldsymbol{\alpha})=\textbf{X}_1^\top(\boldsymbol{\alpha})\textbf{X}_1(\boldsymbol{\alpha})$, $\textbf{v}_n^{(1)}(\boldsymbol{\alpha})=\textbf{X}_1^\top(\boldsymbol{\alpha})\textbf{Y}(\boldsymbol{\alpha})$, and 
\begin{eqnarray*}
\textbf{v}_n(\boldsymbol{\alpha})=\textbf{X}^\top(\boldsymbol{\alpha})\textbf{Y}(\boldsymbol{\alpha})=\begin{pmatrix}
        \textbf{X}_1^\top(\boldsymbol{\alpha})\textbf{Y}(\boldsymbol{\alpha}) \\
        \textbf{X}_2^\top(\boldsymbol{\alpha})\textbf{Y}(\boldsymbol{\alpha})
    \end{pmatrix}.
\end{eqnarray*}
Thus, in \eqref{F1}, we have 
\begin{eqnarray}\label{F2}
    (\boldsymbol{\Omega}_n^{(1)}(\boldsymbol{\alpha}))^{-1}\textbf{v}_n^{(1)}(\boldsymbol{\alpha})-\boldsymbol{\beta}_{s01} &&= (\textbf{X}_1^\top(\boldsymbol{\alpha})\textbf{X}_1(\boldsymbol{\alpha}))^{-1}\textbf{X}_1^\top(\boldsymbol{\alpha})\textbf{Y}(\boldsymbol{\alpha})-\boldsymbol{\beta}_{s01}\nonumber\\
    &&=(\textbf{X}_1^\top(\boldsymbol{\alpha})\textbf{X}_1(\boldsymbol{\alpha}))^{-1}\textbf{X}_1^\top(\boldsymbol{\alpha})\textbf{Y}(\boldsymbol{\alpha})\nonumber\\
    &&-(\textbf{X}_1^\top(\boldsymbol{\alpha})\textbf{X}_1(\boldsymbol{\alpha}))^{-1}\textbf{X}_1^\top(\boldsymbol{\alpha})\textbf{X}_1(\boldsymbol{\alpha})\boldsymbol{\beta}_{s01}\nonumber\\
    &&=(\textbf{X}_1^\top(\boldsymbol{\alpha})\textbf{X}_1(\boldsymbol{\alpha}))^{-1}\textbf{X}_1^\top(\boldsymbol{\alpha})[\textbf{Y}(\alpha)-\textbf{X}_1(\boldsymbol{\alpha})\boldsymbol{\beta}_{s01}].
\end{eqnarray}
Since $\boldsymbol{\beta}_{s02}=0$,  
\begin{eqnarray*}
\textbf{X}(\boldsymbol{\alpha})\boldsymbol{\beta}_{s0}= \begin{pmatrix} \textbf{X}_1(\boldsymbol{\alpha}) & \textbf{X}_2(\boldsymbol{\alpha}) \end{pmatrix}
\begin{pmatrix}
\boldsymbol{\beta}_{s01}\\\boldsymbol{\beta}_{s02}
\end{pmatrix}=\textbf{X}_1(\boldsymbol{\alpha})\boldsymbol{\beta}_{s01},
\end{eqnarray*}
and 
\begin{equation*}
\begin{split}
    \widehat{\textbf{b}}(\boldsymbol{\alpha}) &=\boldsymbol{\Omega}^{-1}_n(\boldsymbol{\alpha})\textbf{v}_n(\boldsymbol{\alpha}) \\
    &=(\textbf{X}^\top(\boldsymbol{\alpha})\textbf{X}(\boldsymbol{\alpha}))^{-1}\textbf{X}^\top(\boldsymbol{\alpha})\textbf{Y}(\boldsymbol{\alpha})\nonumber\\
    &=\textbf{X}^{-1}(\boldsymbol{\alpha})\textbf{Y}(\boldsymbol{\alpha}).
\end{split}
\end{equation*}
Then, from \eqref{F2}, we have 
\begin{eqnarray}\label{F3}
    &&(\boldsymbol{\Omega}_n^{(1)}(\boldsymbol{\alpha}))^{-1}\textbf{v}_n^{(1)}(\boldsymbol{\alpha})-\boldsymbol{\beta}_{s01}\nonumber\\
    &&=(\textbf{X}_1^\top(\boldsymbol{\alpha})\textbf{X}_1(\boldsymbol{\alpha}))^{-1}\textbf{X}_1^\top(\boldsymbol{\alpha})[\textbf{Y}(\boldsymbol{\alpha}) - \textbf{X}(\boldsymbol{\alpha})\boldsymbol{\beta}_{s0}]\nonumber\\
    &&=(\textbf{X}_1^\top(\boldsymbol{\alpha})\textbf{X}_1(\boldsymbol{\alpha}))^{-1}\textbf{X}_1^\top(\boldsymbol{\alpha})\textbf{X}(\boldsymbol{\alpha})[\textbf{X}^{-1}(\boldsymbol{\alpha})\textbf{Y}(\boldsymbol{\alpha})-\boldsymbol{\beta}_{s0}]\nonumber\\
    &&=(\textbf{X}_1^\top(\boldsymbol{\alpha})\textbf{X}_1(\boldsymbol{\alpha}))^{-1}\textbf{X}_1^\top(\boldsymbol{\alpha})\textbf{X}(\boldsymbol{\alpha})[\widehat{\textbf{b}}(\boldsymbol{\alpha})-\boldsymbol{\beta}_{s0}].
\end{eqnarray}
From \eqref{F3}, we obtain
\begin{eqnarray}\label{F4}
    \left\lVert(\boldsymbol{\Omega}_n^{(1)}(\boldsymbol{\alpha}))^{-1}\textbf{v}_n^{(1)}(\boldsymbol{\alpha})-\boldsymbol{\beta}_{s01}\right\lVert\leq \left\lVert(\textbf{X}_1^\top(\boldsymbol{\alpha}
    )\textbf{X}_1(\boldsymbol{\alpha}))^{-1}\right\lVert \cdot \left\lVert\textbf{X}_1^\top(\boldsymbol{\alpha}
    )\textbf{X}(\boldsymbol{\alpha})\right\rVert \cdot \left\lVert\widehat{\textbf{b}}(\boldsymbol{\alpha})-\boldsymbol{\beta}_{s0}\right\rVert.
\end{eqnarray}
Since $\textbf{X}^\top(\boldsymbol{\alpha}
    )\textbf{X}(\boldsymbol{\alpha})=(\textbf{X}_1^\top(\boldsymbol{\alpha}
    ), \textbf{X}^\top_2(\boldsymbol{\alpha}))^\top\textbf{X}(\boldsymbol{\alpha})=\begin{pmatrix}
        \textbf{X}_1^\top(\boldsymbol{\alpha}
    )\textbf{X}(\boldsymbol{\alpha})\\
    \textbf{X}_2^\top(\boldsymbol{\alpha}
    )\textbf{X}(\boldsymbol{\alpha})
    \end{pmatrix}$, then
 \begin{eqnarray*}
        \left\lVert\textbf{X}_1^\top(\boldsymbol{\alpha}
    )\textbf{X}(\boldsymbol{\alpha}) \right\rVert\leq \left\lVert\textbf{X}^\top(\boldsymbol{\alpha}
    )\textbf{X}(\boldsymbol{\alpha})\right\rVert=\left\lVert\boldsymbol{\Omega}_n(\boldsymbol{\alpha})\right\rVert.
 \end{eqnarray*}
Noticing $\boldsymbol{\Omega}_n^{(1)}(\boldsymbol{\alpha})=\textbf{X}_1^\top(\boldsymbol{\alpha}
    )\textbf{X}_1(\boldsymbol{\alpha})$, from \eqref{F4}, we have 
 \begin{eqnarray*}
  &&\sup_{\boldsymbol{\alpha}\in H_{n1}} \left\lVert(\boldsymbol{\Omega}_n^{(1)}(\boldsymbol{\alpha}))^{-1}\textbf{v}^{(1)}_n(\boldsymbol{\alpha}) - \boldsymbol{\beta}_{s01} \right\rVert \\
 &&\leq \sup_{\boldsymbol{\alpha}\in H_{n1}}\left[{\left\lVert \left( \frac{\textbf{X}_1^\top(\boldsymbol{\alpha})
    \textbf{X}_1(\boldsymbol{\alpha})}{n} \right)^{-1} \right\rVert \cdot \left\lVert \frac{\textbf{X}^\top(\boldsymbol{\alpha}
    )\textbf{X}(\boldsymbol{\alpha})}{n}\right\rVert \cdot \left\lVert \widehat{\textbf{b}}(\boldsymbol{\alpha})-\boldsymbol{\beta}_{s0}\right\rVert}\right]\\
    &&\leq \sup_{\boldsymbol{\alpha}\in H_{n1}}\left\lVert \left( \frac{\boldsymbol{\Omega}_n^{(1)}(\boldsymbol{\alpha})}{n} \right)^{-1} \right\rVert \cdot \sup_{\boldsymbol{\alpha}\in H_{n1}}\left\lVert \frac{\boldsymbol{\Omega}_n(\boldsymbol{\alpha})}{n} \right\rVert \cdot \sup_{\boldsymbol{\alpha}\in H_{n_1}}\left\lVert \widehat{\textbf{b}}(\boldsymbol{\alpha})-\boldsymbol{
    \beta
    }_{s0}\right\rVert\\
    &&= \sup_{\boldsymbol{\alpha}\in H_{n1}}\left[ \lambda_{\text{max}} \left\{ \left(\frac{\boldsymbol{\Omega}_n^{(1)}(\boldsymbol{\alpha})}{n} \right)^{-1}\right\} \right] \cdot \sup_{\boldsymbol{\alpha}\in H_{n1}}\left[ \lambda_{\text{max}}\left\{\frac{\boldsymbol{\Omega}_n(\boldsymbol{\alpha})}{n} \right\} \right] \cdot \sup_{\boldsymbol{\alpha}\in H_{n1}}\left[ \left\lVert \widehat{\textbf{b}}(\boldsymbol{\alpha})-\boldsymbol{\beta}_{s0}\right\rVert \right]\\
    &&=\sup_{\boldsymbol{\alpha}\in H_{n1}}\left[ \left\{ \lambda_{\text{min}} \left(\frac{\boldsymbol{\Omega}_n^{(1)}(\boldsymbol{\alpha})}{n} \right) \right\}^{-1} \right] \cdot \sup_{\boldsymbol{\alpha}\in H_{n1}}\left[ \lambda_{\text{max}}\left\{\frac{\boldsymbol{\Omega}_n(\boldsymbol{\alpha})}{n} \right\} \right] \cdot \sup_{\boldsymbol{\alpha}\in H_{n1}}\left[ \left\lVert \widehat{\textbf{b}}(\boldsymbol{\alpha})-\boldsymbol{\beta}_{s0}\right\rVert \right].
\end{eqnarray*}
Then, by condition \textbf{C5}, we have
\begin{eqnarray*}
    && \sup_{\boldsymbol{\alpha}\in H_{n1}}\left[ \left\{ \lambda_{\text{min}} \left(\frac{\boldsymbol{\Omega}_n^{(1)}(\boldsymbol{\alpha})}{n} \right) \right\}^{-1}\right].\sup_{\boldsymbol{\alpha}\in H_{n1}}\left[ \lambda_{\text{max}}\left\{\frac{\boldsymbol{\Omega}_n(\boldsymbol{\alpha})}{n} \right\} \right].\sup_{\boldsymbol{\alpha}\in H_{n1}}\left[ \left\lVert \widehat{\textbf{b}}(\boldsymbol{\alpha})-\boldsymbol{\beta}_{s0}\right\rVert \right]\\
    &&\leq \left[\frac{1}{c}\right]^{-1}\cdot c \cdot \sup_{\boldsymbol{\alpha}\in H_{n1}}\left\lVert\widehat{\textbf{b}}(\boldsymbol{\alpha})-\boldsymbol{\beta}_{s0}
    \right\rVert\\
    &&=c^2 \sup_{\boldsymbol{\alpha}\in H_{n1}}\left\lVert 
    \widehat{\textbf{b}}(\boldsymbol{\alpha})-\boldsymbol{\beta}_{s0} \right\rVert. 
\end{eqnarray*}
By \textbf{\textit{Lemma 1 (i)}}, $\sup_{\boldsymbol{\alpha}\in H_{n1}} \left\lVert \widehat{\textbf{b}}(\boldsymbol{\alpha})-\boldsymbol{\beta}_{s0} \right\rVert=O_p(\sqrt{p_n/n})$. Then,
\begin{eqnarray*}
\sup_{\boldsymbol{\alpha}\in H_{n_1}}\left\lVert (\boldsymbol{\Omega}_n^{(1)}(\boldsymbol{\alpha}))^{-1}\textbf{v}_n^{(1)}(\boldsymbol{\alpha}) -\boldsymbol{\beta}_{s01} \right\rVert=O_p(\sqrt{p_n/n}).
\end{eqnarray*}
Therefore, from \eqref{F1},we obtain
\begin{eqnarray}\label{F5}
 \sup_{\boldsymbol{\alpha}\in H_{n_1}}\left\lVert f(\boldsymbol{\alpha})-\boldsymbol{\beta}_{s01}+\lambda_n(\boldsymbol{\Omega}_n^{(1)}(\boldsymbol{\alpha}))^{-1}\textbf{D}_1(\boldsymbol{\alpha})f(\boldsymbol{\alpha})\right\rVert=O_p(\sqrt{p_n/n}).
\end{eqnarray}
Next, we want to show 
\begin{eqnarray}\label{F6}
 \sup_{\boldsymbol{\alpha}\in H_{n_1}}\left\lVert \lambda_n(\boldsymbol{\Omega}_n^{(1)}(\boldsymbol{\alpha}))^{-1}\textbf{D}_1(\boldsymbol{\alpha})f(\boldsymbol{\alpha})\right\rVert=o_p(\sqrt{q_n/n}).
\end{eqnarray}
Then, from \eqref{F5} and \eqref{F6}, it follows that
\begin{eqnarray*}
 \sup_{\boldsymbol{\alpha}\in H_{n_1}}\left\lVert f(\boldsymbol{\alpha}) - \boldsymbol{\beta}_{s01} \right\rVert=O_p(\sqrt{p_n/n})\longrightarrow 0,
\end{eqnarray*}
which implies with probability tending to 1, that $f(\boldsymbol{\alpha})\in H_{n_1}$, i.e., $f(\boldsymbol{\alpha})$ is a mapping from $H_{n_1}$ to itself.

In order to prove \eqref{F6}, first, we rewrite it as
\begin{eqnarray*}
\sup_{\boldsymbol{\alpha}\in H_{n1}}\left\lVert \frac{\lambda_n}{n}(n^{-1}\boldsymbol{\Omega}_n^{(1)}(\boldsymbol{\alpha}))^{-1}\textbf{D}_1(\boldsymbol{\alpha})f(\boldsymbol{\alpha})\right\rVert=o_p(\sqrt{q_n/n}).
\end{eqnarray*}
Since $\widehat{\textbf{b}}(\boldsymbol{\alpha})=\textbf{X}^{-1}(\boldsymbol{\alpha})\textbf{Y}(\boldsymbol{\alpha})$, $\textbf{D}_1(\boldsymbol{\alpha})=\text{diag}(\alpha_1^{-2},\ldots,\alpha_{q_n}^{-2})$,
\begin{equation*}
\begin{split}
    \textbf{v}_n^{(1)}(\boldsymbol{\alpha}) &= \textbf{X}_1^\top(\boldsymbol{\alpha})\textbf{Y}(\boldsymbol{\alpha}) \nonumber \\
&=\textbf{X}_1^\top(\boldsymbol{\alpha})\textbf{X}(\boldsymbol{\alpha})\left[ \textbf{X}^{-1}(\boldsymbol{\alpha})\textbf{Y}(\boldsymbol{\alpha}) \right]\nonumber\\
    &=\textbf{X}_1^\top(\boldsymbol{\alpha})\textbf{X}(\boldsymbol{\alpha})\widehat{\textbf{b}}(\boldsymbol{\alpha}).
\end{split}
\end{equation*}
As shown before, we have
\begin{equation*}
\begin{split}
    \left\lVert \widehat{\textbf{b}}(\boldsymbol{\alpha})\right\rVert &= \left\lVert \widehat{\textbf{b}}(\boldsymbol{\alpha}) -\boldsymbol{\beta}_{s0} +\boldsymbol{\beta}_{s0} \right\rVert \\
&\leq \left\lVert \widehat{\textbf{b}}(\boldsymbol{\alpha})-\boldsymbol{\beta}_{s0} \right\rVert + \left\lVert \boldsymbol{\beta}_{s0} \right\rVert\\
    & = o_p\left(\sqrt{p_n/n}\right)+O_p(q_n)\\
    & =O_p(q_n),
\end{split}
\end{equation*}
and 
\begin{eqnarray} \label{F7}
    \left\lVert \textbf{v}_n^{(1)}(\boldsymbol{\alpha})\right\rVert && \leq \left\lVert \textbf{X}_1^\top(\boldsymbol{\alpha})\textbf{X}(\boldsymbol{\alpha})\right\rVert \left\lVert \widehat{\textbf{b}}(\boldsymbol{\alpha})\right\rVert \nonumber\\
    &&\leq \left\lVert \textbf{X}^\top(\boldsymbol{\alpha})\textbf{X}(\boldsymbol{\alpha}) \right\rVert\left\lVert\widehat{\textbf{b}}(\boldsymbol{\alpha})\right\rVert \nonumber\\
    &&=n\left\lVert \frac{\boldsymbol{\Omega}_n(\boldsymbol{\alpha})}{n}\right\rVert
    \left\lVert \widehat{\textbf{b}}(\boldsymbol{\alpha})\right\rVert\nonumber\\
    && \leq c \cdot n \left\lVert \widehat{\textbf{b}}(\boldsymbol{\alpha})\right\rVert \nonumber\\
    &&\leq c \cdot n \cdot O_p(q_n)\qquad \text{by condition \textbf{C4}.}
\end{eqnarray}
Then, 
\begin{eqnarray} \label{F8}
    \left\lVert f(\boldsymbol{\alpha})\right\rVert &&= \left\lVert \left(\boldsymbol{\Omega}_n^{(1)}(\boldsymbol{\alpha})+\lambda_n \textbf{D}_1(\boldsymbol{\alpha})\right)^{-1}\textbf{v}_n^{(1)}(\boldsymbol{\alpha})   \right\rVert\nonumber\\
    &&\leq \frac{1}{n}\left\lVert \left(\frac{\boldsymbol{\Omega}_n^{(1)}(\boldsymbol{\alpha})}{n} +\frac{\lambda_n}{n}\textbf{D}_1(\boldsymbol{\alpha})\right)^{-1}\right\rVert\left\lVert \textbf{v}_n^{(1)}(\boldsymbol{\alpha}) \right\rVert\nonumber\\
    &&=\frac{1}{n}\lambda_{\text{max}}\left[ \left(\frac{\boldsymbol{\Omega}_n^{(1)}(\boldsymbol{\alpha})}{n} +\frac{\lambda_n}{n}\textbf{D}_1(\boldsymbol{\alpha})\right)^{-1}\right]\left\lVert \textbf{v}_n^{(1)}(\boldsymbol{\alpha}) \right\rVert\nonumber\\
    &&=\frac{1}{n} \left[\lambda_{\text{min}} \left(\frac{\boldsymbol{\Omega}_n^{(1)}(\boldsymbol{\alpha})}{n} +\frac{\lambda_n}{n}\textbf{D}_1(\boldsymbol{\alpha})\right) \right]^{-1}\left\lVert \textbf{v}_n^{(1)}(\boldsymbol{\alpha}) \right\rVert\nonumber\\
    &&\leq \frac{1}{n}\left[\lambda_{\text{min}}\left(\frac{\boldsymbol{\Omega}_n^{(1)}(\boldsymbol{\alpha})}{n}\right)\right]^{-1}\left\lVert \textbf{v}_n^{(1)}(\boldsymbol{\alpha}) \right\rVert \qquad \text{(since $\frac{\lambda_n}{n}\textbf{D}_1(\boldsymbol{\alpha})$ is positive definite)}\nonumber\\
    &&\leq \frac{1}{n(1/c)}\left\lVert \textbf{v}_n^{(1)}(\boldsymbol{\alpha}) \right\rVert\qquad\text{by condition \textbf{C4}} \text{ and by \eqref{F7}} \nonumber\\
    &&\leq \frac{1}{n} \cdot c^2 \cdot n \cdot O_p(q_n) \nonumber\\
    &&=c^2 \cdot O_p(q_n).
\end{eqnarray}
Since $\boldsymbol{\alpha}\in H_{n_1}$, when $n$ is large enough, $|\alpha_j|\geq a_0/2,~ 1\leq j\leq q_n$ by condition \textbf{C7}, then, 
\begin{eqnarray}
    \label{F9}
    \left \lVert \textbf{D}_1(\boldsymbol{\alpha})\right\rVert =\lambda_{\text{max}}(\textbf{D}_1(\boldsymbol{\alpha}))=\max_{1\leq j\leq q_n}(a_j^{-2})\leq (a_0/2)^{-2}=\frac{4}{a_0^2}.
\end{eqnarray}
Thus, by \eqref{F6}, \eqref{F7}, and \eqref{F8}, we have 
\begin{eqnarray}
    \left\lVert \frac{\lambda_n}{n}\left( n^{-1}\boldsymbol{\Omega}_n^{(1)}(\boldsymbol{\alpha})\right)^{-1}\textbf{D}_1(\boldsymbol{\alpha})f(\boldsymbol{\alpha}) \right \rVert &&\leq \frac{\lambda_n}{n}\left\lVert \left(n^{-1}\boldsymbol{\Omega}_n^{(1)}(\boldsymbol{\alpha})\right)^{-1} \right\rVert \left\lVert \textbf{D}_1(\boldsymbol{\alpha})\right\rVert
    \left\lVert f(\boldsymbol{\alpha})\right\rVert\nonumber\\
    &&\leq \frac{\lambda_n}{n}\left(\frac{1}{1/c}\right)(4a_0^{-2})c^2O_p(q_n)\nonumber\\
    &&=(4c^3a_0^{-2})O_p\left(\frac{\lambda_n\sqrt{q_n}}{\sqrt{n}}\sqrt{\frac{q_n}{n}}\right)\nonumber\\
    &&=\left(4c^3a_0^{-2} \right)o_p\left( \sqrt{\frac{q_n}{n}}\right)\qquad\text{(since by condition \textbf{C6}, $\frac{\lambda_n\sqrt{q_n}}{n}\to 0$)}\nonumber\\
    &&=o_p\left(\sqrt{\frac{q_n}{n}} \right).
\end{eqnarray}
Thus, 
\begin{eqnarray*}
    \sup_{\boldsymbol{\alpha}\in H_{n_1}}\left \lVert \lambda_n(\boldsymbol{\Omega}_n^{(1)}(\boldsymbol{\alpha}))^{-1}\textbf{D}_1(\boldsymbol{\alpha})f(\boldsymbol{\alpha})\right\rVert=o_p\left(\sqrt{\frac{q_n}{n}}\right) \; \text{\eqref{F6} holds.}
\end{eqnarray*}
Recall that $\boldsymbol{\Omega}_n(\boldsymbol{\alpha})=\boldsymbol{\Omega}_n(\boldsymbol{\beta})\Big|_{\boldsymbol{\beta}_{s1}=\boldsymbol{\alpha},~\boldsymbol{\beta}_{s2}=0}$, $\textbf{v}_n(\boldsymbol{\alpha})=\textbf{v}_n(\boldsymbol{\beta})\Big|_{\boldsymbol{\beta}_{s1}=\boldsymbol{\alpha},~\boldsymbol{\beta}_{s2}=0}$, $\boldsymbol{\Omega}_n^{(1)}(\boldsymbol{\alpha})=\boldsymbol{\Omega}^{(1)}_n(\boldsymbol{\beta})\Big|_{\boldsymbol{\beta}_{s1}=\boldsymbol{\alpha},~\boldsymbol{\beta}_{s2}=0}$, and $\textbf{v}_n^{(1)}(\boldsymbol{\alpha})=\textbf{v}^{(1)}_n(\boldsymbol{\beta})\Big|_{\boldsymbol{\beta}_{s1}=\boldsymbol{\alpha},~\boldsymbol{\beta}_{s2}=0}$.  We have proved \eqref{F6}, i.e., $\sup_{\boldsymbol{\alpha}\in H_{n_1}}\left\lVert f(\boldsymbol{\alpha})-\boldsymbol{\beta}_{s01} \right\rVert = o_p\left(\sqrt{q_n/n} \right)$ which implies that with probability tending to one, $f(\boldsymbol{\alpha})$ is a mapping from $H_{n_1}$ to itself.

Multiplying $\boldsymbol{\Omega}_n^{(1)}(\boldsymbol{\alpha})+\lambda_n\textbf{D}_1(\boldsymbol{\alpha})$ on both sides of \eqref{A10}, we obtain
\begin{equation}
    \label{G1}\left(\boldsymbol{\Omega}_n^{(1)}(\boldsymbol{\alpha}) +\lambda_n\textbf{D}_1(\boldsymbol{\alpha})\right)f(\boldsymbol{\alpha})=\textbf{v}_n^{(1)}(\boldsymbol{\alpha}).
\end{equation}
Denote the $j$th row of $\boldsymbol{\Omega}_n^{(1)}(\boldsymbol{\alpha})$ by $\boldsymbol{\omega}_j^\top(\boldsymbol{\alpha})$ and the $j$th row of $\textbf{D}_1(\boldsymbol{\alpha})$ by $\boldsymbol{d}_j^\top(\boldsymbol{\alpha})$. Then, 
\begin{eqnarray*}
\boldsymbol{m}_j^\top(\boldsymbol{\alpha})=\left( \frac{\partial^2 [\sum_{i=1}^n \log f_n(v_{ni},(\boldsymbol{\alpha}^\top, \boldsymbol{0}^\top),\boldsymbol{\phi})]}{\partial \alpha_j \partial\alpha_1},\ldots,\frac{\partial^2 [\sum_{i=1}^n \log f_n(v_{ni},(\boldsymbol{\alpha}^\top, \boldsymbol{0}^\top),\boldsymbol{\phi})]}{\partial \alpha_j \partial\alpha_{q_n}}\right),
\end{eqnarray*}
where $\boldsymbol{d}_j^\top(\boldsymbol{\alpha})=(0,\ldots,0,\alpha_j^{-2},0,\ldots,0)$, $j=1,\ldots,q_n$. We take derivatives on both sides of \eqref{G1} and have 
\begin{eqnarray}\label{G2}
 \frac{\partial}{\partial\boldsymbol{\alpha}^\top}\left[ \left(\boldsymbol{\Omega}_n^{(1)}(\boldsymbol{\alpha})+\lambda_n \textbf{D}_1(\boldsymbol{\alpha})\right)f(\boldsymbol{\alpha}) \right]=\frac{\partial}{\partial\boldsymbol{\alpha}^\top}[\textbf{v}_n^{(1)}(\boldsymbol{\alpha})].
\end{eqnarray}
Since 
\begin{equation*}
\begin{split}
    \textbf{v}_n(\boldsymbol{\alpha}) &= \dot{\ell}_n(\boldsymbol{\alpha}|\widetilde{\boldsymbol{\phi}})+\boldsymbol{\Omega}_n(\boldsymbol{\alpha})\begin{pmatrix}
        \boldsymbol{\alpha}\\
        \boldsymbol{0}
    \end{pmatrix} \\
    &=\dot{\ell}_n(\boldsymbol{\alpha}|\widetilde{\boldsymbol{\phi}})+\begin{pmatrix}
        \boldsymbol{\Omega}_n^{(1)}(\boldsymbol{\alpha})&\boldsymbol{\Omega}_n^{(12)}(\boldsymbol{\alpha})\\
        \boldsymbol{\Omega}_n^{(21)}(\boldsymbol{\alpha})&
        \boldsymbol{\Omega}_n^{(2)}(\boldsymbol{\alpha})
    \end{pmatrix}\begin{pmatrix}
        \boldsymbol{\alpha}\\
        \boldsymbol{0}
    \end{pmatrix} \\
    &=\dot{\ell}_n(\boldsymbol{\alpha}|\widetilde{\boldsymbol{\phi}})+\begin{pmatrix}
        \boldsymbol{\Omega}_n^{(1)}(\boldsymbol{\alpha})\boldsymbol{\alpha}\\
        \boldsymbol{\Omega}_n^{(21)}(\boldsymbol{\alpha})\boldsymbol{\alpha}
    \end{pmatrix},
\end{split}
\end{equation*}
then $\textbf{v}_n^{(1)}(\boldsymbol{\alpha})=\dot{\ell}^{(1)}_n(\boldsymbol{\alpha}|\widetilde{\boldsymbol{\phi}})+\boldsymbol{\Omega}_n^{(1)}(\boldsymbol{\alpha})\boldsymbol{\alpha}$, and by \textbf{Lemma 2}, we have
\begin{eqnarray} \label{G3}
    \frac{\partial \textbf{v}_n^{(1)}(\boldsymbol{\alpha})}{\partial \boldsymbol{\alpha}^\top} &&= -\boldsymbol{\Omega}_n^{(1)}(\boldsymbol{\alpha})+\boldsymbol{\Omega}_n^{(1)}(\boldsymbol{\alpha}) \textbf{I}_{q_n}+ \begin{pmatrix}
        \boldsymbol{\alpha}^\top &\ldots&0\\
        \vdots&\ddots&\vdots\\
        0&\ldots&\boldsymbol{\alpha}^\top
    \end{pmatrix} \begin{pmatrix}
        \left( \frac{\partial \omega_{1}^\top(\boldsymbol{\alpha})}{\partial \boldsymbol{\alpha}} \right)^\top\\
        \vdots\\
        \left( \frac{\partial \omega_{q_{n}}^\top(\boldsymbol{\alpha})}{\partial \boldsymbol{\alpha}} \right)^\top
    \end{pmatrix} \nonumber \\
    &&=\begin{pmatrix}
        \boldsymbol{\alpha}^\top &\ldots&0\\
        \vdots&\ddots&\vdots\\
        0&\ldots&\boldsymbol{\alpha}^\top
    \end{pmatrix}
    \begin{pmatrix}
        \left( \frac{\partial \omega_{1}^\top(\boldsymbol{\alpha})}{\partial \boldsymbol{\alpha}} \right)^\top\\
        \vdots\\
        \left( \frac{\partial \omega_{q_{n}}^\top(\boldsymbol{\alpha})}{\partial \boldsymbol{\alpha}} \right)^\top
    \end{pmatrix}.
\end{eqnarray}
Applying \textbf{Lemma2} to the left-hand-side of \eqref{G2} gives
\begin{equation}\label{G4}
\begin{split}
    &\frac{\partial}{\partial\boldsymbol{\alpha}^\top}\left[ \left(\boldsymbol{\Omega}_n^{(1)}(\boldsymbol{\alpha})+\lambda_n \textbf{D}_1(\boldsymbol{\alpha})\right)f(\boldsymbol{\alpha}) \right] \\
    &=\left(\boldsymbol{\Omega}_n^{(1)}(\boldsymbol{\alpha})+\lambda_n \textbf{D}_1(\boldsymbol{\alpha})\right) \frac{\partial}{\partial\boldsymbol{\alpha}^\top}f(\boldsymbol{\alpha})\\
    &+\begin{pmatrix}
        f^\top(\boldsymbol{\alpha})&\ldots&0\\
        \vdots&\ddots&\vdots\\
        0&\ldots&f^\top(\boldsymbol{\alpha})
    \end{pmatrix}    
    \left[\begin{pmatrix}
        \left( \frac{\partial \boldsymbol{\omega}_1^\top (\boldsymbol{\alpha})}{\partial \boldsymbol{\alpha}}\right)^\top\\
        \vdots\\
        \left( \frac{\partial \boldsymbol{\omega}_{q_n}^\top (\boldsymbol{\alpha})}{\partial \boldsymbol{\alpha}}\right)^\top
    \end{pmatrix}+\lambda_n\begin{pmatrix}
        \left( \frac{\partial \textbf{d}_1^\top (\boldsymbol{\alpha})}{\partial \boldsymbol{\alpha}}\right)^\top\\
        \vdots\\
        \left( \frac{\partial \textbf{d}_{q_n}^\top (\boldsymbol{\alpha})}{\partial \boldsymbol{\alpha}}\right)^\top
    \end{pmatrix}\right].
\end{split}
\end{equation}
Since 
\begin{eqnarray*}
    \frac{\partial \textbf{d}_j^\top}{\partial \boldsymbol{\alpha}}=\begin{pmatrix}
        0&\ldots&0&0&0&\ldots&0\\
        &\vdots&&\vdots&&\vdots&\\
        0&\ldots&0&-2\alpha_j^{-3}&0&\ldots&0\\
        &\vdots&&\vdots&&\vdots&\\0&\ldots&0&0&0&\ldots&0
    \end{pmatrix},
\end{eqnarray*}
and
\begin{eqnarray*}
    f^\top(\boldsymbol{\alpha})\left( \frac{\partial \textbf{d}_j^\top}{\partial\boldsymbol{\alpha}} \right)^\top=(0,\ldots,0,-2f_j(\boldsymbol{\alpha})\alpha_j^{-3},0,\ldots,0),
\end{eqnarray*}
then
\begin{eqnarray*}
\label{G5}
    \begin{pmatrix}
        f^\top(\boldsymbol{\alpha})&\ldots&0\\
        \vdots&\ddots&\vdots\\
        0&\ldots&f^\top(\boldsymbol{\alpha})
    \end{pmatrix}\begin{pmatrix}
        \left( \frac{\partial \boldsymbol{\omega}_1^\top (\boldsymbol{\alpha})}{\partial \boldsymbol{\alpha}}\right)^\top\\
        \vdots\\
        \left( \frac{\partial \boldsymbol{\omega}_{q_n}^\top (\boldsymbol{\alpha})}{\partial \boldsymbol{\alpha}}\right)^\top
    \end{pmatrix}=0.
\end{eqnarray*}
Denote $\dot{f}(\boldsymbol{\alpha})=\frac{\partial f(\boldsymbol{\alpha})}{\partial \boldsymbol{\alpha}^\top}$ (which is a $q_n\times q_n$ matrix) and
\begin{eqnarray*}
    \textbf{J}_n(\boldsymbol{\alpha})=\begin{pmatrix}
        (f(\boldsymbol{\alpha})-\boldsymbol{\alpha})^\top&\ldots&0\\
        \vdots&\ddots&\vdots\\
        0&\ldots&(f(\boldsymbol{\alpha})-\boldsymbol{\alpha})^\top
    \end{pmatrix}\left( \frac{\partial\boldsymbol{\Omega}_n^{(1)}(\boldsymbol{\alpha})}{\partial \boldsymbol{\alpha}}\right)^\top=\textbf{F}_n(\boldsymbol{\alpha})\textbf{P}_n(\boldsymbol{\alpha}).
\end{eqnarray*}
Then, \eqref{G5} becomes
\begin{eqnarray*}
    \left(\boldsymbol{\Omega}_n^{(1)}(\boldsymbol{\alpha})+\lambda_n \textbf{D}_1(\boldsymbol{\alpha}) \right)\dot{f}(\boldsymbol{\alpha})+\lambda_n\text{diag}(-2f_1(\boldsymbol{\alpha})\alpha_1^{{-3}},\ldots,-2f_{q_n}(\boldsymbol{\alpha})\alpha_{q_n}^{{-3}})+\textbf{F}_n(\boldsymbol{\alpha})\textbf{P}_n(\boldsymbol{\alpha})=0,
\end{eqnarray*}
\begin{eqnarray} \label{G6}
    \left(\boldsymbol{\Omega}_n^{(1)}(\boldsymbol{\alpha})+\lambda_n \textbf{D}_1(\boldsymbol{\alpha}) \right)\dot{f}(\boldsymbol{\alpha})=2\lambda_n\text{diag}(f_1(\boldsymbol{\alpha})\alpha_1^{-3},\ldots,f_{q_n}(\boldsymbol{\alpha})\alpha_{q_n}^{-3})-\textbf{F}_n(\boldsymbol{\alpha})\textbf{P}_n(\boldsymbol{\alpha}).
\end{eqnarray}
Dividing both sides of \eqref{G6} by $n$, we have
\begin{eqnarray*}
\left(\frac{\boldsymbol{\Omega}_n^{(1)}(\boldsymbol{\alpha})}{n}+\frac{\lambda_n}{n}\textbf{D}_1(\boldsymbol{\alpha})\right)\dot{f}(\boldsymbol{\alpha})=2\left(\frac{\lambda_n}{n}\right)\text{diag}(f_1(\boldsymbol{\alpha})\alpha_1^{-3},\ldots,f_{q_n}(\boldsymbol{\alpha})\alpha_{q_n}^{-3})-\frac{\textbf{F}_n(\boldsymbol{\alpha})\textbf{P}_n(\boldsymbol{\alpha})}{n}.
\end{eqnarray*}
Then
\begin{eqnarray} \label{G7}
&& \sup_{\boldsymbol{\alpha}\in H_{n1}}\left\lVert \left( \frac{\boldsymbol{\Omega}_n^{(1)}(\boldsymbol{\alpha})}{n}+\frac{\lambda_n}{n}\textbf{D}_1(\boldsymbol{\alpha})\right)\dot{f}(\boldsymbol{\alpha})\right\rVert\ \nonumber \\
&&= \sup_{\boldsymbol{\alpha}\in H_{n1}}\left[\frac{2\lambda_n}{n} \left\lVert \text{diag}(f_1(\boldsymbol{\alpha})\alpha_1^{-3},\ldots,f_{q_n}(\boldsymbol{\alpha})\alpha_{q_n}^{-3})-\textbf{F}_n(\boldsymbol{\alpha})\textbf{P}_n(\boldsymbol{\alpha})\right\rVert\right].
\end{eqnarray}
We first show the right-hand-side of \eqref{G7} is $o_p(1)$. This is equivalent to showing 
\begin{eqnarray}\label{G8}
\sup_{\boldsymbol{\alpha}\in H_{n1}}\left[\frac{2\lambda_n}{n} \left\lVert \text{diag}(f_1(\boldsymbol{\alpha})\alpha_1^{-3},\ldots,f_{q_n}(\boldsymbol{\alpha})\alpha_{q_n}^{-3})-\textbf{F}_n(\boldsymbol{\alpha})\textbf{P}_n(\boldsymbol{\alpha})\right\rVert\right]=o_p(1),
\end{eqnarray}
and 
\begin{eqnarray}\label{G9}
\sup_{\boldsymbol{\alpha}\in H_{n1}} \left\lVert  \frac{\textbf{F}_n(\boldsymbol{\alpha})\textbf{P}_n(\boldsymbol{\alpha})}{n}\right\rVert=o_p(1).
\end{eqnarray}
To show \eqref{G8}, since 
\begin{eqnarray*}
    \left\lVert \text{diag}(f_1(\boldsymbol{\alpha})\alpha_1^{-3},\ldots,f_{q_n}(\boldsymbol{\alpha})\alpha_{q_n}^{-3})\right\rVert = \max_{1\leq j\leq q_n}\left\{ |f_j(\boldsymbol{\alpha})\alpha_j^{-3}|\right\},
\end{eqnarray*}
by condition \textbf{C7}, $a_0\leq |\beta_{s0,j}|\leq a_1$, $1\leq j \leq q_n$, when $\boldsymbol{\alpha}\in H_{n1}$, $|\alpha_j-\beta_{s0,j}|\leq \delta \sqrt{p_n/n}$. Then, when $n$ is large enough, 
\begin{eqnarray*}
    |\alpha_j|\geq |\beta_{s0,j}|-\delta\sqrt{p_n/n}\geq |\beta_{s0,j}|-\frac{1}{2}|\beta_{s0,j}|=\frac{1}{2}|\beta_{s0,j}|\geq a_0/2.
\end{eqnarray*}
We obtain $|\alpha_j^{-3}|\leq (a_0/2)^{-3}$.
By 
\begin{eqnarray*}
    \sup_{\boldsymbol{\alpha}\in H_{n_1}}\left\lVert f(\boldsymbol{\alpha})-\boldsymbol{\beta}_{s01} \right\rVert\leq o_p\left(\delta\sqrt{p_n/n} \right),
\end{eqnarray*}
which we showed before, we have
\begin{eqnarray*}
    \sup_{\boldsymbol{\alpha}\in H_{n_1}}\left\lVert f(\boldsymbol{\alpha})-\boldsymbol{\beta}_{s01} \right\rVert \leq o_p\left(\delta\sqrt{p_n/n} \right)=o_p(1),
\end{eqnarray*}
and therefore we obtain 
\begin{eqnarray*}
    \sup_{\boldsymbol{\alpha}\in H_{n_1}}|f_j(\boldsymbol{\alpha})|\leq |\beta_{s0,j}|+o_p(1)\leq a_1+o_p(1).
\end{eqnarray*}
Thus
\begin{eqnarray*}
    \sup_{\boldsymbol{\alpha}\in H_{n_1}}|f_j(\boldsymbol{\alpha})\alpha_j^{-3}|\leq (a_1+o_p(1))(a_0/2)^{-3}=O_p(1),
\end{eqnarray*}
and
\begin{eqnarray*}
    \max_{1\leq j\leq q_n}\left\{|f_j(\boldsymbol{\alpha})\alpha_j^{-3}| \right\}=o_p(1).
\end{eqnarray*}
Since $\lambda_n/n \longrightarrow 0$, then 
\begin{eqnarray} \label{extra}
\sup_{\boldsymbol{\alpha}\in H_{n1}}\left[ \frac{2\lambda_n}{n} \left\lVert \text{diag}(f_1(\boldsymbol{\alpha})\alpha_1^{-3},\ldots,f_{q_n}(\boldsymbol{\alpha})\alpha_{q_n}^{-3})\right\rVert\right]\leq \frac{\lambda_n}{n} O_p(1) = o_p(1).
\end{eqnarray}
\eqref{extra} implies that \eqref{G8} holds. 
Now, we prove \eqref{G9}. Since $\left\lVert \textbf{F}_n(\boldsymbol{\alpha})\textbf{P}_n(\boldsymbol{\alpha}) \right\rVert \leq  \left\lVert \textbf{F}_n(\boldsymbol{\alpha}) \right\rVert \left\lVert \textbf{P}_n(\boldsymbol{\alpha}) \right\rVert$, one can write
\begin{eqnarray*}
    \textbf{F}_n(\boldsymbol{\alpha})\textbf{F}_n^\top(\boldsymbol{\alpha})=\begin{pmatrix}
        \left\lVert f_n(\boldsymbol{\alpha})-\boldsymbol{\alpha} \right\rVert^2 & \ldots & 0\\
        \vdots&\ddots&\vdots\\
        0&\ldots& \left\lVert f_{n}(\boldsymbol{\alpha})-\boldsymbol{\alpha} \right\rVert^2
    \end{pmatrix},
\end{eqnarray*}
where $\left\lVert \textbf{F}_n(\boldsymbol{\alpha})\textbf{F}_n^\top(\boldsymbol{\alpha})\right\rVert=\lambda_{\text{max}}(\textbf{F}_n(\boldsymbol{\alpha})\textbf{F}_n^\top(\boldsymbol{\alpha}))=\left\lVert f_n(\boldsymbol{\alpha})-\boldsymbol{\alpha} \right\rVert\leq \left\lVert f_n(\boldsymbol{\alpha})-\boldsymbol{\beta}_{s01}\right\rVert+\left\lVert \boldsymbol{\alpha}-\boldsymbol{\beta}_{s01}\right\rVert $, and 
\begin{equation*}
\begin{split}
    \sup_{\boldsymbol{\alpha}\in H_{n1}}\left\lVert f_n(\boldsymbol{\alpha})-\boldsymbol{\alpha} \right\rVert &\leq \sup_{\boldsymbol{\alpha}\in H_{n1}}  \left\lVert f_n(\boldsymbol{\alpha})-\boldsymbol{\beta}_{s01}\right\rVert+\sup_{\boldsymbol{\alpha}\in H_{n1}}\left\lVert \boldsymbol{\alpha}-\boldsymbol{\beta}_{s01}\right\rVert \\
    &=O_p\left(\sqrt{p_n/n} \right)+\delta\left(\sqrt{p_n/n} \right)\\
    &=O_p\left(\sqrt{p_n/n} \right).
\end{split}
\end{equation*}
Therefore, 
\begin{eqnarray}\label{G10}
    \sup_{\boldsymbol{\alpha}\in H_{n1}}\left\lVert \textbf{F}_n(\boldsymbol{\alpha}) \right\rVert=O_p\left(\sqrt{p_n/n} \right).
\end{eqnarray}
On the other hand, we have
\begin{eqnarray*}
    \frac{\textbf{P}_n^\top(\boldsymbol{\alpha})\textbf{P}_n(\boldsymbol{\alpha})}{n^2}=\sum_{j=1}^{q_n}\left(\frac{1}{n}\frac{\partial \omega_j^\top(\boldsymbol{\alpha})}{\partial \boldsymbol{\alpha}}\right)\left(\frac{1}{n}\frac{\partial \omega_j^\top(\boldsymbol{\alpha})}{\partial \boldsymbol{\alpha}}\right)^\top,
\end{eqnarray*}
and therefore, we obtain
\begin{equation*}
\begin{split}
    \left\lVert \frac{\textbf{P}_n^\top(\boldsymbol{\alpha})\textbf{P}_n(\boldsymbol{\alpha})}{n^2} \right\rVert&\leq \sum_{j=1}^{q_n}\left\lVert\left(\frac{1}{n}\frac{\partial \omega_j^\top(\boldsymbol{\alpha})}{\partial \boldsymbol{\alpha}}\right)\left(\frac{1}{n}\frac{\partial \omega_j^\top(\boldsymbol{\alpha})}{\partial \boldsymbol{\alpha}}\right)^\top\right\rVert\\
    &=\sum_{j=1}^{q_n}\lambda_{\text{max}}\left[\left(\frac{1}{n}\frac{\partial \omega_j^\top(\boldsymbol{\alpha})}{\partial \boldsymbol{\alpha}}\right)\left(\frac{1}{n}\frac{\partial \omega_j^\top(\boldsymbol{\alpha})}{\partial \boldsymbol{\alpha}}\right)^\top\right].
\end{split}
\end{equation*}
Since the trace of a symmetric matrix is equal to the sum of its eigenvalues, we obtain 
\begin{eqnarray*}
\begin{split}
    \left\lVert \frac{\textbf{P}_n^\top(\boldsymbol{\alpha})\textbf{P}_n(\boldsymbol{\alpha})}{n^2} \right\rVert &\leq \sum_{j=1}^{q_n}\text{trace}\left[\left(\frac{1}{n}\frac{\partial \omega_j^\top(\boldsymbol{\alpha})}{\partial \boldsymbol{\alpha}}\right)\left(\frac{1}{n}\frac{\partial \omega_j^\top(\boldsymbol{\alpha})}{\partial \boldsymbol{\alpha}}\right)^\top\right]\\
    &=\sum_{j=1}^{q_n}\sum_{k=1}^{q_n}\sum_{h=1}^{q_n}\left(\frac{1}{n}\frac{\partial \omega _{jk}(\boldsymbol{\alpha})}{\partial\boldsymbol{\alpha}_h} \right)^2.
\end{split}
\end{eqnarray*}
Noticing that 
\begin{eqnarray*}
    \boldsymbol{w}_{j}^\top(\boldsymbol{\alpha})=\left( \frac{\partial^2 [\sum_{i=1}^n \log f_n(v_{ni},(\boldsymbol{\alpha}^\top, \boldsymbol{0}^\top),\widetilde{\boldsymbol{\phi}})]}{\partial \alpha_j \partial\alpha_1},\ldots,\frac{\partial^2 [\sum_{i=1}^n \log f_n(v_{ni},(\boldsymbol{\alpha}^\top, \boldsymbol{0}^\top),\widetilde{\boldsymbol{\phi}})]}{\partial \alpha_j \partial\alpha_{q_n}}\right),
\end{eqnarray*}
by Cauchy-Schwarz inequality and condition \textbf{C9}, we have 
\begin{eqnarray*}
\begin{split}
    \left[\frac{1}{n}\frac{\partial \omega_{jk}(\boldsymbol{\alpha})}{\partial \boldsymbol{\alpha}_h} \right]^2 &= \left[\frac{1}{n}\frac{\partial^3[\sum_{i=1}^n \log f_n(v_{ni},(\boldsymbol{\alpha}^\top, \boldsymbol{0}^\top),\widetilde{\boldsymbol{\phi}})]}{\partial \alpha_j \partial\alpha_k\partial \alpha_h} \right]^2\\
    &=\frac{1}{n^2}\left[\sum_{i=1}^n\frac{\partial^3 [ \log f_n(v_{ni},(\boldsymbol{\alpha}^\top, \boldsymbol{0}^\top),\widetilde{\boldsymbol{\phi}})]}{\partial \alpha_j \partial\alpha_k\partial \alpha_h} \right]^2\\
    &\leq \frac{n}{n^2}\sum_{i=1}^n\left[\frac{\partial^3 [\log f_n(v_{ni},(\boldsymbol{\alpha}^\top, \boldsymbol{0}^\top),\widetilde{\boldsymbol{\phi}})]}{\partial \alpha_j \partial\alpha_k\partial \alpha_h} \right]^2\\
    &\leq \frac{1}{n}\sum_{i=1}^n {M^2_n}_{jkh}(v_{ni}).
\end{split}
\end{eqnarray*}
Hence, we have
\begin{eqnarray*}
    \sup_{\boldsymbol{\alpha}\in H_{n1}}\left\lVert \frac{\textbf{P}_n^\top(\boldsymbol{\alpha})\textbf{P}_n(\boldsymbol{\alpha})}{n^2} \right\rVert\leq \frac{1}{n}\sum_{j=1}^{q_n}\sum_{k=1}^{q_n}\sum_{h=1}^{q_n}\sum_{i=1}^{n}{M^2_n}_{jkh}(v_{ni}).
\end{eqnarray*}
Since condition \textbf{C9} indicates $E_{(\boldsymbol{\beta},\boldsymbol{\phi})}\left\{{M^2_n}_{jkh}(v_{ni}) \right\} < M_d < \infty$, we have
\begin{eqnarray*}
    E_{(\boldsymbol{\beta},\boldsymbol{\phi})}\left[\frac{1}{n}\sum_{j=1}^{q_n}\sum_{k=1}^{q_n}\sum_{h=1}^{q_n}{M^2_n}_{jkh}(v_{ni})\right]\leq M_{d}\cdot q_{n}^3,
\end{eqnarray*}
which implies $\frac{1}{n}\sum_{j=1}^{q_n}\sum_{k=1}^{q_n}\sum_{h=1}^{q_n}{M^2_n}_{jkh}(v_{ni})=O_p(q_n^3)$. As a result, we deduce that
\begin{eqnarray} \label{G11}
\sup_{\boldsymbol{\alpha}\in H_{n1}}\left\lVert \frac{\textbf{P}_n^\top(\boldsymbol{\alpha})\textbf{P}_n(\boldsymbol{\alpha})}{n^2} \right\rVert=O_p(q_n^3).
\end{eqnarray}
Finally, by \eqref{G10} and \eqref{G11}, we have
\begin{equation*}
\begin{split}
    \sup_{\boldsymbol{\alpha}\in H_{n1}}\left\lVert \frac{\textbf{F}_n(\boldsymbol{\alpha})\textbf{P}_n(\boldsymbol{\alpha})}{n}\right\rVert &\leq O_p\left(q_n^{3/2}\sqrt{p_n/n} \right) \\
&=O_p\left(\sqrt{p_n q_n^{3}/n}\right)\\
    &\leq O_p\left(\sqrt{p_n^2 q_n^{2}/n}\right) \\
&=O_p\left(p_n q_n/\sqrt{n}\right).
\end{split}
\end{equation*}
Consequently, by condition \textbf{C6}, $p_nq_n/\sqrt{n}\longrightarrow 0$, we have 
\begin{eqnarray*}
    \sup_{\boldsymbol{\alpha}\in H_{n1}}\left\lVert \frac{\textbf{F}_n(\boldsymbol{\alpha})\textbf{P}_n(\boldsymbol{\alpha})}{n}\right\rVert=o_p(1),
\end{eqnarray*}
which means that \eqref{G9} holds. By \eqref{G7}, we have
\begin{equation} \label{G12}
    \sup_{\boldsymbol{\alpha}\in H_{n_1}}\left\lVert \left( \frac{\boldsymbol{\Omega}_n^{(1)}(\boldsymbol{\alpha})}{n}+\frac{\lambda_n}{n}\textbf{D}_1(\boldsymbol{\alpha})\right)\dot{f}(\boldsymbol{\alpha})\right\rVert=o_p(1).
\end{equation}
Subsequently, we aim to demonstrate that with probability tending to one, we have
\begin{eqnarray*}
 \sup_{\boldsymbol{\alpha}\in H_{n_1}}\left\lVert \dot{f}(\boldsymbol{\alpha})\right\rVert\longrightarrow 0.
\end{eqnarray*}
Since for any two matrices $\textbf{A}$ and $\textbf{B}$, by the 2-norm properties, we have
\begin{eqnarray*}
    \lambda_{\text{min}}(\textbf{A})\left\lVert \textbf{B}\right\rVert\leq \left\lVert \textbf{AB}\right\rVert\leq \lambda_{\text{max}}(\textbf{A})\left\lVert \textbf{B}\right\rVert.
\end{eqnarray*}
Then, according to condition \textbf{C6}, we can conclude that
\begin{eqnarray*}
    \left\lVert\frac{\boldsymbol{\Omega}_n^{(1)}(\boldsymbol{\alpha})}{n}\dot{f}(\boldsymbol{\alpha}) \right\rVert\geq \frac{1}{c}\left\lVert \dot{f}(\boldsymbol{\alpha}) \right\rVert.
\end{eqnarray*}
By condition \textbf{C7}, when $n$ is large enough, $\forall j \in \{ 1,\ldots,q_n\}$,
\begin{eqnarray*}
    |\alpha_j|\geq |\beta_{s0,j}|-|\alpha_j-\beta_{s0,j}|\geq |\beta_{s0,j}|-\frac{a_0}{2}\geq \frac{a_0}{2}>0.
\end{eqnarray*}
Then, 
\begin{eqnarray*}
    \left\lVert \textbf{D}_1(\boldsymbol{\alpha})\right\rVert=\lambda_{\text{max}}(\textbf{D}_1(\boldsymbol{\alpha}))=\max_{1\leq j\leq q_n}(\alpha_j^{-2})\leq (a_0/2)^{-2},
\end{eqnarray*}
and
\begin{eqnarray*}
    \frac{\lambda_n}{n}\left\lVert \textbf{D}_1(\boldsymbol{\alpha})\dot{f}(\boldsymbol{\alpha})\right\rVert\leq\frac{\lambda_n}{n}\lambda_{\text{max}}(\textbf{D}_1(\boldsymbol{\alpha}))\left\lVert \dot{f}(\boldsymbol{\alpha})\right\rVert \leq  \frac{\lambda_n}{n}(a_0/2)^{-2}\left\lVert \dot{f}(\boldsymbol{\alpha})\right\rVert.
\end{eqnarray*}
Therefore, we have
\begin{equation}\label{G13}
\begin{split}
\left\lVert\left(\frac{\boldsymbol{\Omega}_n^{(1)}(\boldsymbol{\alpha})}{n}+\frac{\lambda_n}{n}\textbf{D}_1(\boldsymbol{\alpha})\right)\dot{f}(\boldsymbol{\alpha}) \right\rVert &\geq  \left\lVert\left(\frac{\boldsymbol{\Omega}_n^{(1)}(\boldsymbol{\alpha})}{n}\right)\dot{f}(\boldsymbol{\alpha}) \right\rVert -\frac{\lambda_n}{n}\left\lVert \textbf{D}_1(\boldsymbol{\alpha})\dot{f}(\boldsymbol{\alpha}) \right\rVert \nonumber\\
   &\geq \frac{1}{c}\left\lVert \dot{f}(\boldsymbol{\alpha})  \right\rVert-\frac{\lambda_n}{n}(\alpha_0/2)^{-2}\left\lVert \dot{f}(\boldsymbol{\alpha})  \right\rVert\nonumber\\
   &= \left[\frac{1}{c}-\frac{\lambda_n}{n}(\alpha_0/2)^{-2} \right]\left\lVert \dot{f}(\boldsymbol{\alpha})\right\rVert.
\end{split}
\end{equation}
By\eqref{G12} and \eqref{G13} we obtain
\begin{eqnarray*}
    \left[\frac{1}{c}-\frac{\lambda_n}{n}(\alpha_0/2)^{-2} \right]\sup_{\boldsymbol{\alpha}\in H_{n_1}}\left\lVert \dot{f}(\boldsymbol{\alpha})\right\rVert \leq o_p(1),
\end{eqnarray*}
and $\sup_{\boldsymbol{\alpha}\in H_{n_1}}\left\lVert \dot{f}(\boldsymbol{\alpha})\right\rVert=o_p(1)$ which implies that $\dot{f}(\cdot)$ is a contraction mapping from $H_{n1}$ to itself with probability tending to one. Hence, according to the contraction mapping theorem, there exists one unique fixed-point $\widehat{\boldsymbol{\alpha}}^*\in H_{n_1}$ such that 
\begin{eqnarray}
    \label{A11}\widehat{\boldsymbol{\alpha}}^*=(\boldsymbol{\Omega}_n^{(1)}(\widehat{\boldsymbol{\alpha}}^*)+\lambda_n \textbf{D}_1(\widehat{\boldsymbol{\alpha}}^*))^{-1}\textbf{v}_n^{(1)}(\widehat{\boldsymbol{\alpha}}^*).
\end{eqnarray}
\textit{Proof of Theorem 1 (i)}. By definition of $\widehat{\boldsymbol{\beta}}$ and $\widehat{\boldsymbol{\beta}}_{s2}^{(m)}$, ${\widehat{\boldsymbol{\beta}}}=\lim_{m\longrightarrow \infty}\widehat{\boldsymbol{\beta}}^{(m)}$, and ${\widehat{\boldsymbol{\beta}}}_{s2}=\lim_{m\longrightarrow\infty}\widehat{\boldsymbol{\beta}}_{s2}^{(m)}$. Since $\widehat{\boldsymbol{\beta}}^{(m)}\in H_n$, by \textbf{Lemma 1(i)}, 
\begin{eqnarray*}
    \widehat{\boldsymbol{\beta}}_{s2}^{(m)}=\boldsymbol{\gamma}^*(\widehat{\boldsymbol{\beta}}_{s2}^{(m-1)})\leq \frac{1}{c_0}\left\lVert(\widehat{\boldsymbol{\beta}}_{s2}^{(m-1)}) \right\rVert\leq \ldots\leq \left(\frac{1}{c_0}\right)^{m}\left\lVert \widehat{\boldsymbol{\beta}}_{s2}^{(0)}\right\rVert.
\end{eqnarray*}
Hence, $\left(\frac{1}{c_0}\right)^m \longrightarrow 0$, $m \longrightarrow \infty$ and $\lim_{m \rightarrow \infty}\widehat{\boldsymbol{\beta}}_{s2}^{(m)}=0$ which implies that $\widehat{\boldsymbol{\beta}}_{s2}=0$ with probability tending to one. 
\vspace*{20px}

\noindent
\textit{Proof of Theorem 1 (ii)}. In \textbf{Lemma 3}, we have shown that the following equation
\begin{eqnarray}\label{R1}
    \boldsymbol{\alpha}=(\boldsymbol{\Omega}_n^{(1)}(\boldsymbol{\alpha})+\lambda_n\textbf{D}_1(\boldsymbol{\alpha}))^{-1}\textbf{v}_n^{(1)}(\boldsymbol{\alpha})
\end{eqnarray}
has a unique fixed-point $\widehat{\boldsymbol{\alpha}}^*$ in the domain $H_{n_1}$ such that 
\begin{eqnarray}\label{R11}
    \widehat{\boldsymbol{\alpha}}^*=(\boldsymbol{\Omega}_n^{(1)}(\widehat{\boldsymbol{\alpha}}^*)+\lambda_n\textbf{D}_1(\widehat{\boldsymbol{\alpha}}^*))^{-1}\textbf{v}_n^{(1)}(\widehat{\boldsymbol{\alpha}}^*),
 \end{eqnarray}
where 
\begin{eqnarray*}
    \boldsymbol{\Omega}_n^{(1)}(\widehat{\boldsymbol{\alpha}}^*)=\boldsymbol{\Omega}_n^{(1)}(\boldsymbol{\beta})\Big|_{\boldsymbol{\beta}_{1}=\widehat{\boldsymbol{\alpha}}^*,\boldsymbol{\beta}_{2}=0},
\end{eqnarray*}
and
\begin{eqnarray*}
    \textbf{v}_n^{(1)}(\widehat{\boldsymbol{\alpha}}^*)=\textbf{v}_n^{(1)}(\boldsymbol{\beta})\Big|_{\boldsymbol{\beta}_{1}=\widehat{\boldsymbol{\alpha}}^*,\boldsymbol{\beta}_{2}=0}.
\end{eqnarray*}
The next part is to show that with probability tending to one, $\widehat{\boldsymbol{\beta}}_{s1}=\widehat{\boldsymbol{\alpha}}^*$, i.e., $P(\widehat{\boldsymbol{\beta}}_{s1}=\widehat{\boldsymbol{\alpha}}^*)=1$ or with probability tending to one, $\widehat{\boldsymbol{\beta}}_{s1}$ is the unique fixed-point of \eqref{R1}.

First, by \eqref{A2} (shown previously) that is 
\begin{eqnarray*}
\begin{pmatrix}
    \boldsymbol{\alpha}^*(\boldsymbol{\beta}) - \boldsymbol{\beta}_{s01} \\
    \boldsymbol{\gamma}^*(\boldsymbol{\beta})
\end{pmatrix}
+ \frac{\lambda_n}{n} 
\begin{pmatrix}
    \textbf{A}(\boldsymbol{\beta})\textbf{D}_1(\boldsymbol{\beta}_{s1})\boldsymbol{\alpha}^*(\boldsymbol{\beta}) + \textbf{B}(\boldsymbol{\beta})\textbf{D}_2(\boldsymbol{\beta}_{s2})\boldsymbol{\gamma}^*(\boldsymbol{\beta}) \\
    \textbf{B}^\top(\boldsymbol{\beta}) \textbf{D}_1(\boldsymbol{\beta}_{s1})\boldsymbol{\alpha}^*(\boldsymbol{\beta}) + \textbf{G}(\boldsymbol{\beta})\textbf{D}_2(\boldsymbol{\beta}_{s2})\boldsymbol{\gamma}^*(\boldsymbol{\beta}) 
\end{pmatrix}
= \widehat{\textbf{b}}(\boldsymbol{\beta}) - \boldsymbol{\beta}_{s0}.
\end{eqnarray*}
We obtain 
\begin{eqnarray*}
    \boldsymbol{\gamma}^*(\boldsymbol{\beta})+\frac{\lambda_n}{n}\left(\textbf{B}^\top(\boldsymbol{\beta}) \textbf{D}_1(\boldsymbol{\beta}_{s1})\boldsymbol{\alpha}^*(\boldsymbol{\beta}) + \textbf{G}(\boldsymbol{\beta})\textbf{D}_2(\boldsymbol{\beta}_{s2})\boldsymbol{\gamma}^*(\boldsymbol{\beta}) \right)=(\widehat{\textbf{b}}(\boldsymbol{\beta}) - \boldsymbol{\beta}_{s0})^{(2)},
\end{eqnarray*}
where $(\widehat{\textbf{b}}(\boldsymbol{\beta}) - \boldsymbol{\beta}_{s0})^{(2)}$ are the elements corresponding to $\boldsymbol{\beta}_{s02}$. We want to show that $\lim_{\boldsymbol{\beta}_{s2}\rightarrow 0}\boldsymbol{\gamma}^*(\boldsymbol{\beta})=0$. By \textbf{Lemma 1(i)}, 
\begin{eqnarray*}
    \left\lVert \boldsymbol{\gamma}^*(\boldsymbol{\beta}) \right\rVert\leq \left\lVert \boldsymbol{\beta}_{s2} \right\rVert \longrightarrow 0. 
\end{eqnarray*}
Therefore, $\lim_{\boldsymbol{\beta}_{s2}\rightarrow 0}\boldsymbol{\gamma}^*(\boldsymbol{\beta})=0$.
By multiplying $(\boldsymbol{\Omega}_n(\boldsymbol{\beta})+\lambda_n \textbf{D}(\boldsymbol{\beta}))$ on both sides of \eqref{A1}, one can get
\begin{eqnarray}
\{\boldsymbol{\Omega}_n(\boldsymbol{\beta}) + \lambda_n \textbf{D}(\boldsymbol{\beta})\}
\begin{pmatrix}
    \boldsymbol{\alpha}^*(\boldsymbol{\beta}) \\
    \boldsymbol{\gamma}^*(\boldsymbol{\beta})
\end{pmatrix} 
= \boldsymbol{\gamma}_n(\boldsymbol{\beta}),
\end{eqnarray}
which can be rewritten as 
\begin{eqnarray*}
\begin{bmatrix}
    \begin{pmatrix}
        \boldsymbol{\Omega}_n^{(1)}(\boldsymbol{\beta})&\boldsymbol{\Omega}_n^{(12)}(\boldsymbol{\beta})\\
        \boldsymbol{\Omega}_n^{(21)}(\boldsymbol{\beta})&\boldsymbol{\Omega}_n^{(2)}(\boldsymbol{\beta})
    \end{pmatrix}+
    \begin{pmatrix}
        \lambda_n\textbf{D}_1(\boldsymbol{\beta}_{s1})& \boldsymbol{0}\\
        \boldsymbol{0} & \lambda_n\textbf{D}_2(\boldsymbol{\beta}_{s2})
    \end{pmatrix}
\end{bmatrix}\begin{pmatrix}\boldsymbol{\alpha}^*(\boldsymbol{\beta})\\
\boldsymbol{\gamma}^*(\boldsymbol{\beta})
    
\end{pmatrix}=\begin{pmatrix}
    \boldsymbol{\gamma}_n^{(1)}(\boldsymbol{\beta})\\
    \boldsymbol{\gamma}_n^{(2)}(\boldsymbol{\beta})
\end{pmatrix}.
\end{eqnarray*}
Consequently,
\begin{eqnarray*}
    \left(\boldsymbol{\Omega}_n^{(1)}(\boldsymbol{\beta})+\lambda_n \textbf{D}_1(\boldsymbol{\beta})\right)\boldsymbol{\alpha}^*(\boldsymbol{\beta})+\boldsymbol{\Omega}_n^{(12)}(\boldsymbol{\beta})\boldsymbol{\gamma}^*(\boldsymbol{\beta})=\boldsymbol{\gamma}_n^{(1)}(\boldsymbol{\beta}).
\end{eqnarray*}
Then, we have
\begin{eqnarray*}
    \boldsymbol{\alpha}^*(\boldsymbol{\beta})=\left( \boldsymbol{\Omega}_n^{(1)}(\boldsymbol{\beta})+\lambda_n\textbf{D}_1(\boldsymbol{\beta})\right)^{-1}\left[\boldsymbol{\gamma}_n^{(1)}(\boldsymbol{\beta})-\boldsymbol{\Omega}_n^{(12)}(\boldsymbol{\beta})\boldsymbol{\gamma}^*(\boldsymbol{\beta}) \right].
\end{eqnarray*}
Since $\lim_{\boldsymbol{\beta}_{s2}\rightarrow 0}\boldsymbol{\gamma}^*(\boldsymbol{\beta})=0$, we have 
\begin{eqnarray*}
    \lim_{\boldsymbol{\beta}_{s2}\rightarrow 0}\left[\boldsymbol{\Omega}_n^{(12)}(\boldsymbol{\beta})\boldsymbol{\gamma}^*(\boldsymbol{\beta})\right]=0.
\end{eqnarray*}
Therefore, 
\begin{eqnarray*}
    \lim_{\boldsymbol{\beta}_{s2}\rightarrow 0}\boldsymbol{\alpha}^*(\boldsymbol{\beta})=\left(\boldsymbol{\Omega}_n^{(1)}(\boldsymbol{\beta})+\lambda_n\textbf{D}_1(\boldsymbol{\beta}) \right)^{-1}\boldsymbol{\gamma}_n^{(1)}(\boldsymbol{\beta}_{s1}) = f(\boldsymbol{\beta}_{s1}).
\end{eqnarray*}
Since $J(\boldsymbol{\beta})=\boldsymbol{\alpha}^*(\boldsymbol{\beta})$ is continuous and thus continuous on the compound set $\boldsymbol{\beta}\in H_n$, hence, as $m\rightarrow\infty$, $\widehat{\boldsymbol{\beta}}_{s2}^{(m)}\rightarrow 0$. We obtain
\begin{eqnarray} \label{R20}
    \eta_m\equiv \sup_{\boldsymbol{\beta}\in H_{n1}}\left\lVert \boldsymbol{\alpha}^*(\boldsymbol{\beta}_{s1},\widehat{\boldsymbol{\beta}}_{s2}^{(m)})-f(\boldsymbol{\beta}_{s1}) \right\rVert \longrightarrow 0.
\end{eqnarray}
Since $f(\cdot)$ is a contract mapping, and $\sup_{\boldsymbol{\alpha}\in H_{n1}}\left\lVert \dot{f}(\boldsymbol{\alpha})\right\rVert \longrightarrow 0$, $n\rightarrow \infty$, then, with probability tending to one, we have 
\begin{eqnarray*}
    \sup_{\boldsymbol{\alpha}\in H_{n1}}\left\lVert \dot{f}(\boldsymbol{\alpha})\right\rVert \leq \frac{1}{c_1}, \quad \text{for some } c_1>1,
\end{eqnarray*}
and we have 
\begin{eqnarray*}
    \left\lVert f(\widehat{\boldsymbol{\beta}}_{s1}^{(m)})-\widehat{\boldsymbol{\alpha}}^*\right\rVert =\left\lVert f(\widehat{\boldsymbol{\beta}}^{(m)})-f(\widehat{\boldsymbol{\alpha}}^*)\right\rVert\leq \frac{1}{c_1}\left\lVert \widehat{\boldsymbol{\beta}}_{s1}^{(m)}-\widehat{\boldsymbol{\alpha}}^*\right\rVert.
\end{eqnarray*}
Note: $\widehat{\boldsymbol{\beta}}^{(m+1)}=\boldsymbol{\alpha}^*(\widehat{\boldsymbol{\beta}}^{(m)})$, i.e., $\widehat{\boldsymbol{\beta}}^{(m+1)}$ updates $\widehat{\boldsymbol{\beta}}^{(m)}$. Let $h_m=\left\lVert \widehat{\boldsymbol{\beta}}_{s1}^{(m)}-\widehat{\boldsymbol{\alpha}}^* \right\rVert$, then
\begin{equation*}
\begin{split}
    h_{m+1} &= \left\lVert \widehat{\boldsymbol{\beta}}_{s1}^{(m+1)}-\widehat{\boldsymbol{\alpha}}^* \right\rVert=\left\lVert \boldsymbol{\alpha}^*(\widehat{\boldsymbol{\beta}}^{(m)})-\widehat{\boldsymbol{\alpha}}^* \right\rVert\\
    &\leq \left\lVert \boldsymbol{\alpha}^*(\widehat{\boldsymbol{\beta}}^{(m)})-f(\widehat{\boldsymbol{\beta}}_{s1}^{(m)}) \right\rVert+\left\lVert f(\widehat{\boldsymbol{\beta}}_{s1}^{(m)})-f(\widehat{\boldsymbol{\alpha}}^*) \right\rVert\\
    &\leq \left\lVert \boldsymbol{\alpha}^*(\widehat{\boldsymbol{\beta}}_{s1}^{(m)},\widehat{\boldsymbol{\beta}}_{s2}^{(m)}) -f(\widehat{\boldsymbol{\beta}}_{s1}^{(m)})\right\rVert+\left\lVert f(\widehat{\boldsymbol{\beta}}_{s1}^{(m)})-f(\widehat{\boldsymbol{\alpha}}^*) \right\rVert\\
    &\leq \eta_m+\frac{1}{c_1}\left\lVert \widehat{\boldsymbol{\beta}}_{s1}^{(m)}-\widehat{\boldsymbol{\alpha}}^*\right\rVert\\
    &\leq \eta_m+\frac{1}{c_1} h_m.
\end{split}
\end{equation*}
By \eqref{R20}, for any $\epsilon>0$, there exists an $N>0$ such that for all $m>N$, $\eta_m<\epsilon$. Therefore, for $m>N$, or $m-N>0$, we have
\begin{equation*}
\begin{split}
    h_{m+1} &\leq \frac{1}{c_1}h_m+\eta_m\\
    &\leq \frac{1}{c_1}(\frac{1}{c_1}h_{m-1}+\eta_{m-1})+\eta_m\\
    &=\frac{1}{c_1^2}h_{m-1}+\frac{1}{c_1}\eta_{m-1}+\eta_m\\
    &\leq \frac{h_1}{c_1^m}+\frac{\eta_1}{c_1^{m-1}}+\frac{\eta_2}{c_1^{m-2}}+\ldots+\frac{\eta_N}{c_1^{m-N}}+\frac{\eta_{N+1}}{c_1^{m-(N+1)}}\\
    &+\ldots+\frac{\eta_{m-1}}{c_1}+\eta_m\\
    &=\frac{h_1}{c_1^m}+\frac{\eta_1}{c_1^{m-1}}+\frac{\eta_2}{c_1^{m-2}}+\ldots+\frac{\eta_N}{c_1^{m-N}}+\left(\frac{\eta_{N+1}}{c_1^{m-(N+1)}}+\ldots+\frac{\eta_{m-1}}{c_1}+\eta_m \right)\\
    &\leq (h_1+\eta_1+\ldots+\eta_N)\frac{1}{c_1^{m-N}}+\left( \frac{1}{c_1^{m-(N+1)}}+\ldots+\frac{1}{c_1}+1\right)\epsilon\\
    &=(h_1+\eta_1+\ldots+\eta_N)\frac{1}{c_1^{m-N}}+\frac{1-(1/c_1)^{m-N}}{1-(1/c_1)},~ \text{by sum of the geometric series}.
\end{split}
\end{equation*}
Since $1/c_1^{m-N}\rightarrow 0$ and $\frac{1-(1/c_1)^{m-N}}{1-(1/c_1)}\rightarrow\frac{c_1}{c_1-1}\epsilon$ when $m\rightarrow\infty$, there exists $N_0>N$ such that when $m>N_0$,
\begin{eqnarray*}
    (h_1+\eta_1+\ldots+\eta_N)\frac{1}{c_1^{m-N}}<\epsilon
\end{eqnarray*}
and 
\begin{eqnarray*}
    \frac{1-(1/c_1)^{m-N}}{1-(1/c_1)}<2\frac{c_1}{c_1-1}\epsilon,
\end{eqnarray*}
which implies 
\begin{eqnarray*}
    h_{m+1}<\left(1+\frac{2c_1}{c_1-1} \right)\epsilon=\frac{3c_1-1}{c_1-1}\varepsilon,
\end{eqnarray*}
and $h_{m+1}\longrightarrow 0$
when $m\rightarrow\infty$.
Hence, with probability tending to one, we have $h_m=\left\lVert \widehat{\boldsymbol{\beta}}_{s1}^{(m)}-\widehat{\boldsymbol{\alpha}}^*\right\rVert\longrightarrow 0$ as $m\longrightarrow\infty$. Since $\widehat{\boldsymbol{\beta}}_{s1}=\lim_{m\rightarrow\infty}\widehat{\boldsymbol{\beta}}_{s1}^{(m)}$ and 
\begin{eqnarray*}
\left\lVert \widehat{\boldsymbol{\beta}}_{s1}-\widehat{\boldsymbol{\alpha}}^* \right\rVert \leq \left\lVert \widehat{\boldsymbol{\beta}}_{s1}-\widehat{\boldsymbol{\beta}}_{s1}^{(m)} \right\rVert+ \left\lVert \widehat{\boldsymbol{\beta}}_{s1}^{(m)} -\widehat{\boldsymbol{\alpha}}^* \right\rVert\longrightarrow 0, 
\end{eqnarray*}
when $m\longrightarrow\infty$. This implies $P(\widehat{\boldsymbol{\beta}}_{s1}=\widehat{\boldsymbol{\alpha}}^*)=1$ and the proof of Theorem 1 (ii) is complete.

\noindent
\textit{Proof of Theorem 1 (iii)}. Based on \eqref{A11}, 
\begin{eqnarray*}
\widehat{\boldsymbol{\alpha}}^*=(\boldsymbol{\Omega}_n^{(1)}(\widehat{\boldsymbol{\alpha}}^*)+\lambda_n\textbf{D}_1(\widehat{\boldsymbol{\alpha}}^*))^{-1}\textbf{v}_n^{(1)}(\widehat{\boldsymbol{\alpha}}^*), 
\end{eqnarray*}
and we have
\begin{eqnarray*}
    \sqrt{n}(\widehat{\boldsymbol{\alpha}}^*-\boldsymbol{\beta}_{s01})=\pi_1+\pi_2,
\end{eqnarray*}
where
\begin{eqnarray*}
    \pi_1 \equiv \sqrt{n}\left[ (\boldsymbol{\Omega}_n^{(1)}(\widehat{\boldsymbol{\alpha}}^*)+\lambda_n(\textbf{D}_1(\widehat{\boldsymbol{\alpha}}^*))^{-1}\boldsymbol{\Omega}_n^{(1)}(\widehat{\boldsymbol{\alpha}}^*)-\textbf{I}_{q_n}\right]\boldsymbol{\beta}_{s01},
\end{eqnarray*}
\begin{eqnarray*}
    \pi_2 \equiv \sqrt{n} \left(\boldsymbol{\Omega}_n^{(1)}(\widehat{\boldsymbol{\alpha}}^*)+\lambda_n(\textbf{D}_1(\widehat{\boldsymbol{\alpha}}^*)\right)^{-1}\left( \textbf{v}_n^{(1)}(\widehat{\boldsymbol{\alpha}}^*)-\boldsymbol{\Omega}_n^{(1)}(\widehat{\boldsymbol{\alpha}}^*)\boldsymbol{\beta}_{s01}\right).
\end{eqnarray*}
Noticing that for any two conformable invertible matrices $\boldsymbol{\zeta}$ and $\boldsymbol{\Psi}$, we have
\begin{eqnarray*}
    (\boldsymbol{\zeta}+\boldsymbol{\Psi})^{-1}=\boldsymbol{\zeta}^{-1}-\boldsymbol{\zeta}^{-1}\boldsymbol{\Psi}(\boldsymbol{\zeta}+\boldsymbol{\Psi})^{-1}.
\end{eqnarray*}
Then
\begin{equation}\label{R25}
    \left(\boldsymbol{\Omega}_n^{(1)}(\widehat{\boldsymbol{\alpha}}^*)+\lambda_n\textbf{D}_1(\widehat{\boldsymbol{\alpha}}^*)\right)^{-1}(\boldsymbol{\Omega}_n^{(1)}(\widehat{\boldsymbol{\alpha}}^*)) = \textbf{I}_{q_n}-\lambda_n(\boldsymbol{\Omega}_n^{(1)}(\widehat{\boldsymbol{\alpha}}^*))^{-1}\textbf{D}_1(\widehat{\boldsymbol{\alpha}}^*)  
\end{equation}
and 
\begin{equation*}
\begin{split}
    \pi_1 &= \sqrt{n}\left[ -\lambda_n(\boldsymbol{\Omega}_n^{(1)}(\widehat{\boldsymbol{\alpha}}^*))^{-1}\textbf{D}_1(\widehat{\boldsymbol{\alpha}}^*)(\boldsymbol{\Omega}_n^{(1)}(\widehat{\boldsymbol{\alpha}}^*)+\lambda_n\textbf{D}_1(\widehat{\boldsymbol{\alpha}}^*))^{-1}\boldsymbol{\Omega}_n^{(1)}(\widehat{\boldsymbol{\alpha}}^*)\boldsymbol{\beta}_{s01}\right]\\
    &=-\frac{\lambda_n}{\sqrt{n}}\left(\frac{1}{n}\boldsymbol{\Omega}_n^{(1)}(\widehat{\boldsymbol{\alpha}}^*) \right)^{-1}\textbf{D}_1(\widehat{\boldsymbol{\alpha}}^*)\left(\frac{1}{n}\boldsymbol{\Omega}_n^{(1)}(\widehat{\boldsymbol{\alpha}}^*)+\frac{\lambda_n}{n}\textbf{D}_1(\widehat{\boldsymbol{\alpha}}^*)\right)^{-1}\frac{1}{n}\boldsymbol{\Omega}_n^{(1)}(\widehat{\boldsymbol{\alpha}}^*)\boldsymbol{\beta}_{s01}.
\end{split}
\end{equation*}
By conditions \textbf{C5} and \textbf{C6}, we have 
\begin{eqnarray}
    \label{R30}
    \left\lVert\pi_1 \right\rVert=O_p(\lambda_n\sqrt{q_n/n})\longrightarrow 0.
\end{eqnarray}
Next, we consider $\pi_2$. It follows from \eqref{R25} and the condition \textbf{C6}, $\lambda_n/\sqrt{n}\rightarrow 0$, that
\begin{equation*}
\begin{split}
    \pi_2 &\equiv \sqrt{n}(\boldsymbol{\Omega}_n^{(1)}(\widehat{\boldsymbol{\alpha}}^*)+\lambda_n\textbf{D}_1(\widehat{\boldsymbol{\alpha}}^*))^{-1}\left(\textbf{v}_n^{(1)}(\widehat{\boldsymbol{\alpha}}^*) -\boldsymbol{\Omega}_n^{(1)}(\widehat{\boldsymbol{\alpha}}^*)\boldsymbol{\beta}_{s01}\right)\\
    &=\sqrt{n}\left[
    (\boldsymbol{\Omega}_n^{(1)}(\widehat{\boldsymbol{\alpha}}^*))^{-1}-\lambda_n(\boldsymbol{\Omega}_n^{(1)}(\widehat{\boldsymbol{\alpha}}^*))^{-1}\textbf{D}_1(\widehat{\boldsymbol{\alpha}}^*)(\boldsymbol{\Omega}_n^{(1)}(\widehat{\boldsymbol{\alpha}}^*)+\lambda_n\textbf{D}_1(\widehat{\boldsymbol{\alpha}}^*))^{-1}\right]\\
    &\left( \textbf{v}_n^{(1)}(\widehat{\boldsymbol{\alpha}}^*)-\boldsymbol{\Omega}_n^{(1)}(\widehat{\boldsymbol{\alpha}}^*)\boldsymbol{\beta}_{s01})\right)\\
    &=\sqrt{n}\left[
    (\frac{1}{n}\boldsymbol{\Omega}_n^{(1)}(\widehat{\boldsymbol{\alpha}}^*))^{-1}-\frac{\lambda_n}{n}(\frac{1}{n}\boldsymbol{\Omega}_n^{(1)}(\widehat{\boldsymbol{\alpha}}^*))^{-1}\textbf{D}_1(\widehat{\boldsymbol{\alpha}}^*)(\frac{1}{n}\boldsymbol{\Omega}_n^{(1)}(\widehat{\boldsymbol{\alpha}}^*)+\frac{\lambda_n}{n}\textbf{D}_1(\widehat{\boldsymbol{\alpha}}^*))^{-1}\right]\\
    &\left( \frac{1}{n}\textbf{v}_n^{(1)}(\widehat{\boldsymbol{\alpha}}^*)-\frac{1}{n}\boldsymbol{\Omega}_n^{(1)}(\widehat{\boldsymbol{\alpha}}^*)\boldsymbol{\beta}_{s01})\right).
\end{split}
\end{equation*}
By assumption (C6), $\lambda_n/n=(\lambda_n / \sqrt{n})(1/\sqrt{n})=o(1)(1/\sqrt{n})=o(1/\sqrt{n})$, we have 
\begin{eqnarray*}
    \pi_2=\sqrt{n}\left[(\frac{1}{n}\boldsymbol{\Omega}_n^{(1)}(\widehat{\boldsymbol{\alpha}}^*))^{-1}-o_p(1/\sqrt{n})
    \right]\left( \frac{1}{n}\boldsymbol{v}_n^{(1)}(\widehat{\boldsymbol{\alpha}}^*)-\frac{1}{n}\boldsymbol{\Omega}_n^{(1)}(\widehat{\boldsymbol{\alpha}}^*)\boldsymbol{\beta}_{s01})\right).
\end{eqnarray*}
By a first-order Taylor expansion, we have 
\begin{eqnarray*}   \textbf{v}_n(\widehat{\boldsymbol{\alpha}}^*)=\textbf{v}_n(\widehat{\boldsymbol{\beta}})\Big|_{\boldsymbol{\beta}_{s1}=\widehat{\boldsymbol{\alpha}}^*,\boldsymbol{\beta}_{s2}=0}=\dot{\ell}_n(\widehat{\boldsymbol{\alpha}}^*|\widetilde{\boldsymbol{\phi}})-\Ddot{\ell}_n(\widehat{\boldsymbol{\alpha}}^*|\widetilde{\boldsymbol{\phi}})\begin{pmatrix}
       \widehat{\boldsymbol{\alpha}}^*\\
       \boldsymbol{0}
   \end{pmatrix}.
\end{eqnarray*}
Then
\begin{equation*}
\begin{split}
    \textbf{v}_n^{(1)}(\widehat{\boldsymbol{\alpha}}^*) &= \dot{\ell}_n^{(1)}(\widehat{\boldsymbol{\alpha}}^*|\widetilde{\boldsymbol{\phi}})+\boldsymbol{\Omega}_n^{(1)}(\widehat{\boldsymbol{\alpha}}^*)\widehat{\boldsymbol{\alpha}}^*\\
    &=\dot{\ell}_n^{(1)}(\widehat{\boldsymbol{\alpha}}^*|\widetilde{\boldsymbol{\phi}})+\Ddot{\ell}^{(1)}_n(\widehat{\boldsymbol{\alpha}}^*|\widetilde{\boldsymbol{\phi}})(\widehat{\boldsymbol{\alpha}}^*-\boldsymbol{\beta}_{s01}),
\end{split}
\end{equation*}
where $\widetilde{\boldsymbol{\alpha}}^*$ is between $\widehat{\boldsymbol{\alpha}}^*$ and $\boldsymbol{\beta}_{s01}$, $\left\lVert \widetilde{\boldsymbol{\alpha}}^*-\boldsymbol{\beta}_{s01}\right\rVert=o_p(1)$, and $\left\lVert \widetilde{\boldsymbol{\alpha}}^*-\widehat{\boldsymbol{\alpha}}^*\right\rVert=o_p(1)$. By condition \textbf{C4}, we have 
\begin{eqnarray*}
    \frac{1}{n}\boldsymbol{\Omega}_n^{(1)}(\widehat{\boldsymbol{\alpha}}^*)- \frac{1}{n}\boldsymbol{\Omega}_n^{(1)}(\widetilde{\boldsymbol{\alpha}}^*)=o_p(1),
\end{eqnarray*}
then 
\begin{equation*}
\begin{split}
    &\frac{1}{n}\boldsymbol{v}_n^{(1)}(\widehat{\boldsymbol{\alpha}}^*)-\frac{1}{n}\boldsymbol{\Omega}_n^{(1)}(\widehat{\boldsymbol{\alpha}}^*)\boldsymbol{\beta}_{s01}\\
    &=\frac{1}{n}\dot{\ell}_n^{(1)}(\boldsymbol{\beta}_{s01}|\widetilde{\boldsymbol{\phi}})-\left(-\frac{1}{n}\Ddot{\ell}_n^{(1)}(\widetilde{\boldsymbol{\alpha}}^*|\widetilde{\boldsymbol{\phi}})\right)(\widetilde{\boldsymbol{\alpha}}^*-\boldsymbol{\beta}_{s01})+\left(\frac{1}{n}\boldsymbol{\Omega}_n^{(1)}(\widehat{\boldsymbol{\alpha}}^*) \right)(\widehat{\boldsymbol{\alpha}}^*-\boldsymbol{\beta}_{s01})\\
    &=\frac{1}{n}\dot{\ell}_n^{(1)}(\boldsymbol{\beta}_{s01}|\widetilde{\boldsymbol{\phi}})+\left(\frac{1}{n}\boldsymbol{\Omega}_n^{(1)}(\widehat{\boldsymbol{\alpha}}^*)-\frac{1}{n}\boldsymbol{\Omega}_n^{(1)}(\widetilde{\boldsymbol{\alpha}}^*)\right)(\widehat{\boldsymbol{\alpha}}^*-\boldsymbol{\beta}_{s01})\\
    &=\frac{1}{n}\dot{\ell}_n^{(1)}(\boldsymbol{\beta}_{s01}|\widetilde{\boldsymbol{\phi}})+o_p(1).
\end{split}
\end{equation*}
Hence, we have 
\begin{equation*}
\begin{split}
    \sqrt{n}(\widehat{\boldsymbol{\alpha}}^*-\boldsymbol{\beta}_{s01})&=\pi_2+\pi_1\\
    &=\sqrt{n}\left[(\textbf{I}^{(1)}(\boldsymbol{\beta}_{s0}))^{-1}-o_p(1/\sqrt{n}) \right]\left[\frac{1}{n}\dot{\ell}_n^{(1)}(\boldsymbol{\beta}_{s01}|\widetilde{\boldsymbol{\phi}})+o_p(1)(\widehat{\boldsymbol{\alpha}}^*-\boldsymbol{\beta}_{s01}) 
    \right]o_p(1)\\
    &=\sqrt{n}\left[(\textbf{I}^{(1)}(\boldsymbol{\beta}_{s0}))^{-1}-o_p(1/\sqrt{n}) \right]\left[\frac{1}{n^2}\dot{\ell}_n^{(1)}(\boldsymbol{\beta}_{s01 }|\widetilde{\boldsymbol{\phi}})\right]+o_p(1)\sqrt{n}(\widehat{\boldsymbol{\alpha}}^*-\boldsymbol{\beta}_{s01})+o_p(1).
\end{split}
\end{equation*}
Further, we obtain
\begin{eqnarray*}
    \sqrt{n}(\widehat{\boldsymbol{\alpha}}^*-\boldsymbol{\beta}_{s01})(1+o_p(1))=\left[(\textbf{I}^{(1)}(\boldsymbol{\beta}_{s0}))^{-1}-o_p(1) \right]\left[\frac{1}{n^2}\dot{\ell}_n^{(1)}(\boldsymbol{\beta}_{s01 }|\widetilde{\boldsymbol{\phi}})\right]+o_p(1),
\end{eqnarray*}
simplifying it, we have 
\begin{eqnarray*}
    \sqrt{n}(\widehat{\boldsymbol{\alpha}}^*-\boldsymbol{\beta}_{s01})=(\textbf{I}^{(1)}(\boldsymbol{\beta}_{s0}))^{-1}\left[\frac{1}{n^2}\dot{\ell}_n^{(1)}(\boldsymbol{\beta}_{s01 }|\widetilde{\boldsymbol{\phi}})\right]+o_p(1).
\end{eqnarray*}
Let $\boldsymbol{\Sigma}=\textbf{I}^{(1)}(\boldsymbol{\beta}_{s0})$, then for any $\textbf{b}_n$ being a $q_n$-vector, assume $\left\lVert \textbf{b}_n\right\rVert=1$ or $\textbf{b}_n^\top \textbf{b}_n=1$, we have
\begin{equation*}
\begin{split}
    \sqrt{n}\textbf{b}_n^\top \boldsymbol{\Sigma}^{-\frac{1}{2}}(\widehat{\boldsymbol{\alpha}}^*-\boldsymbol{\beta}_{s01})&=\textbf{b}_n^\top \boldsymbol{\Sigma}^{-\frac{1}{2}}(\textbf{I}^{(1)}(\boldsymbol{\beta}_{s0}))^{-1}\left[\frac{1}{n^2}\dot{\ell}_n^{(1)}(\boldsymbol{\beta}_{s01 }|\widetilde{\boldsymbol{\phi}})\right]+o_p(1)\\
    &=\textbf{b}_n^\top (\textbf{I}^{(1)}(\boldsymbol{\beta}_{s0}))^{-\frac{1}{2}}\left[\frac{1}{n^2}\dot{\ell}_n^{(1)}(\boldsymbol{\beta}_{s01 }|\widetilde{\boldsymbol{\phi}})\right]+o_p(1).
\end{split}
\end{equation*}
Since $\dot{\ell}_n^{(1)}(\boldsymbol{\beta}_{01 }|\widetilde{\boldsymbol{\phi}})$ is the partial score about $\boldsymbol{\beta}$ and can be considered as the semi-parametric efficient score (see \cite{bickel1993efficient}), we have 
\begin{equation*}
\begin{split}
    Cov\left\{ \textbf{b}_n^\top(\textbf{I}^{(1)}(\boldsymbol{\beta}_{s0}))^{-\frac{1}{2}}\left[\frac{1}{n^2}\dot{\ell}_n^{(1)}(\boldsymbol{\beta}_{s01 }|\widetilde{\boldsymbol{\phi}}) \right]\right\}
    &=\textbf{b}_n^\top(\textbf{I}^{(1)}(\boldsymbol{\beta}_{s0}))^{-\frac{1}{2}}\textbf{I}^{(1)}(\boldsymbol{\beta}_{s0})(\textbf{I}^{(1)}(\boldsymbol{\beta}_{s0}))^{-\frac{1}{2}}\textbf{b}_n\\ 
    &=\textbf{b}_n^\top \textbf{b}_n=1,
\end{split}
\end{equation*}
and therefore, by Central Limit Theorem (CLT) and Slutsky Theorem, we have
\begin{eqnarray*}
    \sqrt{n}\textbf{b}_n^\top \boldsymbol{\Sigma}^{-\frac{1}{2}}(\widehat{\boldsymbol{\alpha}}^*-\boldsymbol{\beta}_{s01})\longrightarrow N(0,1)
\end{eqnarray*}
in distribution, and equivalently, 
\begin{eqnarray*}
    \sqrt{n}\textbf{b}_n^\top \boldsymbol{\Sigma}^{-\frac{1}{2}}(\widehat{\boldsymbol{\beta}}_{s1}-\boldsymbol{\beta}_{s01})\longrightarrow N(0,1)
\end{eqnarray*}
in distribution.

The proof of Theorem 1 (iii) is complete.

%Note: our  $\boldsymbol{\Sigma}=I^{(1)}(\boldsymbol{\beta}_{0})$ is different from that in \cite{zhao2019simultaneous} which is \begin{eqnarray*}
%\boldsymbol{\Sigma}=n\left(\boldsymbol{\Omega}_n^{(1)}(\boldsymbol{\beta}_0)\right)^{-1}\textbf{I}^{(1)}(\boldsymbol{\beta}_{0})\left(\boldsymbol{\Omega}_n^{(1)}(\boldsymbol{\beta}_0)\right)^{-1}
%\end{eqnarray*}
%although the original is 
%\begin{eqnarray*}
%\boldsymbol{\Sigma}=n^2\left(\boldsymbol{\Omega}_n^{(1)}(\boldsymbol{\beta}_0)\right)^{-1}\textbf{I}^{(1)}(\boldsymbol{\beta}_{0})\left(\boldsymbol{\Omega}_n^{(1)}(\boldsymbol{\beta}_0)\right)^{-1}
%\end{eqnarray*}
%which is not correct.
%Our result is truly an oracle property. $I^{(1)}(\boldsymbol{\beta}_{0})$ can be considered as a semiparametric information matrix. 

\end{document}